\documentclass[letterpaper,11pt,leqno]{article}
\usepackage{paper}
\pdfoutput=1
\newcommand{\bib}{bibliography.bib}
\hypersetup{colorlinks, citecolor=blue, filecolor=blue, linkcolor=blue, urlcolor=blue}
\usepackage{graphicx}
\usepackage{url}
\usepackage{engord}
\usepackage{float}
\usepackage{pdflscape}
\usepackage{booktabs}
\usepackage{array}
\usepackage{pgfplots}   
\pgfplotsset{compat=1.14}
\pgfplotsset{every axis label/.append style={font=\tiny}}
\usepackage[labelsep=period]{caption} 

\usepackage{multirow} 
\usepackage{accents}
\usepackage{mathtools}
\usepackage{makecell}
\usepackage{xr}
\usepackage{adjustbox}
\usepackage{colortbl}
\usepackage{threeparttable}
\usepackage{hyperref}

\definecolor{crimsonglory}{rgb}{0.75, 0.0, 0.2}

\newcommand{\vect}[1]{\ensuremath{\mathbf{#1}}}

\newcommand{\specialcell}[2][c]{\begin{tabular}[#1]{@{}l@{}}#2\end{tabular}}
\newcommand{\ubar}[1]{\underaccent{\bar}{#1}}
\makeatletter
\renewcommand*{\@fnsymbol}[1]{\ensuremath{\ifcase#1\or *\or \dagger\or \ddagger\or
   \mathsection\or \mathparagraph\or \|\or **\or \dagger\dagger
   \or \ddagger\ddagger \else\@ctrerr\fi}}
\makeatother

\begin{document}

\title{Reputation-Driven Adoption and Avoidance of Algorithmic Decision Aids in Credence Goods Markets}

\author{Alexander Erlei* \and Lukas Meub
%
\thanks{*Corresponding author. Chair for Economic Policy and SME Research, Georg-August-University Goettingen. 
\href{mailto:alexander.erlei@wiwi.uni-goettingen.de}{alexander.erlei@wiwi.uni-goettingen.de}.\\ Lukas Meub: Institute for Small Business Economics (ifh), Georg‐August‐University Goettingen.\\
We gratefully acknowledge financial support by the Federal Ministry of Education and Research, project "Handwerk mit Zukunft (HaMiZu)", grant number 02K20D001.\\
For this article's data, programming and screenshots, please refer to the online appendix \url{https://osf.io/shcmp/?view_only=e06a7d8315c14999a59c58754dfea809}.}}

\date{ \vspace*{-1cm}}   


\begin{titlepage}
\maketitle

In credence goods markets such as health care or repair services, consumers rely on experts with superior information to adequately diagnose and treat them. Experts, however, are constrained in their diagnostic abilities, which hurts market efficiency and consumer welfare. Technological breakthroughs that substitute or complement expert judgments have the potential to alleviate consumer mistreatment. This article studies how competitive experts adopt novel diagnostic technologies when skills are heterogeneously distributed and obfuscated to consumers. We differentiate between novel technologies that increase expert abilities, and algorithmic decision aids that complement expert judgments, but do not affect an expert's personal diagnostic precision. When consumers build up beliefs about an expert's type through repeated interactions, we show that high-ability experts may strategically forego the decision aid in order to escape a pooling equilibrium by differentiating themselves from low-ability experts. Without future visits, signaling concerns cause all experts to randomize their investment choice, leading to under-utilization from low-ability experts and over-utilization from high-ability experts. Results from two online experiments support our hypotheses. High-ability experts are significantly less likely than low-ability experts to invests into an algorithmic decision aid if reputation building is possible. Otherwise, there is no difference, and experts who believe that consumers play a signaling game randomize their investment choice.

\noindent
\textbf{JEL codes.} C91, C92, C72, D82

\noindent
\textbf{Keywords.} Credence goods, expert ability, diagnostic uncertainty, technology, decision aid, experiment

\end{titlepage}

\section{Introduction}\label{s:introduction}
 
\noindent Credence goods markets are characterized by severe information asymmetries between experts and consumers. This often translates into market inefficiencies, with experts resorting to insufficient treatments or using their information advantage to overcharge consumers, e.g. for unnecessary services. Due to the ubiquity of expert services throughout important economic decision domains such as medicine, law, financial advice or accounting, these patterns are highly consequential, and very costly for both consumers and the overall economy. The theoretical literature often attributes these adverse outcomes to expert fraud. That is, given the right incentives, experts would behave ``honestly'' and treat consumers perfectly. However, in reality, a second factor that can severely inhibit the provision of high-quality services is the expert's \textit{diagnostic ability}. The information asymmetry in credence goods markets generally precludes many consumers from adequately judging the quality of a performed service ex post. This reduces their ability to effectively discipline experts, who in turn have little incentive to invest into their abilities. For example, \cite{zhi2013landscape} document an under-utilization rate of 44.8\% for 46 of the most commonly used diagnostic rests between 1997 and 2012. Simultaneously, around 30\% of health expenditures in the US are used for non-beneficial interventions \citep{brody2010medicine}. Several studies show the detrimental effect of insufficient skills on misdiagnoses as well as faulty prescription of aggressive medication such as antibiotics \citep{chan2022selection,xue2019diagnostic,currie2017diagnosing}. Field experiments in the car repair industry point to under-treatment rates of around 75\% \citep{schneider2012agency}, as well as a negative correlation between competency and low-quality services \citep{rasch2018drives}. In auditing, heterogeneity in expert ability forces consumers to pay high premiums for skilled service providers \citep{aobdia2021heterogeneity}. 

Possibly the most promising solution to these entrenched quality issues are technological breakthroughs, specifically in algorithmic decision systems. Even today, prediction models complement or substitute human judgments in many professional fields, including legal search and bail decisions \citep{kleinberg2018human}, clinical diagnosis, medical scans \citep{topol2019deep} or financial advice \citep{brenner2020robo}. Large language models such as GPT are increasingly able to assist lawyers in discovering relevant information, generate and check legal documents, or predict court outcomes \citep{alarie2018artificial}. According to a study by the Michigan Legal Help Program, clients with access to a legal advice software were equally successful as clients with an attorney when filing for divorce \citep{sandefur2020legal}. Machine learning models have been shown to substantially exceed physicians in predicting heart attacks by avoiding faulty human heuristics \citep{mullainathan2022diagnosing}, and provide improved cancer screening \citep{daysal2022economic}. However, in order to actually realize these potentially large welfare gains, experts need to adopt algorithmic decision systems at scale. If that happens, or how long it might take, is an open, but very consequential question. Humans appear to have persistent biases against algorithmic decisions, especially once they (inevitably) observe errors \citep{dietvorst2015algorithm}. There are also reputational concerns. Relying on an algorithmic decision system precludes the expert from sending a competence signal, and may well indicate the opposite \citep{arkes2007patients,dai2020conspicuous}. 

In this paper, we experimentally investigate the adoption of novel decision aids following a technological shock on verifiable credence goods markets with diagnostic uncertainty and obfuscated, heterogeneous (high vs. low) expert abilities. Our experimental setup differentiates between two different phases. In a first phase, expert abilities are exogenous, and cannot be improved through investment. Contrary to prior experiments on credence goods, experts diagnose consumers problems by completing a short prediction task. High-ability experts are able to utilize more input variables than low-ability experts. All diagnoses are uncertain. Consumers can never observe an expert's ability type. This traps high-ability experts in a pooling equilibrium with low-ability experts, precluding them from exploiting their skill advantage. The second phase simulates an external technological shock. In Experiment 1, depending on treatment, experts receive the option to increase their diagnostic precision by either (1) investing into their own abilities (\textit{Skill}) or (2) investing into an algorithmic decision aid (\textit{Algorithm}). This allows us to capture some distinct features of algorithmic decision aids, and compare them with a baseline where technological improvements are perfectly predictable and quantifiable. Technological investments are always transferred onto consumers via prices, representing e.g. many medical situations in which certain tests are not covered by insurance. We hypothesize that high-ability experts may be incentivized to forego the algorithmic decision aid, but not the skill investments, in order to signal consumers their personal ability type. The intuition is that with decision aids, an expert's personal ability type can still play a role in the consumer's choice process. Decision aids carry uncertainty, because consumers cannot perfectly predict whether an expert is gonna (correctly) utilize the system's information. Consumers also know that high-ability experts derive relatively lower benefits from the decision aid, and may therefore rather compete through lower prices. Low-ability experts, on the other hand, cannot afford to imitate non-investing high-ability experts in the presence of Bayesian consumers. This allows high-ability experts to escape the prior equilibrium, and then exploit their skill advantage. In Experiment 2, we replicate the \textit{Algorithm} condition, and add a \textit{one-shot} condition in which the second phase only lasts one round. There is no future reputation building, leading to a coordination game in which all experts with signaling concerns share the same best-response function and should therefore randomize their investment choice. As a consequence, low-ability experts under-utilize the decision aid, whereas high-ability experts over-utilize it.

Experimental results confirm the importance of obfuscated expert ability types for the costly adoption of algorithmic decision aids. In line with our predictions for Experiment 1, there are no differences in investment behavior between ability types for \textit{Skill}, whereas high-ability experts are much less likely to purchase the algorithmic decision aid than low-ability experts. Furthermore, high-ability experts who forego the algorithm exhibit different price-setting patterns from those who decide to invest, suggesting signaling behavior. Experts generally under-invest, which may be partially driven by consumer choices. Investments tend to shift the expert's price menus towards more self-serving ones that reward fraudulent undertreatment.

Experiment 2 largely replicates the results from Experiment 1, and provides evidence from incentivized beliefs elicitation that the separation between expert types is driven by high-ability experts who think that consumers interpret investments as a signal for low-ability type. Following theory, we find that these same experts tend to randomize their investment choice in \textit{one-shot}, irrespective of ability type. Overall, there are no differences between expert types for algorithm adoption in the \textit{one-shot} setup. Compared to a situation in which reputation building through future interactions is possible, this leads to higher rates of under-investment from low-ability experts and more over-investment from high-ability experts, which hurts efficiency.

To the best of our knowledge, this is the first paper to analyze obfuscated expert abilities and reputation concerns as a potential explanation for and barrier to technological adoption in professional services. It is also the first paper to analyze the importance of future reputation building in the presence of Bayesian consumers for experts' technology adoption behavior. Finally, our experiments offer empirical contributions about the efficiency of credence goods markets when experts are heterogeneous and diagnosis is uncertain, as well as the concurrent efficacy of expert investments after a technological shock.

Our paper relates to the literature on credence goods markets with diagnostic uncertainty. Most articles are theoretical, showing how price competition may negatively affect diagnostic precision when experts need to exert effort \citep{pesendorfer2003second}, that separating diagnosis and treatment can achieve first-best outcomes when consumer evaluations are subjective \citep{bester2018credence}, or that fraud-sensitive penalties can discipline experts whose diagnoses require proper incentives \citep{chen2022efficient}. In recent empirical work, \cite{balafoutas2020diagnostic} show in a laboratory experiment that diagnostic uncertainty decreases efficient service provision and consumer market entry. Insurances reduce investments into diagnostic precision, while pro-social experts invest more. There is also evidence that diagnostic uncertainty does not affect expert dishonesty or consumer trust, while reputation mechanisms remain beneficial for overall market efficiency \citep{tracy2023uncertainty}.

A second relevant string is the literature on heterogeneous expert abilities on credence goods markets. Prior theoretical work shows that efficient equilibria are always possible with heterogeneous, obfuscated expert abilities, but inefficient ones also exist \citep{liu2020role}. Incentives to invest into expert abilities may break down in the presence of discounters \citep{dulleck2009experts}, and heterogeneous cost functions generally prevent first-best solutions \citep{hilger2016don,frankel2014experts}. \cite{schneider2017effects} find that in the presence of low-ability experts who always misdiagnose consumers without exerting effort, a sufficient number of high-ability experts allows for a second-best equilibrium without policy intervention. Empirically, they show with a laboratory experiment that high-ability experts invest less than theoretically predicted, that the market is more efficient than predicted, and that price competition hurts overall welfare \citep{schneider2017expert}.

Third and finally, this article contributes to the literature on decision aids in professional contexts. Especially in the medical literature, it is a widespread sentiment that experts who strongly rely on complementary decision expertise e.g., through laboratory testing, tend to be less competent \citep{schroeder1974variation,daniels1977variation,yeh2014clinician,hall2019doctors,miyakis2006factors,doi2021perception,groopman2007doctors}. There is evidence that both patients and peers derogate the diagnostic ability of physicians who rely on diagnostic aids \citep{cruickshank1985patient,arkes2007patients,promberger2006patients,shaffer2013patients,wolf2014students}. Patients have formulated preferences for physicians using intuition-based diagnoses versus statistical models \citep{eastwood2012people} and general aversions towards statistical methods \citep{dawes1989clinical}. While we do not measure consumers preferences or attitudes towards decision aids, we do study how experts may strategically avoid algorithmic support systems in order to influence consumer beliefs about the expert's ability type. Rather than projecting a general distrust of technology, we thereby explore another possibility why the adoption of algorithmic systems may lag behind the optimal investment path. Skilled experts use the appearance of a novel diagnostic tool to reduce consumer uncertainty about their ability and thereby differentiate themselves from low-ability experts. Depending on the nature of consumer-expert interactions as well as the possibility of reputation building, this may either lead to pooling equilibria without technology adoption, separation between expert types, or a randomization of adoption behavior.

\section{The Market 
\label{sec:model}}

We study a credence goods market with competing experts who are not liable for offering insufficient treatments. Consumers know that they have a big problem with probability \textit{h} or a small problem with probability (1-\textit{h}). Once consumers approach an expert, they are committed to receive the recommended treatment under the offered price, meaning there is no separation of diagnosis and treatment, and consumers cannot consult other information sources. Afterwards, consumers can verify the treatment and observe their payoff. If the problem is solved, consumers receive a payoff of \textit{v}, if it is not solved, they earn nothing. We assume that indifferent consumers choose to visit the expert. Similarly, experts who are indifferent between cheating and treating the consumer honestly always choose the correct treatment. There are two commonly known expert types, high-ability (H, with probability $\gamma$) and low-ability (L, with probability $1-\gamma$) experts, who differ in the probability to perform a costless correct diagnosis: $z \in [0.75, 1]$ vs. $q \in [0.5, 0.75]$. Let $i \in \{H, L\}$ and $k \in \{z, q\}$. Experts always charge for the implemented treatment (verifiability), and a high quality treatment HQT (low quality treatment LQT) treatment induces costs of $\bar{c}$ ($\ubar{c}$), with $\bar{c} > \ubar{c}$. The HQT solves both problems, the LQT only solves the small problem. Expert type ability probabilities are common knowledge. Consumers therefore consider two possible types of misdiagnosis: diagnosing a major problem when they actually have a minor problem, and diagnosing a minor problem in case of a major problem. Consumers cannot observe an expert's ability type. They can, however, identify each individual expert, allowing for reputation. It is by assumption always profitable for the expert to treat the consumer because $v > \bar{c}$. Experts choose a price menu $\vect{P} = (\bar{p}, \ubar{p})$ after learning about their type, with $\bar{p} \text{\,(HQT)} \geq \ubar{p}$. Consumers and experts are assumed to be risk-neutral. \\

\noindent
\textbf{Investments.} After $n$ rounds, experts can choose to increase their diagnostic precision $k$ for a fixed fee $d$. Investments automatically increase $\vect{P}$ by $d$. We assume that the low-ability expert matches the high-ability expert's skill by investing in the new technology. Thus, the low-ability expert improves their diagnostic precision by an additional $z - q$ for any chosen precision level. There are two reasons for that. First, in many cases, low-ability experts presumably make mistakes that are comparatively easy to eliminate. This also implies low costs. Second, investing specifically in tangible new technology like an algorithmic decision system does not only complement the expert's diagnosis, but often serves as an imperfect substitute for it \citep{doyle2010returns,dai2020conspicuous}. Assuming that the low-ability expert's mistakes are relatively easy to fix, we'd expect algorithmic decision aids to mostly eliminate them.\footnote{Even if the assumption that low-ability experts completely catch up to high-ability experts might be too strong for many real-life cases, it is almost certainly true that the marginal benefits of algorithmic systems will on average be larger for low-ability experts.}

\section{Experiment 1 -- Experimental Design
\label{sec:design}}

We conduct two treatments of a pre-registered between-subject (\textit{Skill} vs. \textit{Algorithm}) online group experiment where three experts compete over three clients. The experimental parameters are fixed across treatments. Consumers have a big problem with probability $h = 0.4$, receive $v = 150$ if their problem is solved, and earn $\sigma = 15$ when choosing the outside option. The cost of providing the LQT is $\ubar{c} = 20$, the cost for the HQT is $\bar{c} = 60$. Experts choose between three price vectors. The price for the HQT is $\bar{p} = 100$ and fixed. For $\vect{P}^m$, we set $\ubar{p} = 40$; for $\vect{P}^e$, we set $\ubar{p} = 60$; and for $\vect{P}^s$, we set $\ubar{p} = 80$. Experts always have a markup of 40 for the HQT, and markups of 20/40/60 for the LQT when choosing $\vect{P}^m/\vect{P}^e/\vect{P}^s$. In each group, there are 2 low-ability experts and 1 high-ability expert. Low-ability (High-ability) experts receive the correct signal with a probability of 50\% (75\%). Investing into additional diagnostic precision costs $d = 10$ and increases an expert's diagnostic precision to $k = 0.9$. Importantly, investments also automatically increase $\ubar{p}$ and $\bar{p}$ by 10 Coins, meaning that consumers pay for improved diagnoses. We apply the following conversion rate: 100 Coins = 60 cents. Data from 300 participants (40\% female) was collected on Amazon Mechanical Turk (MTurk) using oTree \citep{chen2016otree} and CloudResearch \citep{litman2017turkprime}. All participants are vetted for quality by CloudResearch, completed at least 50 prior tasks on MTurk with a minimal approval rating of 90\%, and reside in the US. The base payment was \$4.50.

\subsection{Procedure}

Subjects play 25 rounds in groups of six. The first 10 rounds are constant across treatments and always follow the same sequence. Experts choose their price vector, and then diagnose all three consumers by completing a short prediction task. Here, high-ability experts use 3 out of 5 possible input factors, and low-ability experts use 2 input factors to identify the problem. Input numbers are pre-filled for each expert, who automatically make a prediction by clicking on a "diagnose" button. Then, experts receive a diagnostic signal that depends on their ability type. Finally, experts choose for each consumer whether they want to implement the HQT or the LQT, and proceed to a summary screen that shows how many consumers approached them and how much money they made that round. Consumers first decide whether they want to leave the market, or choose one of the three experts. Consumers do not observe an expert's ability level, but they can identify each expert, which allows for reputation building. After choosing an expert or leaving the market, consumers wait to be treated and proceed to a summary screen that shows which expert they approached in this round (as well as in all previous rounds), what prices the expert chose, and how much money they earned.

\begin{figure}[t]
    \centering
    
\begin{tikzpicture}
\draw[->] (-1,0) -- (14,0);

\foreach \x in {0,4,8,12}
\draw (\x cm,15pt) -- (\x cm,1pt);

\foreach \x in {2,6,10,13.5}
\draw (\x cm,-1pt) -- (\x cm,-15pt);

\draw (0,0) node[above=18pt, align = center] {Nature draws \\problem \textcolor{crimsonglory}{$\textbf{h}$}};
\draw (2,0) node[below=18pt, align = center] {Experts make\\ investment decision\\ \textcolor{crimsonglory}{$\textbf{d}$}};
\draw (4,0) node[above=18pt, align = center] {Experts set\\ prices \textcolor{crimsonglory}{$\vect{P}$}};
\draw (6,0) node[below=18pt, align = center] {Consumers observe\\ \textcolor{crimsonglory}{$\textbf{d}$} and \textcolor{crimsonglory}{$\vect{P}$} };
\draw (8,0) node[above=18pt, align = center] {Consumers choose\\ Expert or $\sigma$};
\draw (10,0) node[below=18pt, align = center] {Experts receive\\ diagnostic signal \textcolor{crimsonglory}{$\textbf{k}$} };
\draw (12,0) node[above=18pt, align = center] {Experts choose\\ HQT or LQT};
\draw (13.5,0) node[below=18pt, align = center] {Payoffs\\ are realized };
\end{tikzpicture}

    \caption{Timing of events in each round for \textit{Phase 2}. In \textit{Phase 1}, experts skip the investment decision.}
    \label{fig:procedure}
\end{figure}

After 10 rounds, all subjects are informed that experts now have the opportunity to invest into their diagnostic precision (Figure \ref{fig:procedure}). There are two treatments. In \textit{Skill}, subjects learn that experts can pay 10 Coins each round to increase their diagnostic precision to 90\%. In \textit{Algorithm}, experts can pay 10 Coins each round to rent an algorithmic decision aid that increases the expert's maximum diagnostic precision to 90\% if used correctly. For the algorithmic decision aid, consumers also learn that experts are not forced to use the system, but can choose to ignore it. All subjects know that consumers pay for the investment by automatically paying 10 Coins more per treatment if they choose to approach an investing expert. Then, subjects complete another 15 rounds. Experts first decide whether they want to invest, then choose their price vector, and proceed to diagnosis and treatment. During the diagnosis, investments allow experts to utilize four input numbers, and those in \textit{Algorithm} can forego the decision aid with the click of a button. Consumers observe each expert's investment decision. Otherwise, nothing changes. Upon completing the credence goods experiment, subjects proceed to a short post-experimental questionnaire and answer a battery of demographic questions as well as a question about their risk attitudes \citep{dohmen2011individual}.

\section{Predictions}

\subsection{Phase 1 Equilibria}
To derive equilibria, we differentiate between the three possible markup scenarios: (i) $\bar{p} - \bar{c} > \ubar{p} - \ubar{c}$; (ii) $\bar{p} - \bar{c} < \ubar{p} - \ubar{c}$; (iii) $\bar{p} - \bar{c} = \ubar{p} - \ubar{c}$. First note that in any equilibrium where the expert chooses $\vect{P}^s$ or $\vect{P}^m$ and consequently always implements only the high quality or low quality treatment, diagnostic precision does not matter and prices as well as profits are the same as in the model with transparent expert abilities. That is because consumers anticipate experts to follow their monetary self-interest. Under $\vect{P^e}$, experts signal honesty, because their markups for both treatments are the same. We first derive consumer profits under $\vect{P}^s$ and $\vect{P}^m$.

When (i) the profit margin for the HQT is larger than for the LQT, all expert types always recommend the HQT. Therefore, consumer problems are always solved. Profits are
\begin{equation}
    \pi^m = v - \bar{p}.
\end{equation}

When (ii) the profit margin for the LQT is larger than for the HQT, experts always recommend the low quality treatment, irrespective of ability. Consumers infer the markups from prices and anticipate undertreatment. Profits are
\begin{equation}
\pi^s = (1-h)v - \ubar{p}.
\end{equation}

Finally, there are equal markup scenarios. Here, heterogeneous expert abilities affect consumer payoffs, because lower diagnostic precision leads to more mistakes when treating consumers honestly. Under transparency, consumers anticipate different problems depending on expert type: 
\begin{equation}
    \pi^e = (1-h+hk)v - \bar{p} + (h+k-2hk) \Delta p
\end{equation}

Which markup scenario do experts choose? With perfect diagnosis and full transparency, \cite{dulleck2006doctors} show that under verifiability, experts always efficiently serve consumers by setting equal markups. With diagnostic uncertainty, that is no longer viable, because credible commitment to honesty does not imply that a consumer is sufficiently treated. Under competition, experts choose prices that maximize expected consumer income. Hence, the boundaries determining experts price strategy are:

\begin{equation}
    \pi^m \geq \pi^e \,\text{when}\, h \geq \frac{k \Delta p}{(1 - k)v - (1 - 2k) \Delta p} \coloneqq h^m_i.
\end{equation}

and

\begin{equation}
    \pi^s \geq \pi^e \,\text{when}\, h \leq \frac{(1 - k) \Delta c}{kv + (1 - 2k) \Delta p} \coloneqq h^s_i.
\end{equation}

Under full obfuscation, however, consumers do not know whether to condition their expected payoff on $z$ or $q$. Therefore, they need to incorporate both possibilities, and profits change to

\begin{equation}
    \pi^e = (1 - h + hq (1 - \gamma) + \gamma hz)v - \bar{p} + (h - 2hq(1 - \gamma) - \gamma 2hz + q(1 - \gamma) + \gamma z) \Delta p.   
\end{equation}

The boundaries determining experts price strategy are:

\begin{equation}
    \pi^m \leq \pi^e \,\text{when}\, h \leq \frac{\Delta p(q - \gamma q + \gamma z)}{v(1 - q(1 - \gamma) - \gamma z) - \Delta p(1 - 2q(1 - \gamma) - \gamma 2z)} \coloneqq h^m_o.
\end{equation}

and

\begin{equation}
    \pi^s \leq \pi^e \,\text{when}\, h \geq \frac{\Delta p(q - \gamma q + \gamma z -1)}{v(q(\gamma - 1) - \gamma z) - \Delta p(1 - 2q(1 - \gamma) - \gamma 2z)} \coloneqq h^s_o.
\end{equation}

Note that in comparison to the transparent scenario, expected consumer profits from approaching the high-ability expert decrease. This makes a straightforward argument why high-ability experts are incentivized to signal their true ability type. 

We now turn towards three possible equilibria. Without reputation or repeated interactions, given full obfuscation, both expert types always charge the same prices and earn the same profits in equilibrium (see also \citealp{liu2020role}). If the high-ability expert would post a price vector that increases expected consumer profits, the low-ability expert could simply copy it. This works as long as obfuscation is strong enough such that consumers are sufficiently unlikely to believe that an expert following the high-ability expert's optimal pricing path is, in fact, a high-ability expert. Furthermore, in any equilibrium where experts choose $\vect{P}^s$ or $\vect{P}^m$ and consequently always implement only the high quality or low quality treatment, diagnostic precision never matters, irrespective of reputation. Hence, from the expert's perspective, their personal ability type is inconsequential.

In a first equilibrium, experts may always choose the HQT and charge $\vect{P}^m$, irrespective of type. This is the equilibrium we focus on in the experiment. Consumer beliefs $\mu(\vect{P})$ that the expert is a high ability expert must be low enough. Otherwise, the high-ability expert would be able to exploit their skill advantage by committing to treat consumers truthfully and offer higher profits. Hence, the expected loss from inadvertently approaching the low-ability expert who charges $\vect{P}^e$ (as compared to approaching an expert who charges $\vect{P}^m$) must be greater than or equal to the expected benefit from approaching the high-ability expert who charges $\vect{P}^e$.\footnote{This only works for values of $h$ and $z$ under which the transparent high-ability expert would be able to charge $\vect{P}^e$. Otherwise, signalling better diagnostic precision has no impact on the expert's price setting.} Then, consumer beliefs that the expert is a high-ability expert do not exceed:
\begin{equation}
    \tilde{\gamma}^m \coloneqq \frac{vh(1-q) - (h-2hq+q) \Delta p}{(z-q) (vh - 2h \Delta p + \Delta p)}.
\end{equation}
\textbf{Equilibrium 1.} Both expert types charge $\vect{P}^m$, always perform the major treatment, consumers believe $\mu(\vect{P}) \in [0, \tilde{\gamma^m}]$ if the expert charges $\vect{P}^e$, meaning $\ubar{p}-\ubar{c} = \bar{p} - \bar{c}$, and consumers always visit the expert. Note that this also implies that the high-ability expert \textit{could} attract more consumers and thus increase profits in a scenario with only partial obfuscation by increasing consumer beliefs such that $\tilde{\gamma} > \tilde{\gamma}^m$ for $h \in [h^m_o,h_{H}^m]$.\\

\textbf{Equilibrium 2.} The analogous equilibrium for the case where experts always choose the minor treatment implies:

\begin{equation}
    \tilde{\gamma}^s \coloneqq \frac{-vhq - (h-2hq+q-1) \Delta p}{(z-q) (vh - 2h \Delta p + \Delta p)}.    
\end{equation}
 
 Both expert types charge $\vect{P}^s$, always perform the minor treatment, consumers believe $\mu(\vect{P}) \in [0, \tilde{\gamma^s}]$ if the expert charges $\vect{P}^e$, meaning $\ubar{p}-\ubar{c} = \bar{p} - \bar{c}$, and consumers always visit the expert. The high-ability expert could improve their profits in a scenario with only partial obfuscation by changing consumer beliefs such that $\tilde{\gamma} > \tilde{\gamma}^s$ for $h \in [h^s_o,h_{H}^s]$.\\

\begin{figure}[!t]
    \centering
\includegraphics[width=0.45\textwidth]{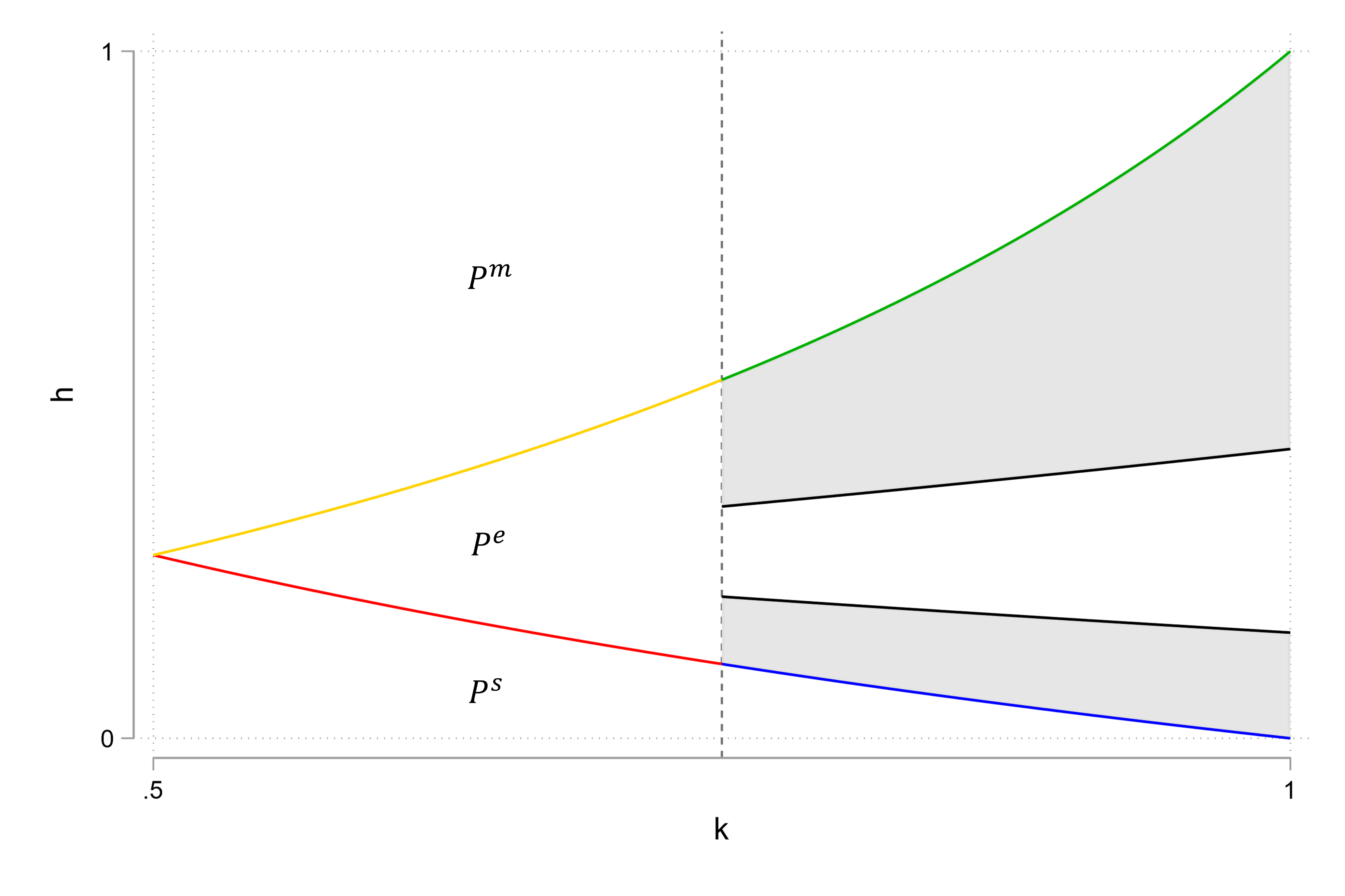}
\includegraphics[width=0.47\textwidth]{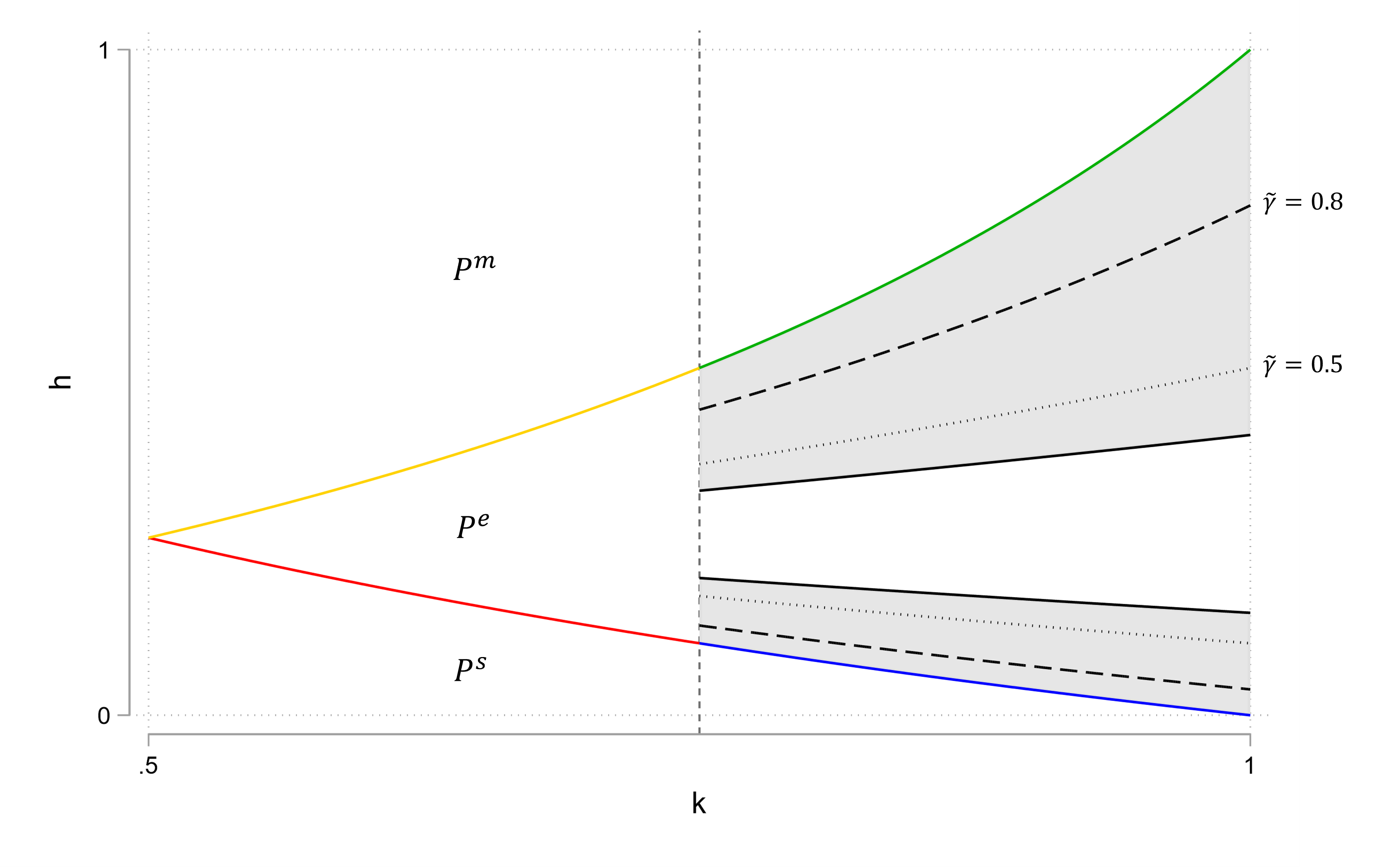}
    \caption{Left: Price setting under fully transparent and fully obfuscated expert abilities. The shaded areas represents the high-ability expert's loss. Due to consumer uncertainty, the high-ability expert cannot choose the profit-maximizing price vector $\vect{P}^e$ as often as they would want to. Variables for this figure: $q = 0.5$, $\gamma = 1/3$. Right: Variations for different consumer beliefs $\tilde{\gamma}$.}
    \label{fig:prices_obfus_p1}
\end{figure}

 \textbf{Equilibrium 3.} Finally, there are equal mark-up equilibria. We briefly describe them for completeness, but do not analyze them empirically. Both expert types charge $\vect{P}^e$ and without reputation or repeated interactions, consumers believe $\mu(\vect{P}) \in [0, \tilde{\gamma}$] because abilities are fully obfuscated and beliefs are fully determined by nature. Consumers always visit the expert. Contrary to the other two equilibria, treatment behavior is not obvious. We know that for $h \in [h_{L}^s, h_{L}^m]$, both expert types always follow their own diagnosis. However, within the gray area (see Figure \ref{fig:prices_obfus_p1}) as well as at any point between $h_{L}$ and $h_{H}$, the incentives of the two expert types to follow their diagnosis differ. That is because the high-ability expert may have options to deliver better treatments for the consumer, whereas the low-ability expert, knowing about their small diagnostic precision, could be incentivized to completely rely on one treatment and thereby minimize diagnostic mistakes. For the consumer, treatment choices do not make a difference, because they cannot infer anything from prices or experience.

 If consumers repeatedly approach identifiable experts, their beliefs about their expert's type $\mu(\vect{P})$ change over time. That is because under $\vect{P}^e$, consumers expect honest expert behavior. On a verifiable credence goods market, this implies an outcome sequence with three distinct observations: undertreatment ($-\ubar{p}$), efficient treatment of the small problem ($V - \ubar{p}$), and overtreatment or the efficient treatment of the big problem ($V - \bar{p}$). The difference in diagnostic precision between expert types precludes low-ability experts from offering the same sequence of outcomes as high-ability experts do. Therefore, in the long run, consumers who use Bayes rule to update their beliefs will necessarily identify their expert's ability type. Low-ability experts cannot deviate from charging $\vect{P}^e$ because consumers know that $\vect{P}^e$ is optimal and will always opt for such an expert. Hence, while low-ability experts may be able to compete with high-ability experts for a limited number of interactions, in the long run, expert type becomes transparent. 

  The consequences of consumer belief updating depend on the low-ability expert's diagnostic precision, the market's reputation mechanism, the number of competitors, and consumer beliefs about competitor abilities. If there is a sufficiently accurate and complete public reputation system, all consumers observe each expert's ability type over time and consequently opt for high-ability experts. Therefore, consumers always visit an expert, consumer surplus increases compared to equilibria without reputation, and low-ability experts cannot compete. 
  
  If reputation is the private information of the consumer, there are two different cases. One, $h \in [h^s_L, h^m_L]$ and the low-ability expert maximizes expected consumer income by choosing $\vect{P}^e$ and committing to honesty. Here, the consumer always switches away from any expert $i$ to another expert $j$ as long as $\mu_i(\vect{P}^e) < \mu_j(\vect{P}^e)$. There are no opportunity costs to switching, since all experts always follow the same strategy. Due to costless exploration, consumers necessarily identify high-ability experts in the long run, and therefore achieve the same surplus as in the public reputation case.
  
  Second, $h \in [h^m_L, h^m_o] \vee h \in [h^s_o, h^s_L]$, and the low-ability expert maximizes expected consumer income by choosing $\vect{P}^m$ or $\vect{P}^s$. Here, the low-ability expert imitates the high-ability expert ($\vect{P}^e$) at the expense of consumer welfare. If low-ability experts maximize consumer surplus by always choosing the HQT or the LQT, consumers face no switching costs because all low-ability experts offer them the same expected value. Hence, consumers always switch away from expert $i$ when $\mu_i(\vect{P}^e) < \mu_j(\vect{P}^e)$ until they find a high-ability expert. This incentivizes low-ability experts to (i) first imitate high-ability expert treatment behavior and then (ii) switch towards their personal optimal treatment strategy once the Bayesian consumer's beliefs $\mu(\vect{P})$ are such that they would prefer the expert to always prescribe the HQT or the LQT, because they are certain enough that the expert is not a high-ability expert. In that situation, low-ability experts make switching costly by allowing consumers to ``discipline'' them over time. Switching away from a disciplined low-ability expert is costly because the consumer may switch towards an undisciplined low-ability expert who imitates the high-ability expert's treatment behavior by following their diagnostic signal, and thereby reduces the consumer's expected income. Therefore, it may be beneficial for the consumer to stay with the disciplined low-ability expert, depending on (1) the likelihood of switching to an undisciplined low-ability expert, (2) the number of future interactions (and thus the value of exploration), and (3) consumer aversions towards being lied to.

 The three equilibria types show that high-ability experts are trapped in a pooling equilibrium for $h \in [h^s_H, h^s_o] \vee h \in [h^m_o, h^m_H]$. When $h \in [h^s_o,\, h^m_o]$, identifiable high-ability experts out-compete low-ability experts in the long-run. However, for $h \in [h^m_L, h^m_o] \vee h \in [h^s_o, h^s_L]$, low-ability experts may retain some consumers by making switching costly. Low-ability experts always imitate high-ability expert prices, because they do not attract any consumer otherwise. Regarding price setting, we propose:
 
 \textbf{Proposition 1.} \textit{When expert abilities are fully obfuscated, the following scenarios arise:}

\begin{equation}
    \left\{\begin{array}{lr}
        \vect{P}^s & \,\text{if}\, h \in [0, h^s_o],\\
        \vect{P}^e & \,\text{if}\, h \in [h^s_o,\, h^m_o],\\
        \vect{P}^m & \, \text{if}\, h \in [h^m_o, 1]
        \end{array}\right\}
\end{equation}\\

\subsection{Phase 1 Hypotheses}
We can now insert the experimental parameters to derive our hypothesis. In Phase 1, consumers believe that an expert is of high ability type with $\gamma = 1/3$. Expected consumer income equals $\pi^m = 50 $ under $\vect{P}^m$, $\pi^s = 10$ under $\vect{P}^s$, and $\pi^e = 45.67$ under $\vect{P}^e$. This follows from consumers observing the mark-up incentives for experts and adjusting their expectations accordingly. Therefore, experts always charge $\vect{P}^m$ in equilibrium irrespective of type, with $ h = 0.4 > h^m_o = 0.34$ and $\gamma = 1/3 < \tilde{\gamma}^m = 0.59$. There is no incentive to deviate from implementing the high-quality treatment, meaning experts never under-treat clients, who therefore always enter the market. Experts over-treat clients with a probability of $1 - h = 0.6$. High-ability experts cannot differentiate themselves from low-ability experts, and price-setting therefore does not differ between types. 

\medskip
\textbf{Hypothesis 1:} Experts maximize expected consumer income by choosing $\vect{P}^m$ irrespective of type.

\medskip
\textbf{Hypothesis 2:} There is no difference between high-ability expert and low-ability expert behavior.

\medskip
\textbf{Hypothesis 3:} Consumers always enter the market.

\subsection{Phase 2 Hypotheses: Skill}
In \textit{Skill}, experts can increase their diagnostic precision to $k = 0.9$ by paying $d = 10$ before choosing prices. Consumers observe whether an expert invested. Hence, by investing, an expert's prior personal ability type becomes obsolete, and all investing experts are the same type. Expected client income from approaching an expert who invested into their diagnostic precision equals $\pi_{inv}^e = 57$. Comparing $\pi_{inv}^e$ to $\pi^m$, $\pi^s$ and $\pi^e$ reveals a straightforward incentive for experts to invest. Under full transparency, the low ability experts offers consumers an expected income of $40$ under $\vect{P^e}$, and the high-ability experts offers an expected income of $57$. Hence, even if consumers infer something about the expert's type when they invest, no expert type can out-compete an investing expert. At most, consumers are indifferent between approaching the investing and the not-investing transparent high-ability expert, when $\tilde{\gamma} = 1$. Otherwise, they prefer any expert who invested, regardless of their prior ability type. Assuming that high-ability experts cannot signal their ability type with certainty by foregoing investments, consumer beliefs about an expert's type are inconsequential, and we propose:

\medskip
\textbf{Hypothesis 4:} Low-ability experts in \textit{Skill} always invest to improve their diagnostic ability.
\medskip

\textbf{Hypothesis 5:} Low-ability experts in \textit{Skill} maximize expected consumer income by choosing $\vect{P}^e$.
\medskip

\textbf{Hypothesis 6:} High-ability experts in \textit{Skill} always invest to improve their diagnostic ability.
\medskip

\textbf{Hypothesis 7:} High-ability experts in \textit{Skill} maximize expected consumer income by choosing $\vect{P}^e$.
\medskip

\subsection{Phase 2 Hypotheses: Algorithm}
In \textit{Algorithm}, experts can rent an algorithmic decision aid that increases their diagnostic precision to $k = 0.9$ by paying $d = 10$ before choosing prices. Compared to \textit{Skill}, this setup exhibits some crucial differences. First, consumers know that experts can choose to forego the decision aid's information even after paying for the algorithm. Second, because consumers do not have insight into the expert's diagnostic process, they are uncertain about the efficacy of the algorithm's information. Hence, consumers still consider an expert's personal precision when anticipating their expected income. We argue that this allows high-ability experts to differentiate themselves from low-ability experts, and escape the pooling equilibrium derived earlier.

Looking at expert incentives, first note that once uncertainty is reduced and consumer beliefs exceed certain thresholds $\tilde{\gamma}$, pooling equilibria break down because high-ability experts can, e.g., change their pricing strategy in return for their increased estimated diagnostic precision. The same is true for the low-ability expert, but reversed. Lower uncertainty decreases their potential profits. Thus, as long as there is uncertainty, high-ability experts experience a penalty, whereas low-ability experts can charge more. Figure \ref{fig:prices_obfus_p1} illustrates how different levels of $\tilde{\gamma}$ change price setting. Without investments, high-ability experts have obvious incentives to persuade consumers of their diagnostic type.

With investments, we have shown above that a transparent high-ability expert would be indifferent between investing and not investing because in both cases, they offer consumers an expected income of $\pi^e = 57$ after choosing $\vect{P^e}$. Introducing the aforementioned uncertainty around decision aids changes that. Because consumers cannot be certain that experts utilize the algorithmic decision aid (i) at all or (ii) effectively, expected consumer income after investing changes to $\pi_{alg,inv}^e \leq 57$. Then, depending on the level of uncertainty as well as consumer beliefs about an expert's type, it may be profitable for high-ability experts to forego investments, if they can plausibly signal their type to consumers who judge the benefits of the decision aid as uncertain. Yet, if that is the case, low-ability experts may also have incentives to once again imitate high-ability experts, which is initially hidden to consumers because of obfuscated expert abilities. Note, however, that in order to imitate high-ability experts, low-ability experts now also rely on $\vect{P^e}$. In a one-shot game, this is irrelevant, because there is no consumer belief formation. In \textit{Phase 1}, it is irrelevant because the pooling equilibrium implies $\vect{P^m}$, which masks the low-ability expert's abilities. In \textit{Phase 2}, however, imitating low-ability experts cannot hide their relative diagnostic imprecision due to consumer expectations of being treated honestly (the high-ability expert's strategy). Therefore, we now have to ask whether it is profitable for low-ability experts to imitate high-ability experts despite Bayesian consumers who update their beliefs.

First, to allow high-ability experts to try and differentiate themselves from low-ability experts by foregoing the algorithm, and thereby to allow low-ability experts to imitate them by foregoing the algorithm, consumers need to believe that an expert who does not invest in round 11 is of high-ability type with $\mu(\vect{P}) \in [\tilde{\gamma}^m,1]$ where $\tilde{\gamma}^m = 0.59$. Consumers then use Bayes's rule to update their beliefs $\mu(\vect{P})$ each round. There are three possible observations: (a) undertreatment ($-\ubar{p}$ Coins), (b) overtreatment or efficient treatment of big problem (50 Coins), and (c) efficient treatment of small problem (90 Coins). An honest high-ability (low-ability) expert has respective observation probabilities of 0.1 (0.2), 0.45 (0.5) and 0.45 (0.3). Let $n$ be the number of relevant a-signals, $m$ be the number of relevant b-signals, and $o$ be the number of relevant c-signals. Approaching a low-ability expert is an L-event, approaching a high-ability expert is a H-event. Given any sample of observations, consumers calculate the posterior probability:
\begin{equation}
    Pr(L | n, m, o) = \frac{Pr(n,m,o|L)Pr(L)}{Pr(n,m,o|L)Pr(L) + Pr(n,m,o|H)Pr(H)}.
\end{equation}

\begin{figure}[htpb!]
    \centering
\includegraphics[width=0.9\textwidth]{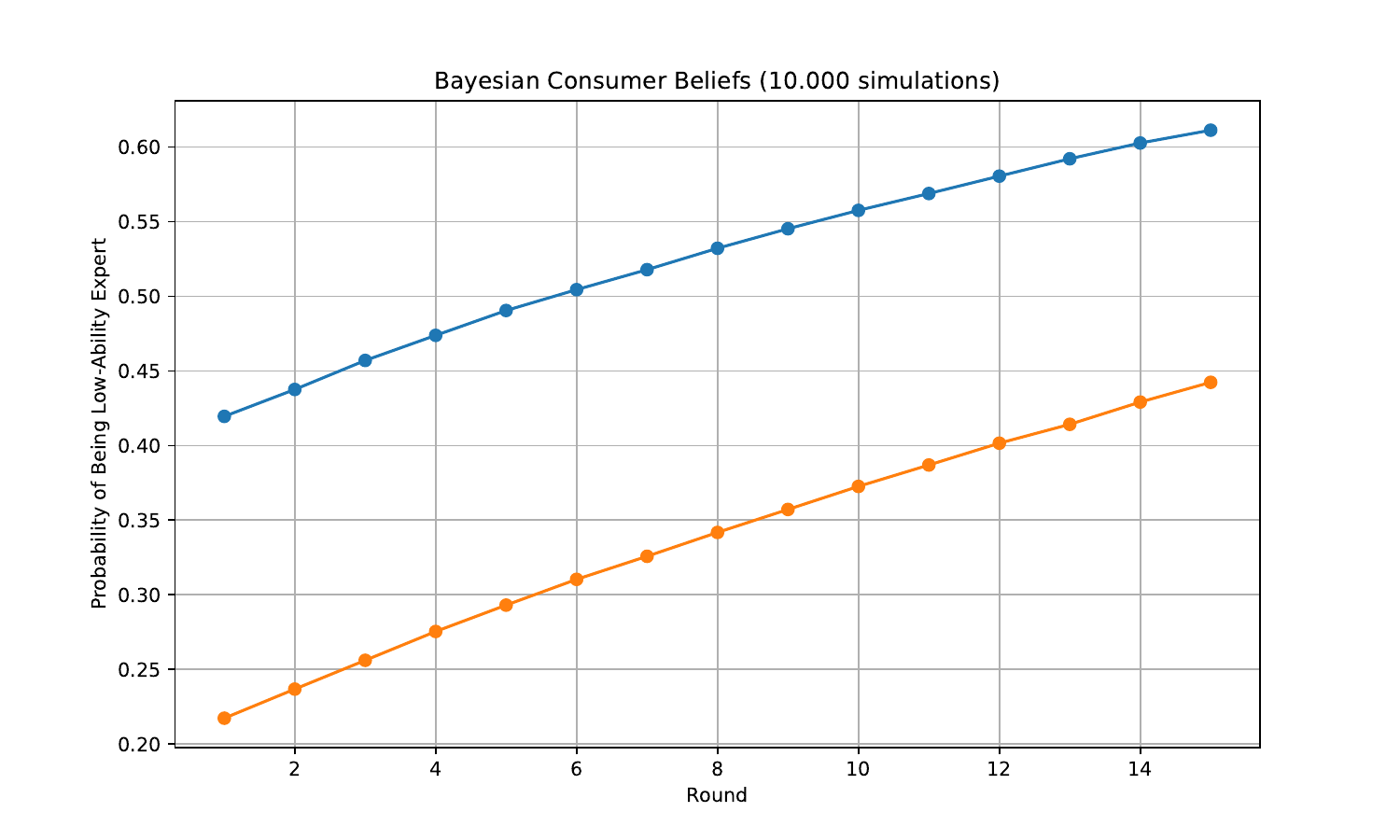}
    \caption{Consumer Belief Updating with $Pr(L) = 0.4$ and $Pr(L) = 0.2$.}
    \label{fig:bay_belief}
\end{figure}

$Pr(L)$ is the prior probability for event L that is determined by consumer beliefs at the beginning of round 11 $\mu(\vect{P}) \in [\tilde{\gamma}^m,1]$. Using this, we can simulate a probability distribution of consumer beliefs. To demonstrate, let $\mu(\vect{P}) = \tilde{\gamma} \in \{0.6, 0.8\}$. That is, consumers believe that their expert has high abilities with a probability of 60\% or 80\%. At this point, consumers would rather approach a not-investing expert with $\vect{P^e}$ under full obfuscation than a $\vect{P^m}$-expert who always successfully treats the problem because $\tilde{\gamma} > \tilde{\gamma}^m$. Figure \ref{fig:bay_belief} shows average consumer beliefs that a representative expert is of low-ability type over time, using 10.000 simulations. Here, we assume that low-ability experts fully imitate high-ability experts by choosing $\vect{P^e}$ and following their diagnosis.

It is clear that consumer beliefs change substantially over time, which constraints the imitation of high-ability experts.\footnote{On the equilibrium path, consumers do not switch away from an expert in Phase 1 because there is no undertreatment, and expect some undertreatment in Phase 2 as long as it is optimal to approach a $\vect{P^e}$-expert. Therefore, consumers gather repeated observations from an individual expert.} A second possibility is that low-ability experts imitate high-ability experts in price setting and investment behavior, but always choose the HQT in order to somewhat compensate their diagnostic disadvantage. As shown in Figure \ref{fig:bay_belief_hqt} in the appendix, this strategy very quickly reveals the expert's ability type. Having demonstrated the limits for low-ability experts to imitate high-ability experts, we now turn towards experts' optimal strategies and the conditions under which high-ability experts can escape the pooling equilibrium by signaling their type.

First, when does the high-ability expert deliberately forego the decision aid? Let $w \in [0, 1]$ be the estimated probability from the consumer that an investing expert does not (correctly) utilize the decision aid after renting it. Then, expected consumer profit equals $\pi_{alg, inv}^e = (1-w)*\pi^e_{inv} + w*(\tilde{\gamma} \pi^e_{ninv,ha} + (1-\tilde{\gamma})\pi_{ninv,la} - d) = (1 - w)*57 + w*(30 + \tilde{\gamma}*17)$. In the event of $w$, consumers pay an additional $d$ without gaining diagnostic improvements. Because experts invest, consumers receive an information signal about the expert's type, which is incorporated in $\tilde{\gamma}^{inv} \in [0, \frac{1}{3}]$. If the expert does not invest, they influence consumer beliefs $\tilde{\gamma}^{ninv} \in [\frac{1}{3}, 1]$. Expected consumer profits are $\pi_{alg, ninv}^e = 40 + 17*\tilde{\gamma}$. High-ability experts forego the decision aid along:
\begin{equation}
    \pi_{alg, ninv}^e \geq \pi_{alg, inv}^e \, \text{when} \, w \geq \frac{17(1-\tilde{\gamma}^{ninv})}{27 + 17 \tilde{\gamma}^{ninv}} \coloneqq w_{ninv}, \, \text{or}
\end{equation}
    
\begin{equation}
  \pi_{alg, ninv}^e \geq \pi_{alg, inv}^e \, \text{when} \, \tilde{\gamma} \geq \frac{17 - 27w}{17(1+w)} \coloneqq \tilde{\gamma}_{ninv}   
\end{equation}

If $\tilde{\gamma}^{ninv} \geq \tilde{\gamma}_{ninv}$ after one of the three experts chose to forego the investment, then signaling is a potential strategy for experts. The incentive is c.p. substantially larger for the high-ability expert, because they do not profit from the investment under equilibrium prices $\vect{P}^e$, and do not have to worry about consumer learning as the latter's confidence in their beliefs only increases over time. In a one-shot setting, low-ability experts can imitate high-ability experts, and there is no separation of expert types. We will come back to that case in Section \ref{sec:e2}. With repeated interactions, imitation is limited, due to Bayesian consumer updating. 

To search for equilibria in which expert behavior is not pooled, we must first derive what happens if one expert chooses to \textit{invest}, while the other two choose to \textit{not invest}. Consumers believe with $\tilde{\gamma}^inv$ that either of the not-investing experts is the high-ability expert. The expected income total income of approaching the sole investing expert is $\pi^e_{alg,inv} = R(57 - 27w + 17 \tilde{\gamma}^{inv}w)$, where $R$ is the total number of future expert--consumer interactions (15 in our experiment). Consumers expected income from switching in round 11 to a non-investing $\vect{P}^e$ expert is described by: $\pi^e_{alg,ninv} = (1-\tilde{\gamma}^{inv}) [0.5 * (r_i \pi^e_L + (R - r_i)\pi^e_H) + 0.5 *(R\pi^e_H)] + \tilde{\gamma}^{inv}(R\pi^e_L)$, where $r_i$ is the number of rounds until the Bayesian consumer $i$ identifies the imitating low-ability expert's $j$ true type.\footnote{Here, we assume that means $Pr(L) = 1 - \tilde{\gamma}_j$ must be such that a Bayesian consumer's belief $\tilde{\gamma}_l$ about second not-investing expert $l$ is high enough to prefer them over the investing expert, i.e., $\tilde{\gamma}_l \geq 1 - \frac{27}{17}w$.} Consumers estimate a 50\% probability to accidentally approach the second low-ability expert, whose true type they learn after $r_i$. Note that $\pi^e_{alg,ninv}$ is decreasing in $r_i$, because better imitation increases the cost of switching towards the imitating second low-ability expert. Consumers stay with the investing expert if $\pi^e_{alg,inv} \geq \pi^e_{alg,ninv}$ holds, which gives $r_i \geq \frac{2R [ \pi^e_{alg,inv} - ((1-\tilde{\gamma}^{inv})\pi^e_H + \tilde{\gamma}^{inv}\pi^e_L) ]}{(1-\tilde{\gamma}^{inv})(\pi^e_L - \pi^e_H)}$ or $r_i \geq \frac{30[w(17\tilde{\gamma}^{inv} - 27) + 17\tilde{\gamma}^{inv})]}{-17(1-\tilde{\gamma}^{inv})} \coloneqq \bar{r_i}$ in the current setup. Figure \ref{fig:contour} in the Appendix illustrates this condition.

If a consumer $i$ realizes the expert's type after $r_i < \bar{r_i}$ rounds, they always leave the investing expert in round 11 and search for the high-ability expert. If that holds for all consumers, then no expert ever invests, and all low-ability experts maximize their income by imitating the high-ability expert for $r < \bar{r}$ rounds. Afterwards, consumers switch away because they have been deceived, and switching cannot make them worse off. Hence, there are pooling equilibria in which no expert invests if $r < \bar{r}$ holds for all consumers.

Now, we go back to expert strategies under signaling concerns. Throughout, we assume that not-investing is a sufficiently strong signal such that, if only one expert invests, consumers think that this is the high-ability expert with enough certainty to always want to approach them (see $\tilde{y}_{ninv}$ above). Otherwise, signaling is not a valid strategy. The various payoffs associated with the investment choice in round 11 are illustrated in the \hyperref[sec:app]{Appendix}. Given that $r \geq \bar{r}$ may not hold for all consumers due to differences in belief formation and priors, we define $\alpha \in [0, 1]$ as the share of consumers for whom $r \geq \bar{r}$ holds. Furthermore, because (1) subjects make decisions under high levels of uncertainty, (2) signaling strategies may unravel quickly if only a minority of players follows them, and (3) the experimental setup is quite complex, we use two methods to analyze whether high-ability experts can escape the pooling equilibrium. One, we look for separating and mixed Nash equilibria. Two, we assume bounded rationality where each expert deploys Level-1 thinking and assumes that the other players are Level-0 (randomization) \citep{stahl1995players,camerer2004cognitive}. This approach has been shown in some strategic settings to be a better predictor of behavior \citep{stahl1994experimental,nagel1995unraveling}, particularly for initial play \citep{costa2001cognition,crawford2007level}.

\noindent
\textbf{Non-Pooling Equilibria.} There are two potential separating profiles. Either only the high-ability expert invests, or only the two low-ability experts invest. In addition, there can be mixed equilibria in which only one expert opts for the decision aid, thereby acting as the ``safe option'' for consumers. Optimal strategies depend on how consumers are distributed among experts, and whether experts can observe that. Here, we focus on the case in which an expert cannot observe the distribution of consumers who did not approach them. The results for each individual case are shown in the appendix, along with a more detailed explanation of the results. Tables \ref{tab:alg_ha_3_text} and \ref{tab:alg_ha_3_la_text} illustrate the general payoff idea for the simple case in which all experts ex ante coordinate on the high-ability expert. If the high-ability expert foregoes the algorithm, they keep all consumers as long as the two low-ability experts either both invest or do not invest, because consumers by assumption do not switch when they are indifferent. If only one low-ability expert invests, the high-ability expert loses all $\alpha$ consumers ($r_i \geq \bar{r_i}$) who prefer the safe investing option. On the other hand, if the high-ability expert chooses to rent the decision aid, they only keep all consumers if all experts do so. If both low-ability experts forego investments, the high-ability expert keeps $\alpha$, and may re-attract consumers when $r_i < 0.5R$. Finally, any expert who stands out as the only non-investing expert initially attracts all consumers. The low-ability expert retains consumers for $\frac{r_i}{R}$ rounds, i.e. until imitation is revealed. 

Regarding strategic separation of experts, the only separating equilibrium in which the high-ability expert foregoes the decision aid is an equilibrium by indifference if $x_{ha} = 3$. Otherwise, separation is not stable, because the transparent pivotal high-ability expert always monopolizes the market, which incentivizes low-ability experts to imitate.

\begin{table}[t]
\centering
\small
\caption{High-Ability Expert: Expected consumer attraction conditional on the three experts' investment choices in round 11 for \textit{Algorithm} if all consumers coordinated on the high-ability expert, i.e. $x_{ha} = 3$.}
\label{tab:alg_ha_3_text}
\renewcommand{\arraystretch}{1.2}
\setlength{\tabcolsep}{4pt}
\begin{tabular}{@{}c@{\hspace{0.5em}}c@{}}
\begin{tabular}{cc|c|c|}
  & \multicolumn{1}{c}{} & \multicolumn{2}{c}{LA$_0$}\\
  & \multicolumn{1}{c}{} & \multicolumn{1}{c}{Invest} & \multicolumn{1}{c}{Not Invest} \\\cline{3-4}
  \multirow{2}{*}{LA$_0$} & Invest & 1 & $(1-\alpha)$ \\\cline{3-4}
  & Not Invest & $(1-\alpha)$  & 1 \\\cline{3-4}
\end{tabular}
&
\begin{tabular}{cc|c|c|}
  & \multicolumn{1}{c}{} & \multicolumn{2}{c}{LA$_0$}\\
  & \multicolumn{1}{c}{} & \multicolumn{1}{c}{Invest} & \multicolumn{1}{c}{Not Invest} \\\cline{3-4}
  \multirow{2}{*}{LA$_0$} & Invest & 1 & $0.5\frac{R-r_i}{R}$ \\\cline{3-4}
  & Not Invest & $0.5\frac{R-r_i}{R}$ & $\begin{array}{c} \alpha + \\ (1-\alpha)\frac{R-2r_i}{R} \end{array}$ \\\cline{3-4}
\end{tabular}
\\[4.5em]
HA$_3$ Expert Chooses \textit{Not Invest} & HA$_3$ Expert Chooses \textit{Invest}
\end{tabular}
\end{table}
\begin{table}[t]
\centering
\footnotesize
\caption{Low-Ability Expert: Expected consumer attraction conditional on on the three experts' investment choices in round 11 for \textit{Algorithm} if all consumers coordinated on the high-ability expert, i.e. $x_{ha} = 3$.}
\label{tab:alg_ha_3_la_text}
\renewcommand{\arraystretch}{1.5}
\begin{tabular}{cc}
\begin{tabular}{cc|c|c|}
  & \multicolumn{1}{c}{} & \multicolumn{2}{c}{HA$_3$}\\
  & \multicolumn{1}{c}{} & \multicolumn{1}{c}{Invest} & \multicolumn{1}{c}{Not Invest} \\\cline{3-4}
  \multirow{2}{*}{LA$_0$} & Invest & $\frac{r_i}{R}$ &$0$ \\\cline{3-4}
  & Not Invest & $\begin{array}{c} (1-\alpha)0.5\frac{r_i}{R} + \\ (1-\alpha)0.5\frac{R-r_i}{R} \end{array}$ & 0 \\\cline{3-4}
\end{tabular}
&
\begin{tabular}{cc|c|c|}
  & \multicolumn{1}{c}{} & \multicolumn{2}{c}{HA$_3$}\\
  & \multicolumn{1}{c}{} & \multicolumn{1}{c}{Invest} & \multicolumn{1}{c}{Not Invest} \\\cline{3-4}
  \multirow{2}{*}{LA$_0$} & Invest & 0 &  0 \\\cline{3-4}
  & Not Invest & $0.5\frac{R-r_i}{R}$ & $\alpha$ \\\cline{3-4}
\end{tabular}
\\[3em]
LA$_0$ Expert Chooses \textit{Not Invest} & LA$_0$ Expert Chooses \textit{Invest}
\end{tabular}
\end{table}

\begin{figure}[t] 
    \centering 

    \begin{minipage}[b]{0.45\textwidth} 
        \centering
        \includegraphics[width=\textwidth]{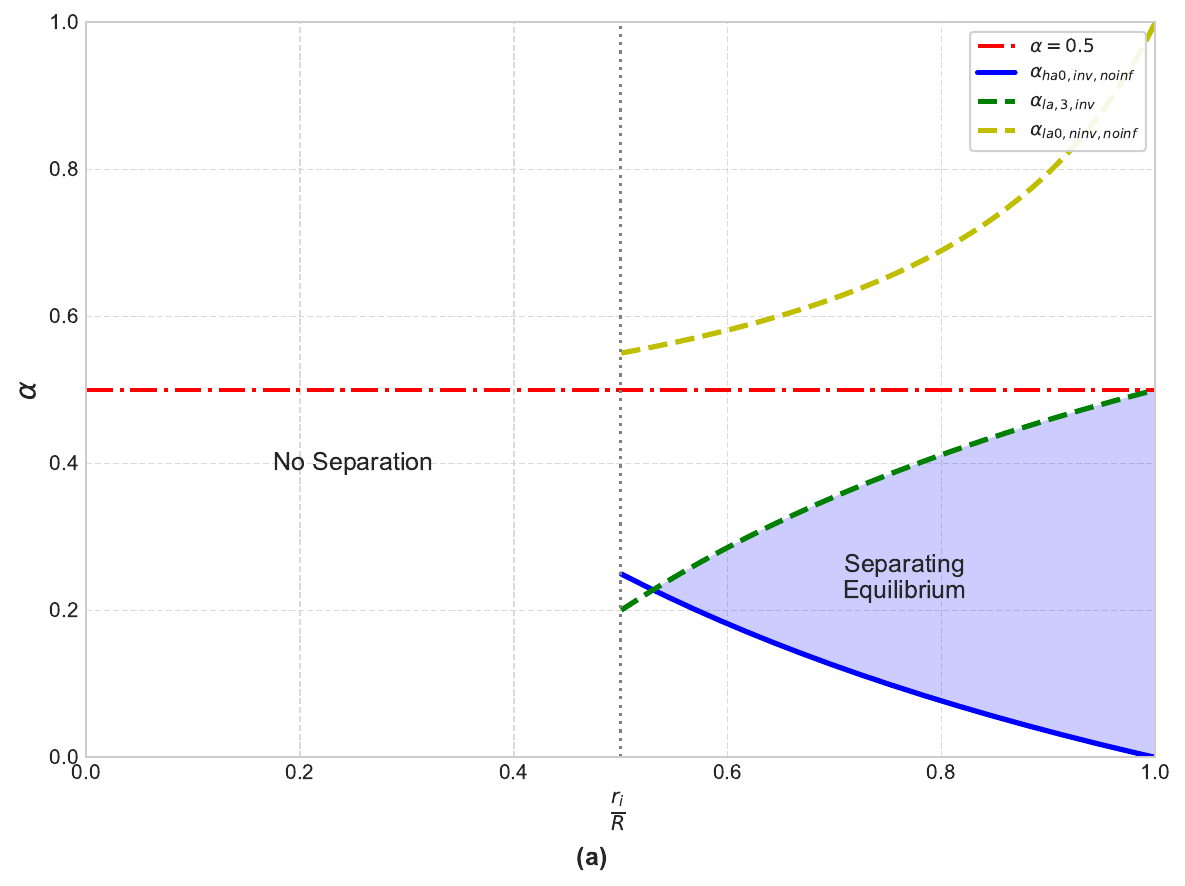}
    \end{minipage}
    \hfill 
    \begin{minipage}[b]{0.45\textwidth}
        \centering
        \includegraphics[width=\textwidth]{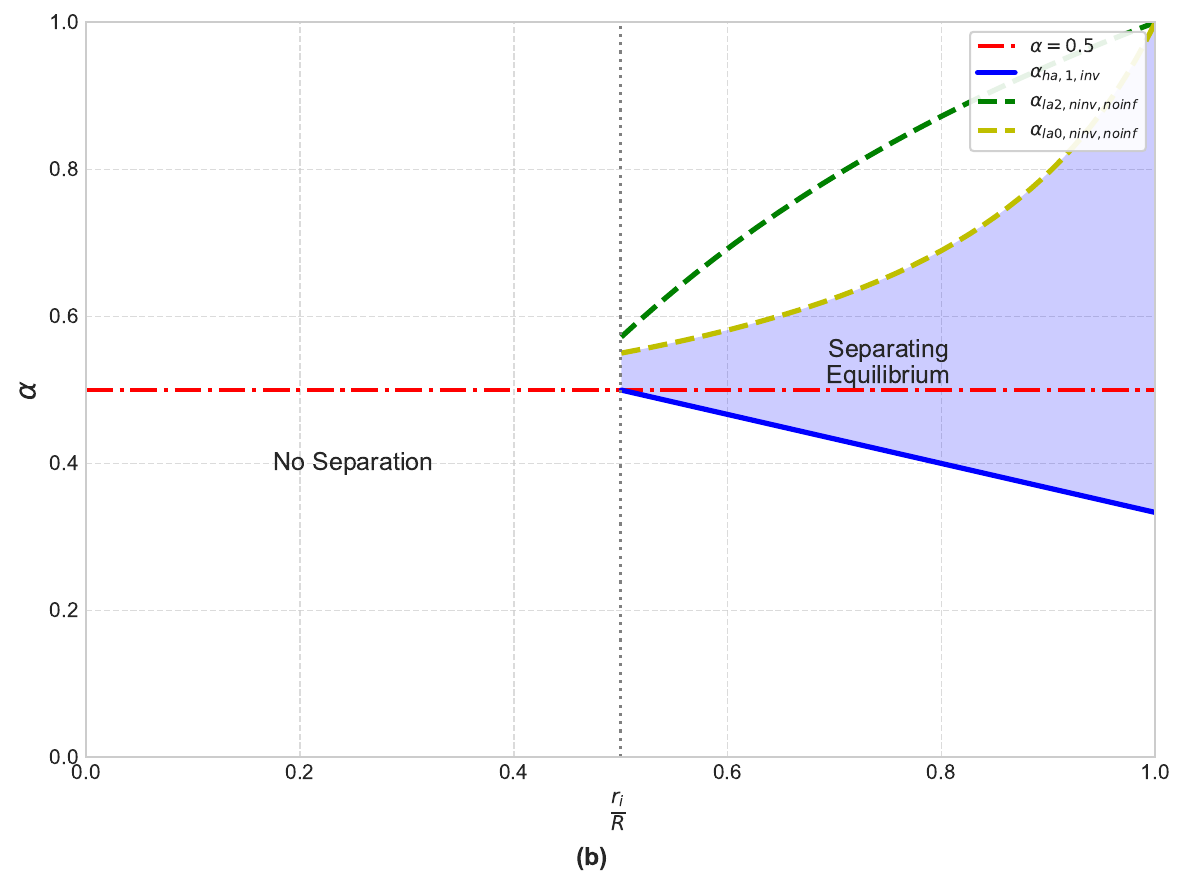}
    \end{minipage}

    \vspace{0.5cm} 

    \begin{minipage}[b]{0.45\textwidth}
        \centering
        \includegraphics[width=\textwidth]{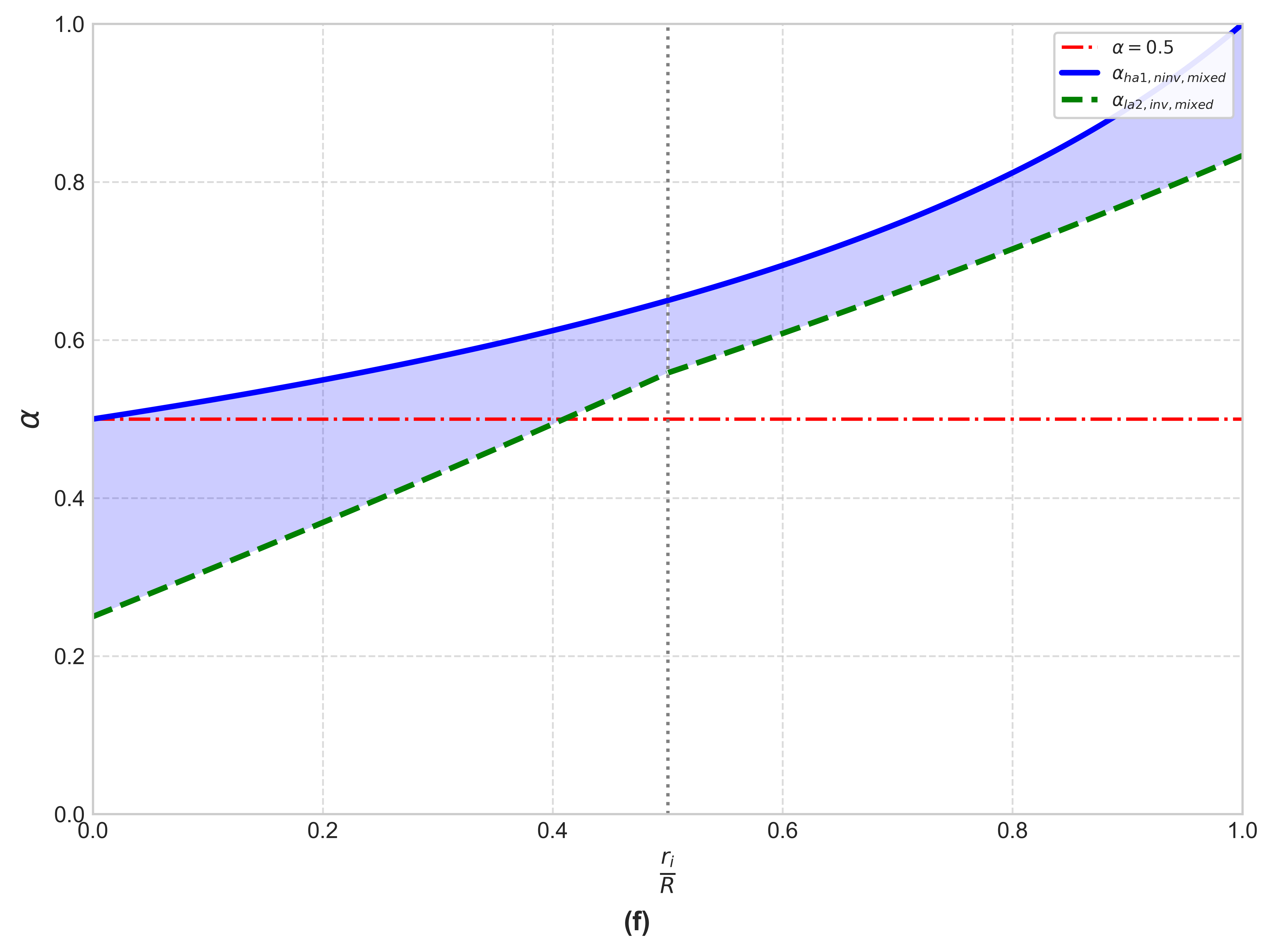}
    \end{minipage}
    \hfill 
    \begin{minipage}[b]{0.45\textwidth}
        \centering
        \includegraphics[width=\textwidth]{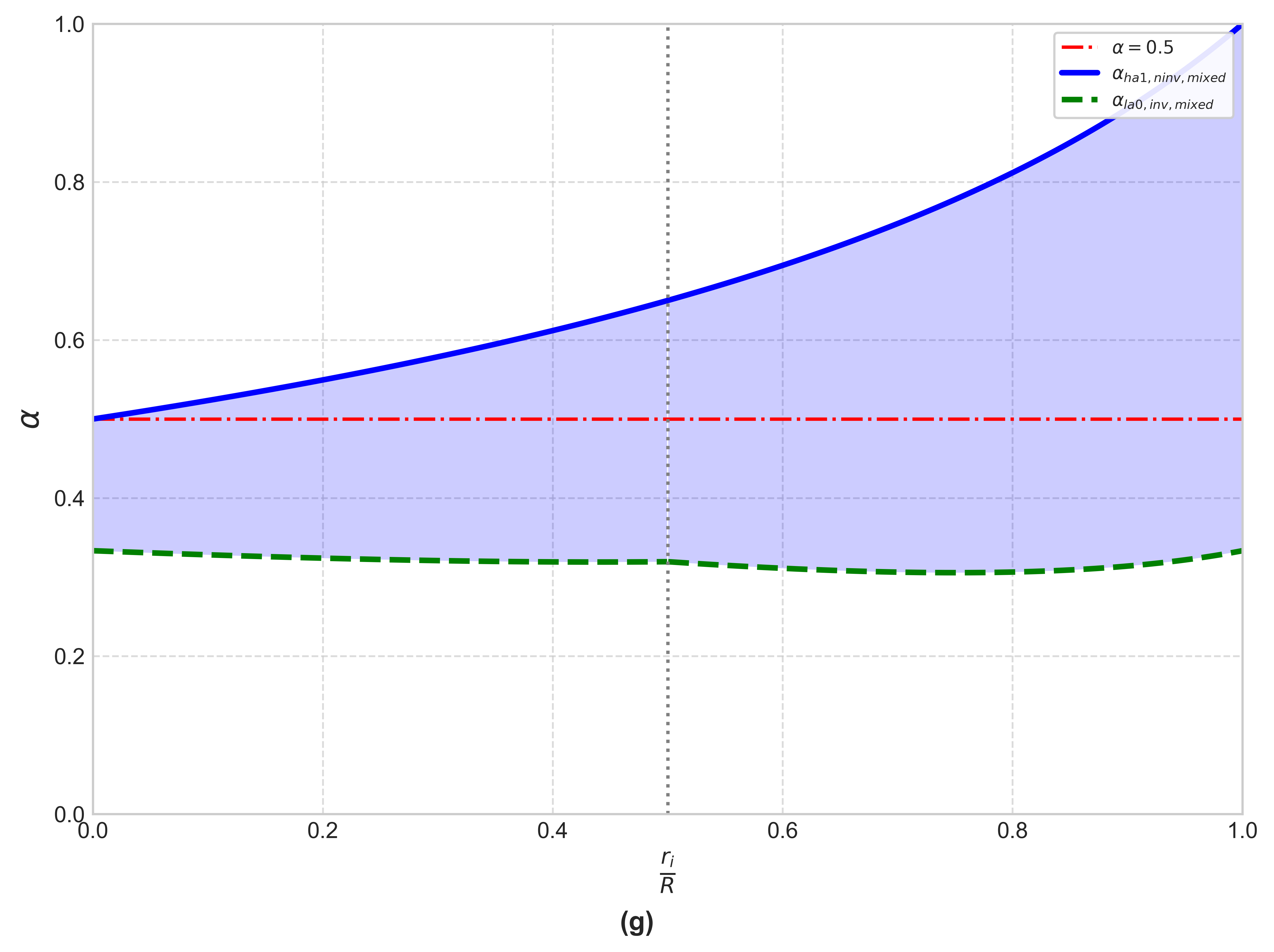}
    \end{minipage}

    \caption{Panels \textbf{(a)} and  \textbf{(b)} show the separating equilibrium space in which \textbf{only the high-ability expert invests} if (i) $x_{ha} = 0$, $x_{la,i} = 3$, $x_{la,j} = 0$ or (ii) $x_{ha} = 1$, $x_{la,i} = 2$, $x_{la,j} = 0$. Panels  \textbf{(c)} and \textbf{(d)} illustrate mixed equilibrium space where \textbf{only one low-ability expert invests} for $x_{ha} = 1$, $x_{la,i} = 2$, $x_{la,j} = 0$. In Panel \textbf{(d)}, LA$_0$ invests. For all cases, experts do not observe the distribution of ``other'' consumers who do not approach them.}
    \label{fig:separating_fig_text}
\end{figure}

In contrast, we find that strategic separation where only the high-ability expert invests -- which is very inefficient -- can be a stable outcome under certain circumstances. Specifically, the high-ability expert must have attracted a small share of consumers at the time of the technological shock (here $x_{ha} \in \{0, 1\}$) and the ratio $\frac{r}{R}$ has to be relatively large (here, with one exception, $\frac{r}{R} > 0.5R$), implying either a small number of expected future interactions or a long imitation horizon. Figure \ref{fig:separating_fig_text} shows two illustrative cases for $x_{ha} = 0$ and $x_{ha} = 1$, and Figures \ref{fig:separating_la3}, \ref{fig:separating_noinf} in the Appendix illustrate all possibilities conditional on the distribution of consumers. These confirm that even high-ability experts can profit from acting as the sole safe option if ability differences are obfuscated, resulting in investment paths that run opposite to the optimal ones. In the current experiment, we do not expect them to be empirically relevant, if only because experts and consumers complete 15 interactions after the introduction of the decision aid, leaving ample room for learning and experimentation. Nevertheless, this suggests that, while there is no clean separation for the most desirable diagnostic path in which only the high-ability expert foregoes the technology, expert separation can occur for the least desirable path. All thresholds and respective hyperbolas are listed in Table \ref{tab:conditions_ha_only_invest_app}.

Beyond clean separation, there may also be mixed equilibria which allow the high-ability expert to escape the pooling equilibrium from Phase 1. In those, one low-ability expert invests, while the other two experts do not invest. Tables \ref{tab:mixed_equilibria} and \ref{tab:mixed_equilibria_noinf} illustrate the respective conditions for different consumer distributions and expert information. In general, as long as the high-ability expert does not attract all consumers in Phase 1, there are parameter constellations in which one low-ability expert can find it profitable to act as the safe option, while the second continues to forego the decision aid. Figure \ref{fig:separating_fig_text} shows an example for $x_{ha} = 1$ and $x_{la} = 2$. The high-ability expert's hyperbola across potential scenarios always suggest a strong inclination to forego the decision aid, such that $\alpha$ must consistently be very high to motivate high-ability investments. Both low-ability experts can find it profitable to be the sole technology adopter, but the parameter space is much larger for the low-ability expert with few (here 0) consumers. This is very intuitive, since low-ability experts with a sizeable consumer stock can profit from imitation. For an illustration of all hyperbolas under different conditions, refer to Figures \ref{fig:mixed_inf} and \ref{fig:mixed_noinf} in the Appendix. Overall, there are ample opportunities for high-ability experts to escape the pooling equilibrium through signaling, depending on consumer preferences $\alpha$, the speed of learning $r_i$ and the number of total future interactions $R$. However, if consumers either do not consider learning ($\alpha = 1$), heavily rely on safe alternatives, or learn very quickly relative to the total number of interactions, there is less (or no) room for separation. Moreover, pure separation in expert strategies is only possible when the high-ability expert (inefficiently) chooses to invest, not vice versa.

\begin{figure}[t] 
    \centering 

    \begin{minipage}[b]{0.45\textwidth} 
        \centering
        \includegraphics[width=\textwidth]{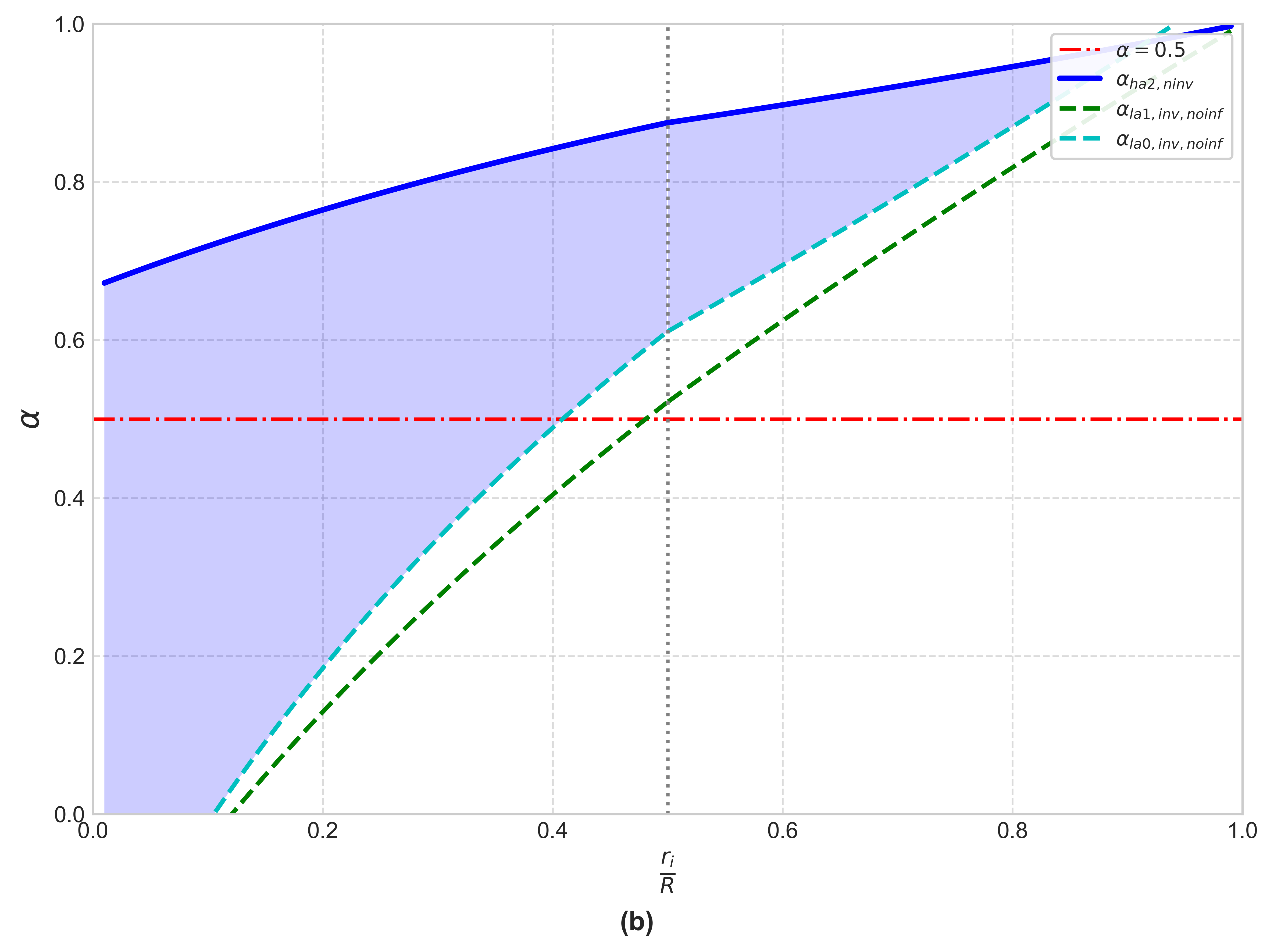}
    \end{minipage}
    \hfill 
    \begin{minipage}[b]{0.45\textwidth}
        \centering
        \includegraphics[width=\textwidth]{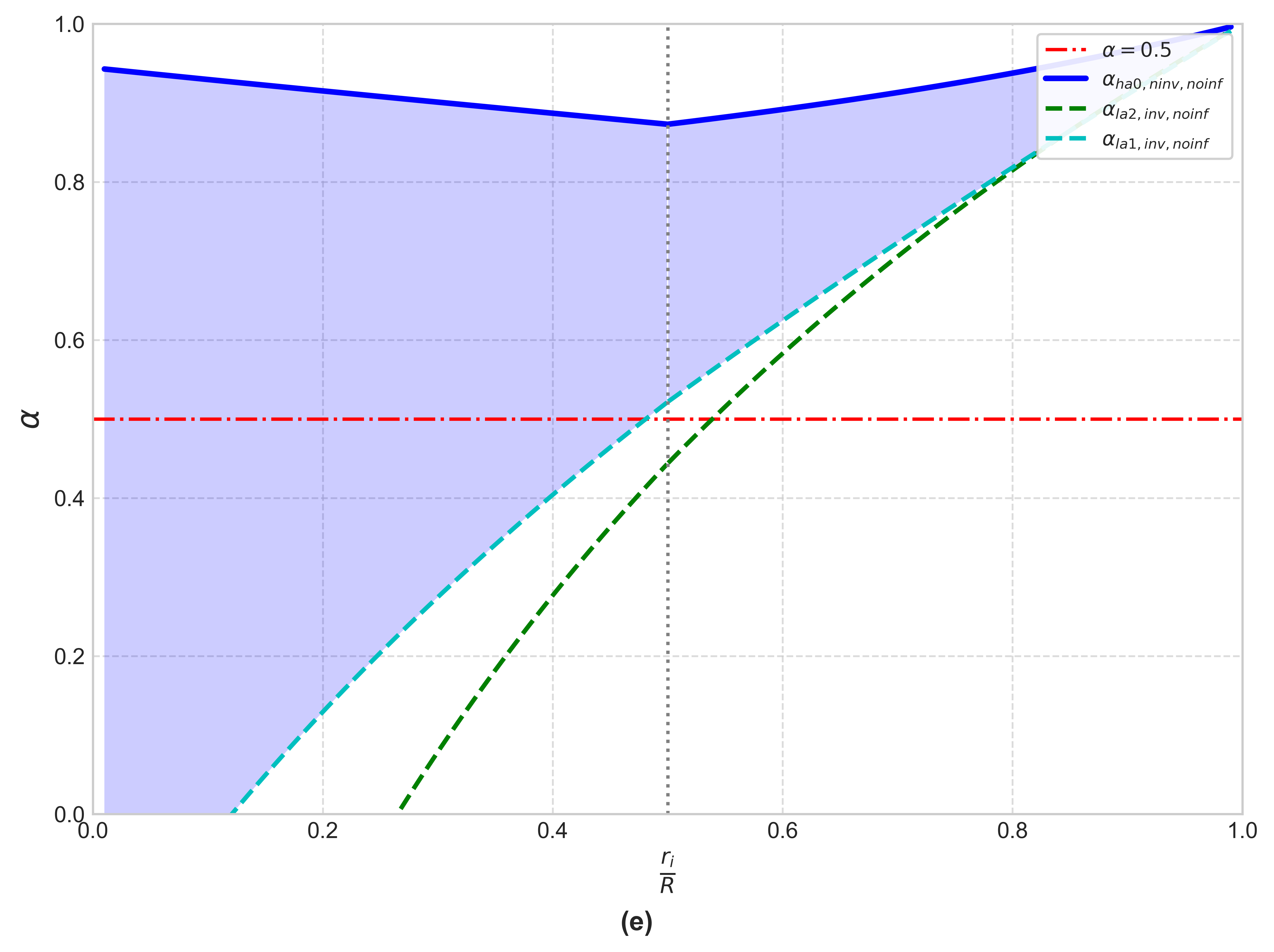}
    \end{minipage}
    \caption{Strategic expert separation in which only the high-ability expert foregoes the decision aid. Experts invest above their respective hyperbola, and do not invest below it. Panel \textbf{(b)} shows the separating space when $x_{ha} = 2$, Panel \textbf{(e)=} for $x_{ha} = 0$, $x_{la,i} = 2$ and $x_{la,j} = 1$.}
    \label{fig:level1_separating_text}
\end{figure}

\begin{figure}[htbp]
    \centering 

    \begin{minipage}[b]{0.45\textwidth} 
        \centering
        \includegraphics[width=\textwidth]{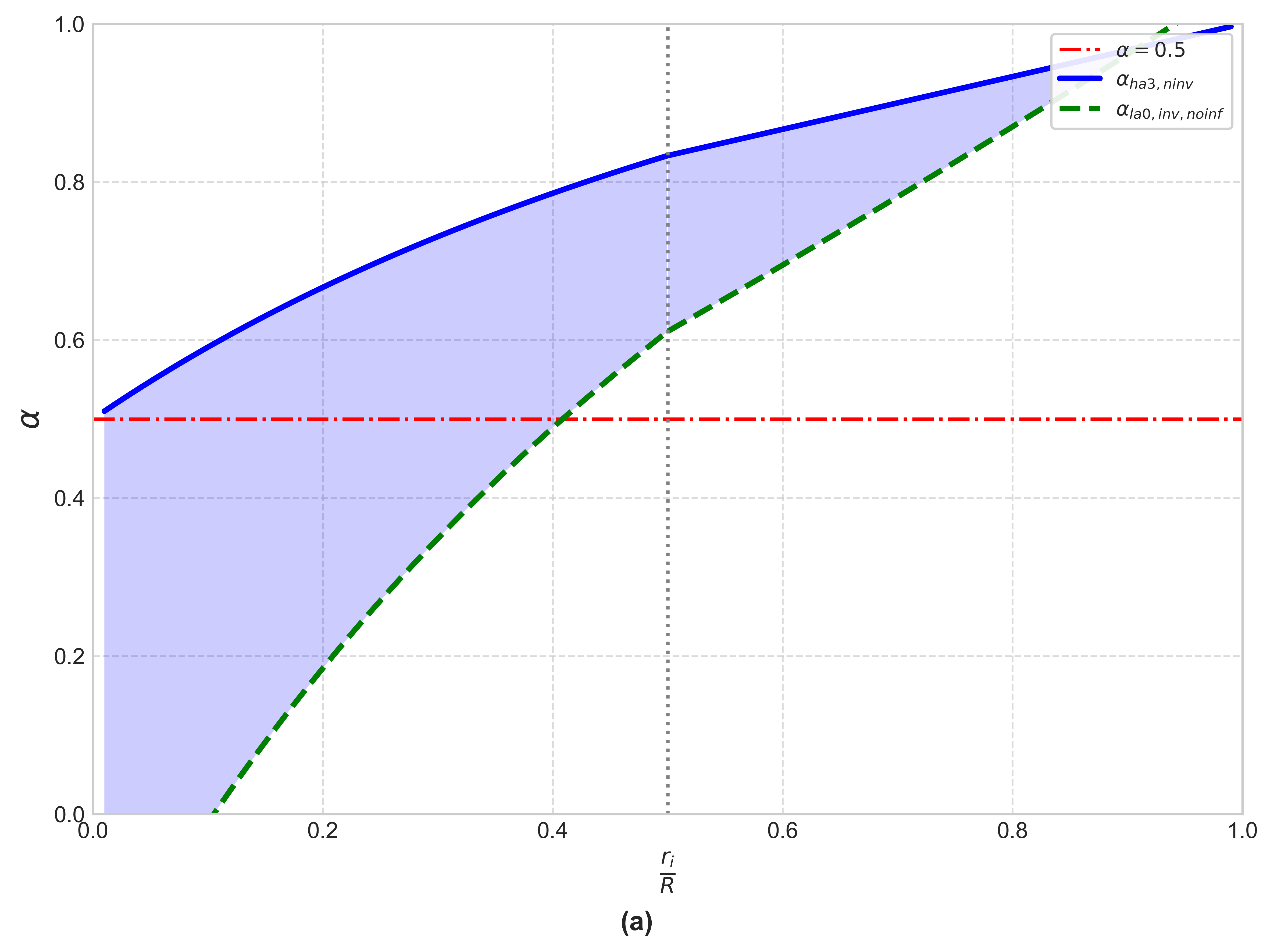}
    \end{minipage}
    \hfill 
    \begin{minipage}[b]{0.45\textwidth}
        \centering
        \includegraphics[width=\textwidth]{fig/level1_ha2.png}
    \end{minipage}

    \vspace{0.5cm} 

    \begin{minipage}[b]{0.45\textwidth}
        \centering
        \includegraphics[width=\textwidth]{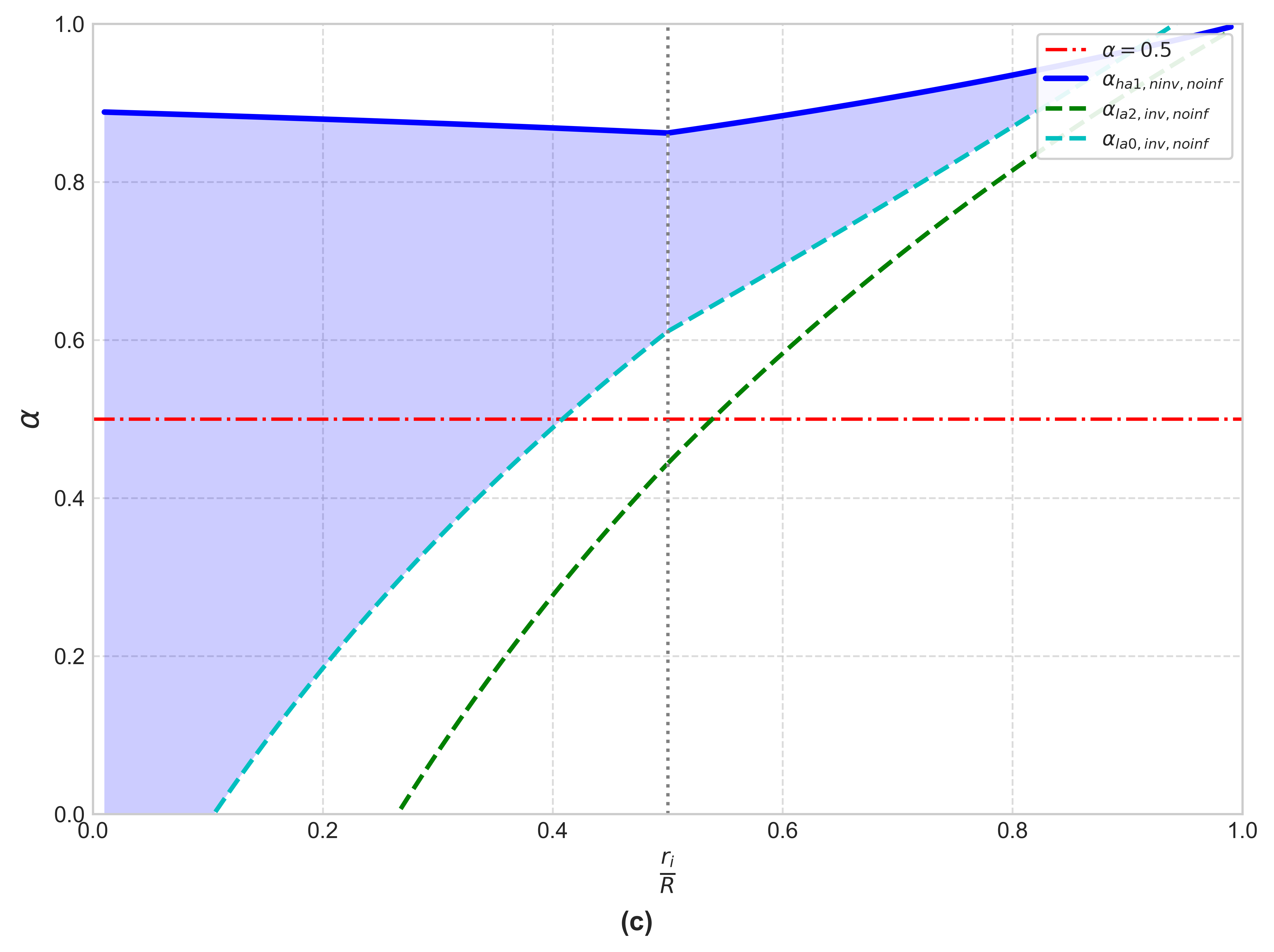}
    \end{minipage}
    \hfill 
    \begin{minipage}[b]{0.45\textwidth}
        \centering
        \includegraphics[width=\textwidth]{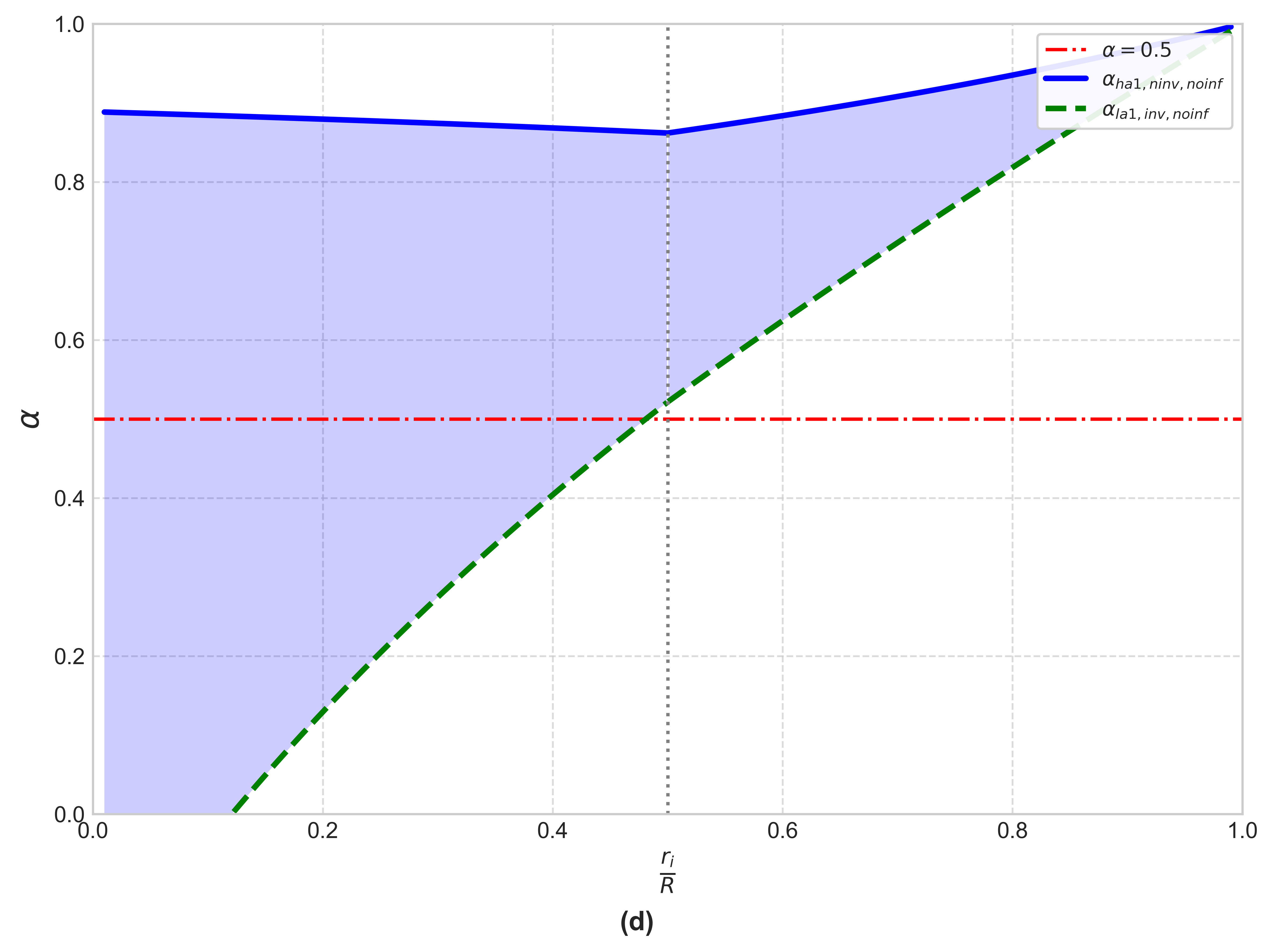}
    \end{minipage}

    \vspace{0.5cm} 

    \begin{minipage}[b]{0.45\textwidth}
        \centering
        \includegraphics[width=\textwidth]{fig/level1_ha0la21.png}
    \end{minipage}
    \hfill 
    \begin{minipage}[b]{0.45\textwidth}
        \centering
        \includegraphics[width=\textwidth]{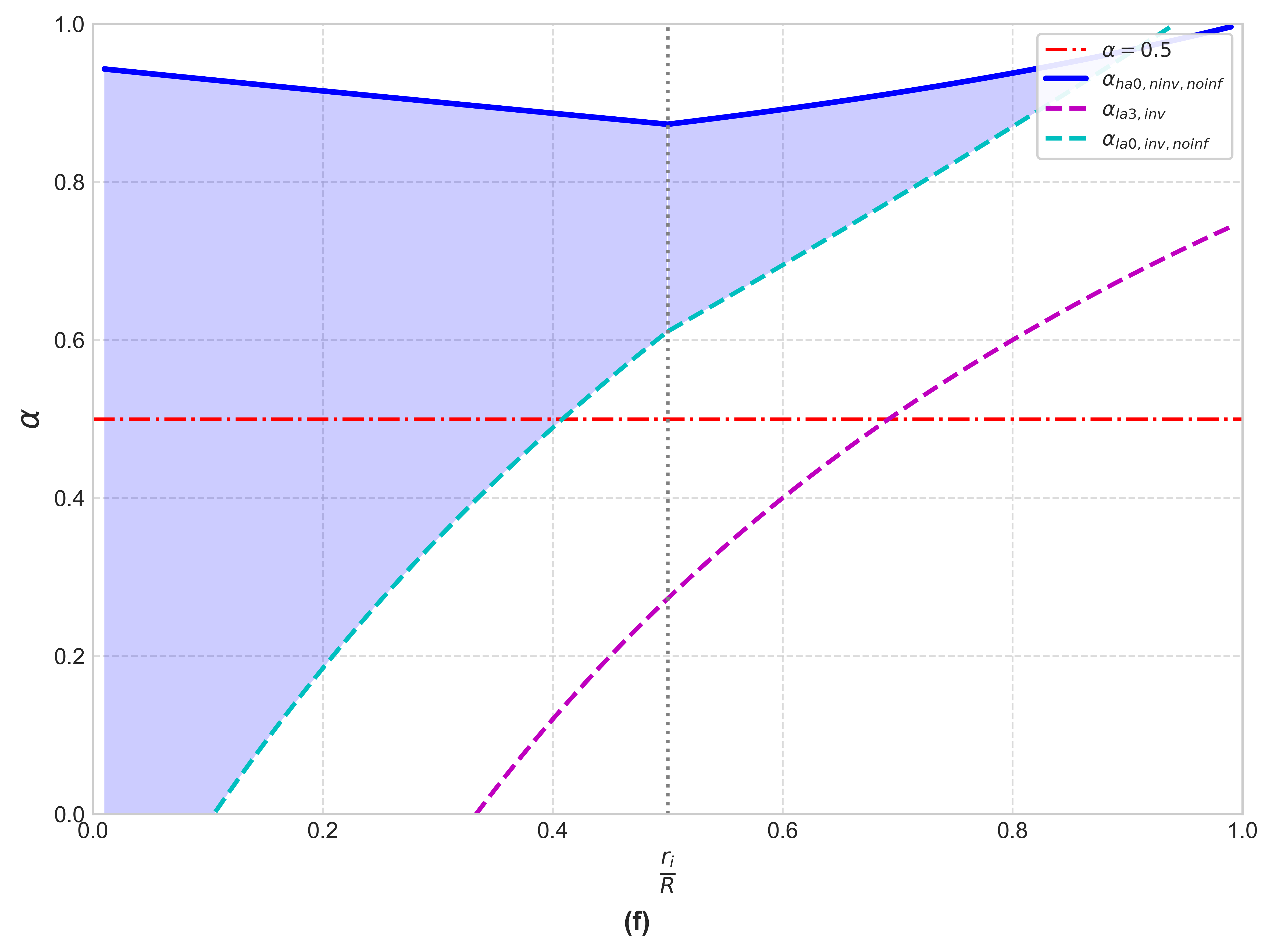}
    \end{minipage}

    \caption{Expert type strategic separation under level-1 thinking (bounded rationality) when the distribution of other consumers is unknown. Only the high-ability expert does not invests. Panel \textbf{(a)} shows the separating space when $x_{ha} = 3$. Panel \textbf{(b)} shows the same for $x_{ha} = 2$ and $x_{la} = 1$. Panel \textbf{(c)} shows separation for $x_{ha} = 1$ and $x_{la} = 2$, \textbf{(d)} when $x_{ha} = 1$, $x_{la,i} = 1$ and $x_{la,j} = 1$. Panel \textbf{(e)} illustrates separation under x$_{ha} = 0$, $x_{la,i} = 2$, $x_{la,j} = 1$ and Panel \textbf{(f)} under $x_{la} = 3$.}
    \label{fig:separating_noinf_level1}
\end{figure}

\noindent
\textbf{Bounded Rationality.} We look at predicted expert behavior based on level-1 thinking experts. This is primarily - but not only -- to predict initial behavior once the novel technology arrives. Our focus lies on pure expert separation where the high-ability expert is the only one to forego the decision aid. This is the most efficient diagnostic pathway, and the one most in line with the idea of ability signaling to escape a pooling situation. Tables \ref{tab:conditions_separations_level1_noinf} and \ref{tab:conditions_separations_level1} in the Appendix summarize all conditions for experts to strategically separate. In contrast to the behavioral predictions from Nash equilibria, assuming that experts simply maximize their expected payoff while being naive about the other experts choices strongly predicts strategic ability-based separation for a large space of parameters. In particular, the high-ability expert almost always chooses to forego the decision aid (see Figure \ref{fig:level1_separating_text} for a selection and Figure \ref{fig:separating_noinf_level1} for a full overview). Low-ability experts' inclination to rent the algorithm increases in $\alpha$, but decreases in $\frac{r_i}{R}$ -- resulting in a high likelihood for pooling behavior without any technology adoption if $\frac{r_i}{R} \, \text{roughly} \geq 0.5$. Consumers being relatively quick in identifying low-ability experts and switching between experts makes investments more attractive for low-ability experts through (i) worsening imitation of the expert themselves and (ii) worsening imitation of other low-ability experts, thereby increasing the low-ability expert's likelihood to poach other consumers. 

Overall, if bounded rationality holds, high-ability experts almost always forego the decision aid. Moreover, there is a large space for pure strategic expert separation in which the high-ability expert escapes the previous pooling equilibrium through signaling. Hence, we predict that, especially right after the appearance of the novel diagnostic technology, high-ability experts will be less likely to invest into it. However, it is unclear how stable that strategy is. While there are no Nash equilibria in which pure separation holds long-term, there are stable mixed equilibria in which only one low-ability expert invests, and their parameter spaces do intersect with the ones derived for bounded rationality. Hence, even if bounded rationality does not explain behavior in equilibrium, high-ability experts may still be able to exploit their skill advantage by foregoing the decision, albeit less effectively. Finally, we should note that signaling still predicts pooling behavior for a wide arrange of parameters. This would cause under-investments from low-ability experts. Moreover, if signaling does not play any role, then theoretical predictions for \textit{Algorithm} do not differ from predictions for \textit{Skill}, meaning all experts should invest (as long as $w \leq 0.48$). Hence, it is difficult to make fine-grained predictions. We hypothesize:  

\medskip
\textbf{Hypothesis 8:} Less high-ability experts than low-ability experts in 
\textit{Algorithm} invest to rent the algorithmic decision aid.
\medskip

\textbf{Hypothesis 9:} Both expert types in \textit{Algorithm} maximize expected consumer income by choosing $\vect{P^e}$.
\medskip

\section{Results \label{sec:results}}
Table \ref{tab:summary_statistics} provides summary statistics for the two periods. We first analyze subject behavior and market efficiency in the first 10 rounds without diagnostic investments. 

\begin{table}[h]
\centering
\begin{tabular}{@{}>{\raggedright\arraybackslash}p{4cm}>{\centering\arraybackslash}p{2.5cm}>{\centering\arraybackslash}p{3.5cm}>{\centering\arraybackslash}p{3.5cm}@{}}
\toprule  
\toprule
 & Phase 1 & \makecell{Phase 2 \\ Skill} & \makecell{Phase 2 \\ Algorithm} \\
\midrule[\heavyrulewidth]  
Market Entry Rate & 0.95 & 0.98 & 0.98 \\[5pt]
Efficiency (relative)  & 0.61 & 0.65 (0.75) & 0.63 (0.74) \\[5pt]
Undertreatment & 0.29 & 0.23 & 0.25 \\[5pt]
Overtreatment & 0.55 & 0.43 & 0.49 \\[5pt]
Efficient Treatment & 0.52 & 0.64 & 0.60 \\[5pt]
Efficient Provision & 0.40 & 0.59 & 0.53 \\[5pt]
Total Expert Surplus & 132 & 133 & 133 \\[5pt]
Total Client Surplus & 131 & 138 & 133 \\[5pt]
Price Vectors & $\vect{P}^m$ 0.28 &$\vect{P}^m$ 0.33 &  $\vect{P}^m$ 0.26 \\[5pt]
 & $\vect{P}^s$ 0.41 & $\vect{P}^s$ 0.41 &  $\vect{P}^s$ 0.40 \\[5pt]
 & $\vect{P}^e$ 0.31 & $\vect{P}^e$ 0.26 & $\vect{P}^e$ 0.34\\[5pt]
\midrule
N & 300 & 150 & 150 \\
\bottomrule  
\end{tabular}
\caption{Summary statistics by phase and treatment. \textit{Efficiency} is calculated as (average actual profit - outside option)/(maximum possible average profit - outside option). \textit{Relative Efficiency} refers to the market's efficiency without correcting for the efficiency loss due to investment costs. \textit{Undertreatment} is the share of treatment choices where the consumer has a big problem but receives the LQT. \textit{Overtreatment} is the share of treatment choices where the consumer has a small problem but receives the HQT. \textit{Efficient Treatment} equals 1 when an expert makes the efficient treatment decision. \textit{Efficient Provision} equals 1 when the expert receives the correct diagnostic signal and consequently chooses the efficient treatment.}
\label{tab:summary_statistics}
\end{table}

In contrast to the model's prediction, experts do not generally maximize expected consumer income by choosing $\vect{P}^m$. The most popular price vector $\vect{P}^s$ has the highest prices and lowest expected income for consumers (see also Figure \ref{fig_prices_1} in the Appendix). This replicates a common effect from prior credence goods experiments \textit{without} diagnostic uncertainty and obfuscated ability heterogeneity: under verifiability, experts rarely choose equal mark-up price vectors, and instead opt for the price vector that splits the gains of trade most equally if expert behave honestly \citep{dulleck2011economics,kerschbamer2017social}. This leads to efficiency losses, as roughly 30\% of big problems are treated with the LQT. 

\begin{figure}[h]
    \centering
\includegraphics[width=0.3\textwidth]{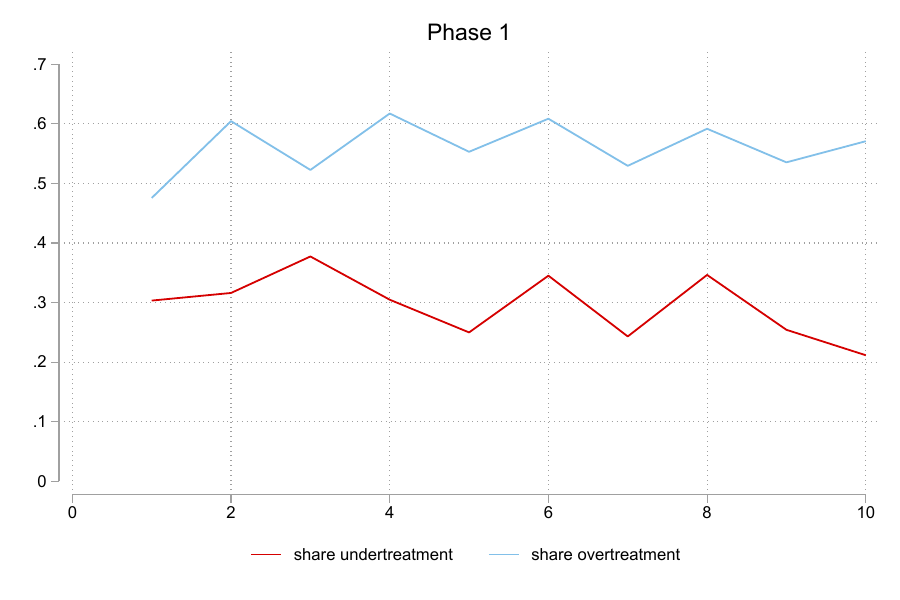}
\includegraphics[width=0.3\textwidth]{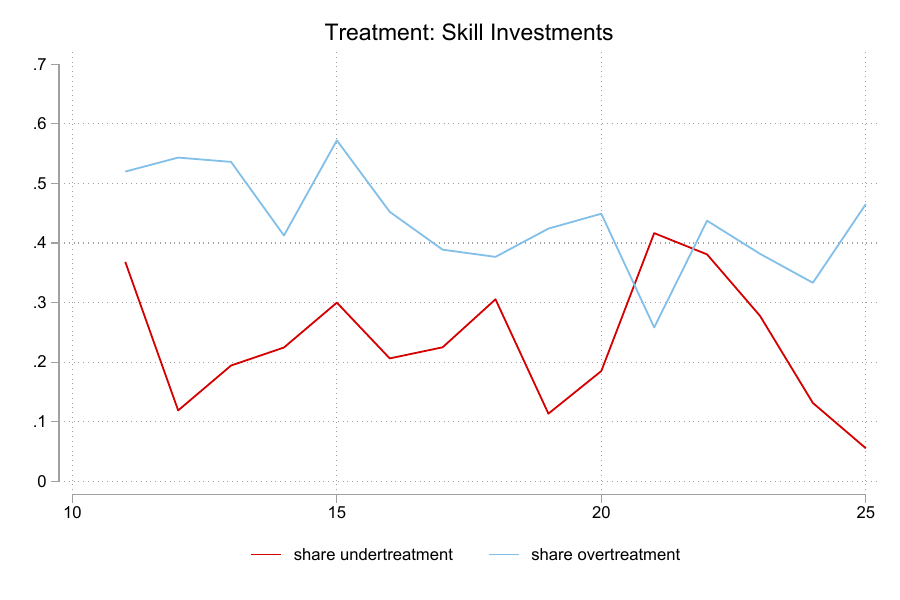}
\includegraphics[width=0.3\textwidth]{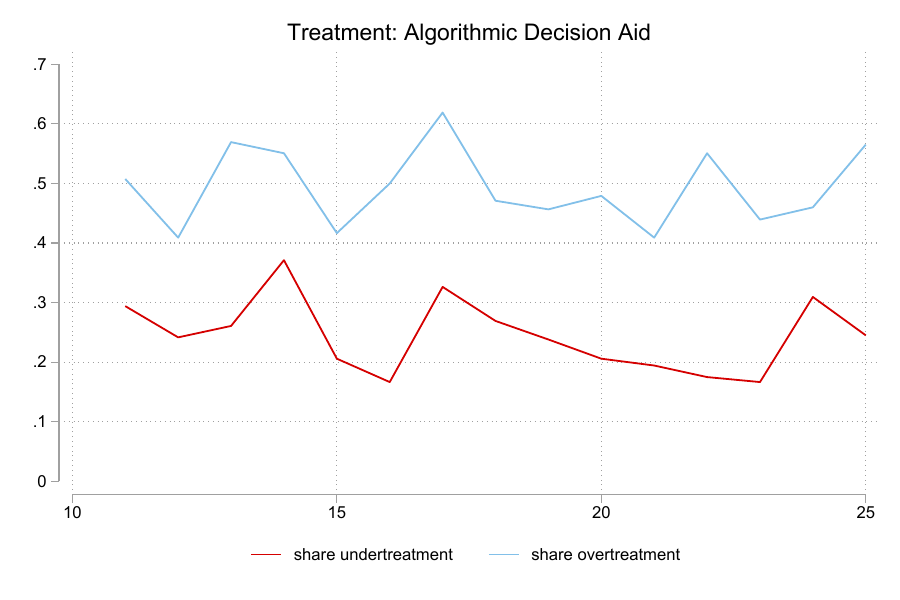}
\caption{Share of undertreatment and overtreatment in (left) the first 10 rounds (middle) Phase 2 of \textit{Skill} and (right) Phase 2 of \textit{Algorithm}}.
\label{fig:fig_over_under}
\end{figure}

Still, compared to other experiments with expert price setting, both $\vect{P}^m$ and $\vect{P}^e$ are relatively common. In fact, experts shift towards the predicted price vector $\vect{P}^m$ over time (logistic panel regression odds ratio: 1.08, p = 0.02, CI (95\%): [1.01, 1.14]). Furthermore, as predicted by the model, high-ability experts ($\vect{P}^m$/$\vect{P}^s$/$\vect{P}^e$ = 0.29/0.39/0.31) set the same prices as low-ability experts ($\vect{P}^m$/$\vect{P}^s$/$\vect{P}^e$ = 0.28/0.41/0.31). There are no differences in the number of consumers an expert can attract in round 10 on average between expert types ($t = 1.2, p = 0.214)$. High-ability experts cannot exploit their diagnostic advantage due to ability obfuscation. Overall, inefficiencies from over- and undertreatment are relatively common (see Figure \ref{fig:fig_over_under}), but do not deter consumers from entering the market. Only 5\% of consumers choices fall on the outside option. This makes the market much more efficient than standard laboratory experiments with perfect diagnostic, suggesting a potential upside of uncertainty for market outcomes (see e.g. \cite{tracy2023uncertainty}).

\subsection{Phase 2 -- Expert Investments}

\begin{figure}[t]
    \centering
    \caption{Expert investment shares for \textit{Skill} and \textit{Algorithm} over the 15 rounds of Phase 2.}

    \includegraphics[width=0.75\textwidth]{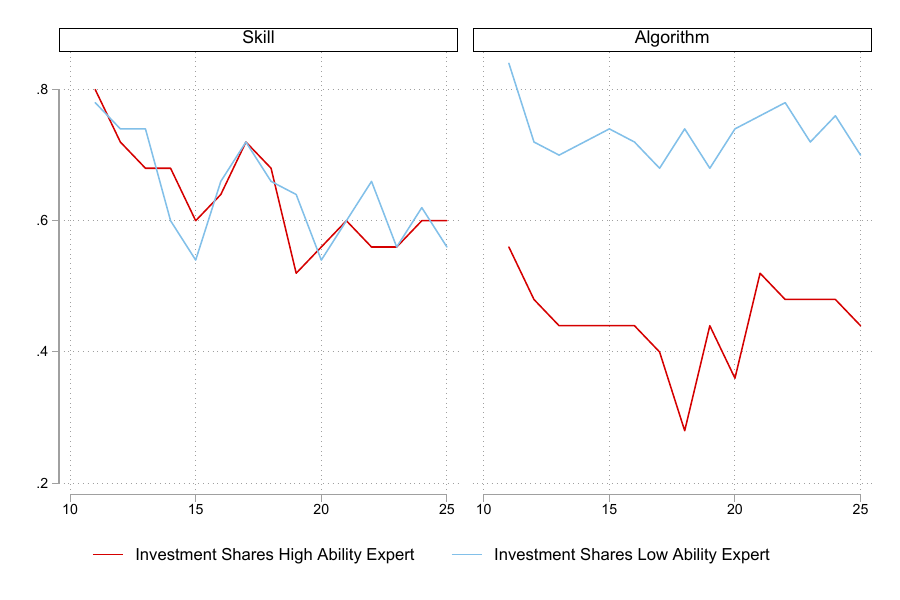}
    \label{fig:investments}
\end{figure}

Round 11 introduces either \textit{skill investments} or an \textit{algorithmic decision aid} as an external technological shock. Each round, experts can increase their maximum diagnostic precision to 90\% by paying 10 Coins, which also automatically increases their prices by 10 Coins. Figure \ref{fig:investments} shows investment shares for the two treatments and expert types in Phase 2. As predicted, there are no differences between expert types for skill investments. Over all 15 rounds, experts invest 64\% of the time. Investments start at around 80\%, but deteriorate significantly (see Table \ref{tab:reg_invest}). Thus, while popular, investment shares lie substantially below the predicted (and optimal) 100\%.

\begin{table}[t!]\centering
\def\sym#1{\ifmmode^{#1}\else\(^{#1}\)\fi}
\caption{Expert Investment Behavior in Phase 2}
\begin{adjustbox}{max width=\textwidth}
\begin{tabular}{l*{5}{c}}
\toprule
&\multicolumn{1}{c}{\specialcell{Skill Investment}}&\multicolumn{1}{c}{\specialcell{Skill Investment}}&\multicolumn{1}{c}{\specialcell{Algorithm Investment}} &\multicolumn{1}{c}{\specialcell{Algorithm Investment }}\\
\midrule

High-Ability &  -0.022&  -0.035&  -0.257\sym{***}  & -0.259\sym{***} \\
 & (0.081)         & (0.079)         & (0.067)  & (0.071)  \\
Round &  &  -0.011\sym{**} & & -0.001   \\
 &     &  (0.004)      &  &  (0.004)  \\
Risk &  &  0.024&   & -0.003\\
 &       & (0.013)    &   & (0.016) \\
Female &  &  0.005&  & -0.016 \\
 &        & (0.083)      &   & (0.061)  \\
Age &  &  -0.004&   & 0.003\\
 &         & (0.004)      & & (0.003)  \\ 
\\

\midrule
$N$          &1125       &1125       &1125   & 1125\\
\bottomrule
\end{tabular}
\label{tab:reg_invest}
\end{adjustbox}
\begin{minipage}{0.95\textwidth} 
\vspace*{0.09cm}
{\footnotesize 
This table reports marginal effects of panel logistic regressions using subject-level random effects and a cluster–robust VCE estimator at the matched group level (standard errors in parentheses). The dependent variable is a binary variable that equals 1 if the expert invests into the new diagnostic technology.
 \sym{*} \(p<0.05\), \sym{**} \(p<0.01\), \sym{***} \(p<0.001\)}
\end{minipage}
\end{table}

In \textit{Algorithm}, investment shares clearly diverge between expert types. In round 11, 84\% of low-ability and 56\% of high-ability experts rent the algorithm ($\tilde{\chi}^2 = 6.91, p = 0.009$). Across all 15 rounds, 73\% of low-ability expert choose to invest, whereas high-ability experts opt for the decision aid with a probability of only 45\%. Logit panel regression with subject-level random effects and clustered standard errors at the matched group level confirm a large and significant negative effect of being a high-ability expert on investment shares (Table \ref{tab:reg_invest}). This provides direct support for our main hypothesis. When an expert's initial ability type becomes inconsequential through skill investments, there are no differences between high- and low-ability experts. With technological decision aids, however, some high-ability experts try to signal their ability type by foregoing investments and offering lower prices. In line with that hypothesis, the algorithmic decision aid does not appear to be less desirable overall. In both treatments, aggregated investment shares are 64\%. That is because, while high-ability experts exhibit a 19 percentage point drop in technology adoption, low-ability experts are qualitatively more likely choose the decision aid -- a pattern likely generated by increased polarization through strategic expert separation. Furthermore, investments into the algorithmic decision aid do not deteriorate over time. This suggests that subjects' do not predict consumers to be generally averse towards algorithms.

\begin{figure}[h]
    \centering
\caption{Price setting depending on treatment, expert type and investment decision.}
\includegraphics[width=0.3\textwidth]{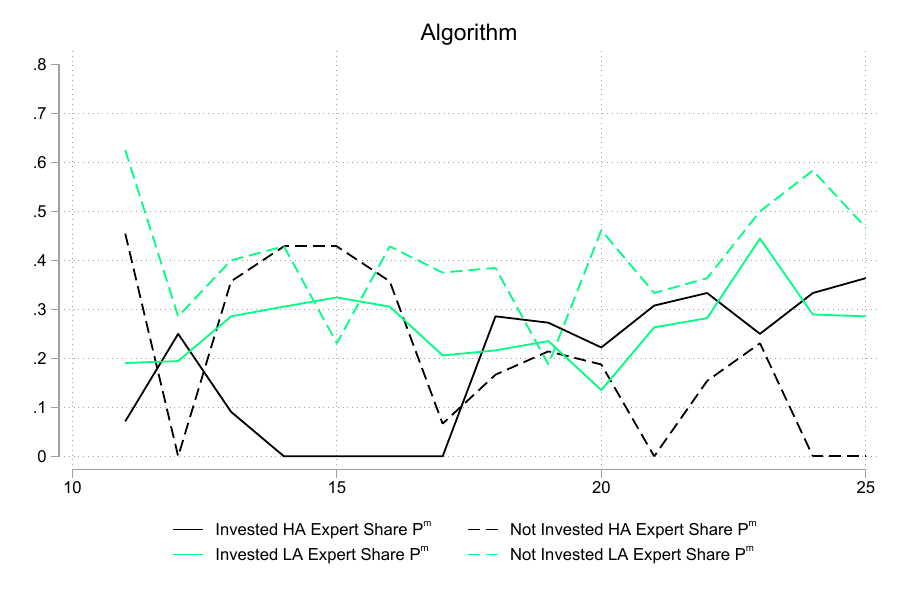}
\includegraphics[width=0.3\textwidth]{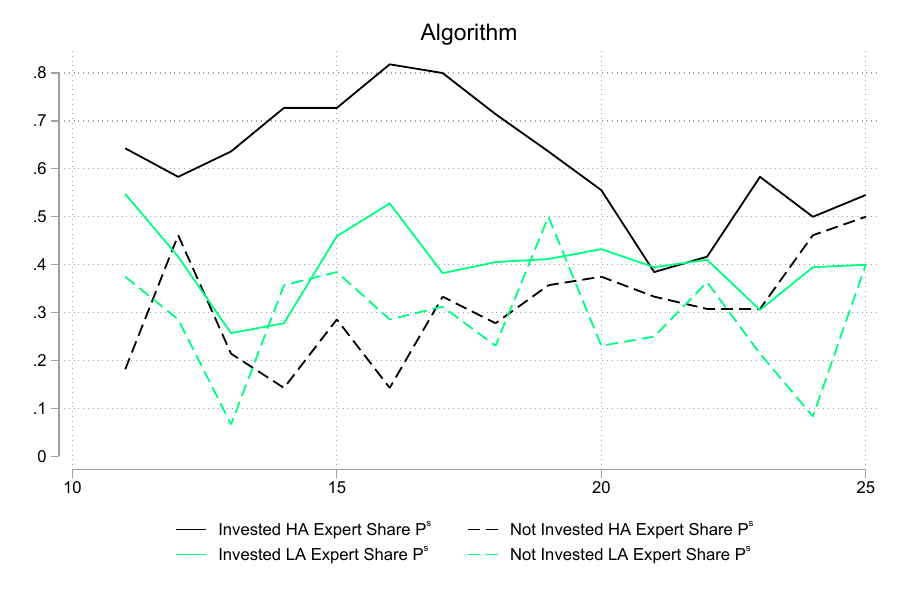}
\includegraphics[width=0.3\textwidth]{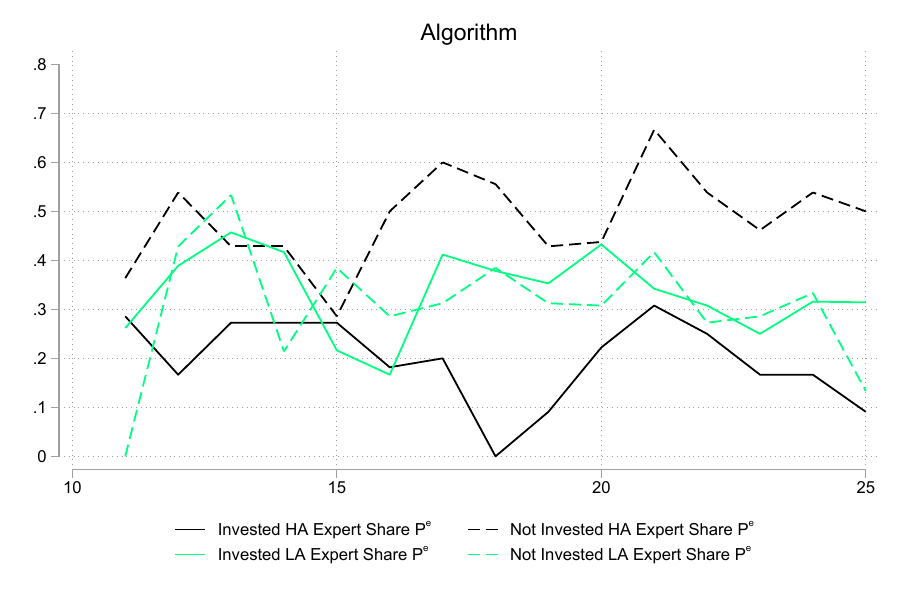}
\includegraphics[width=0.3\textwidth]{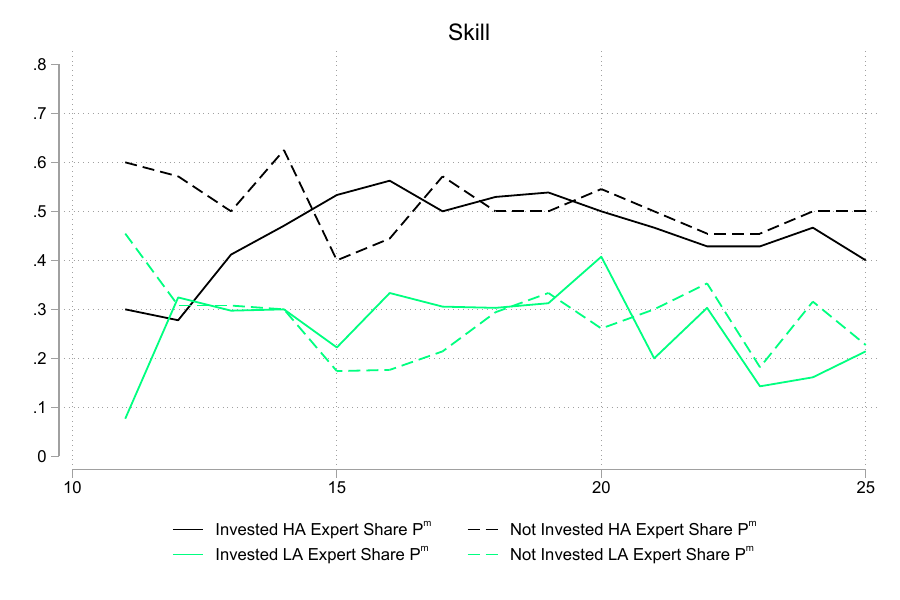}
\includegraphics[width=0.3\textwidth]{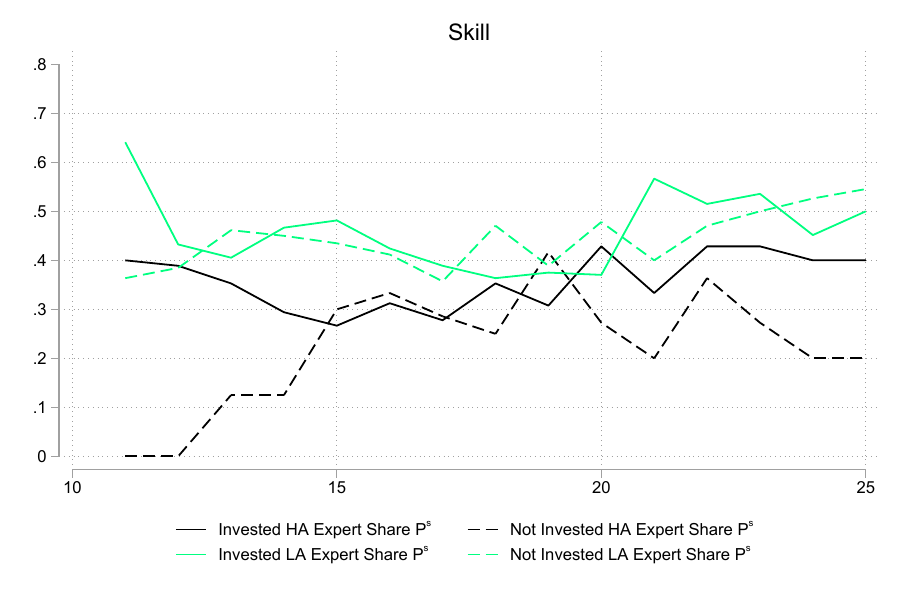}
\includegraphics[width=0.3\textwidth]{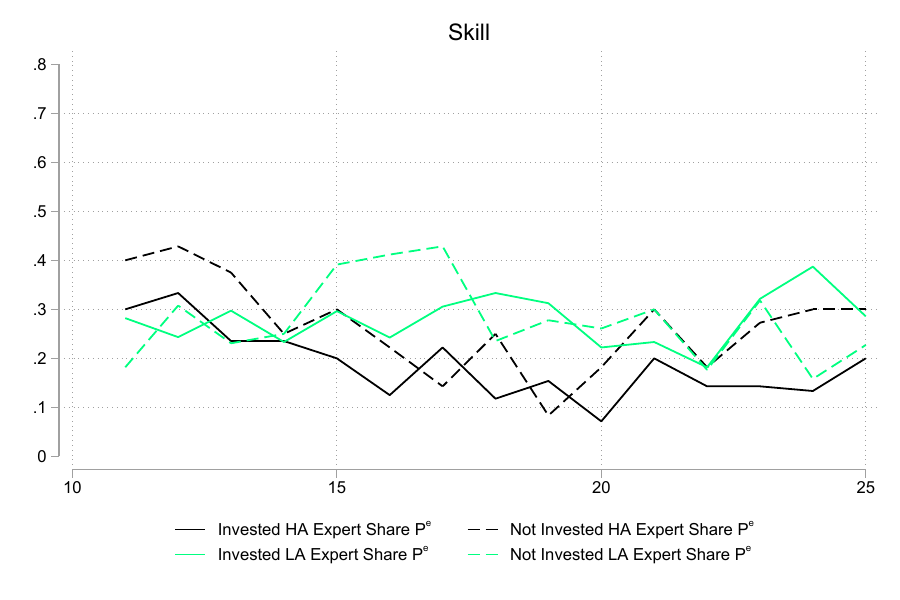}

\label{fig:fig_prices_p2}
\end{figure}

\noindent
\textbf{Expert Price Setting.} Figure \ref{fig:fig_prices_p2} shows the share of experts choosing a specific price vector depending on type, treatment and investment. We are primarily interested in potential strategic differences between high-ability experts who invest and those who forego investments. There are two potential effects. One, an expert may condition their pricing strategy on their current investment choice (within-subject variance). For instance, the standard model predicts a low-ability expert to change their pricing strategy from $\vect{P^m}$ to $\vect{P^e}$ after investing. However, generating the necessary variance implies that an expert switches between different investment and pricing strategies, which is largely inconsistent with signaling behavior. Therefore, we would expect high-ability experts in \textit{Algorithm} to exhibit lower within-subject variance. Second, experts who invest a lot may differ from expert who do not invest or only invest occasionally (between-subject variance). This would, e.g., be consistent with some proportion of high-ability experts foregoing the investment choice to signal their type. We first focus on the second case. Table \ref{tab:betsub_var_prices} shows random effects panel regression results for price setting conditional on the number of rounds an expert chooses to invest. For low-ability experts, and \textit{Skill} treatments, there are no differences. The number of investment choices does not predict price setting in either condition. For high-ability experts, there are large and significant effects of investing frequency on $\vect{P^s}$ and $\vect{P^e}$ in the \textit{Algorithm} treatment. Those who often rent the algorithmic decision aid are more likely to choose $\vect{P^s}$ -- the price menu that (1) has the highest consumer prices, (2) exhibits the highest expected value for experts and (3) splits the gains of trade relatively equally if experts behave honestly. On the other hand, high-ability experts who never or rarely invest into the decision aid are more likely to choose $\vect{P^e}$ -- the price menu that can only be imitated for a short period of time and maximizes expected consumer income if they approach a non-investing high ability expert.

\begin{table}[t]\centering
\def\sym#1{\ifmmode^{#1}\else\(^{#1}\)\fi}
\caption{Expert Price Setting Conditional on Investment Frequency}
\begin{adjustbox}{max width=\textwidth}
\begin{tabular}{@{}lcccccc|cccccc@{}}
& \multicolumn{6}{c}{Low-Ability Expert} & \multicolumn{6}{c}{High-Ability Expert} \\
\midrule

& \multicolumn{3}{c}{Skill} & \multicolumn{3}{c}{Algorithm} & \multicolumn{3}{c}{Skill} & \multicolumn{3}{c}{Algorithm} \\
\cmidrule(r){2-7} \cmidrule(l){8-13}
& $\vect{P^m}$ & $\vect{P^s}$ & $\vect{P^e}$ & $\vect{P^m}$ & $\vect{P^s}$ & $\vect{P^e}$ & $\vect{P^m}$ & $\vect{P^s}$ & $\vect{P^e}$ & $\vect{P^m}$ & $\vect{P^s}$ & $\vect{P^e}$ \\
\cmidrule(r){2-7} \cmidrule(l){8-13}
Investments & 0.004 & -0.002 & -0.002 & -0.009 & 0.001 & 0.013 & 0.001 & 0.006 & -0.003 & -0.004 & 0.045\sym{***} & -0.044\sym{***} \\
& (0.065) & (0.009) & (0.008) & (0.013) & (0.011) & (0.014) & (0.013) & (0.012) & (0.012) & (0.009) & (0.006) & (0.008) \\
\addlinespace
$N$ & 750 & 750 & 750 & 750 & 750 & 750 & 375 & 375 & 375 & 375 & 375 & 375 \\
\bottomrule
\end{tabular}
\label{tab:betsub_var_prices}
\end{adjustbox}
\begin{minipage}{0.95\textwidth} 
\vspace*{0.09cm}
{\footnotesize 
This table reports marginal effects of panel logistic regressions using subject-level random effects and a cluster–robust VCE estimator at the matched group level (standard errors in parentheses). The dependent variable is a binary variable that equals 1 if the expert chooses the respective price vector.
 \sym{*} \(p<0.05\), \sym{**} \(p<0.01\), \sym{***} \(p<0.001\)}
\end{minipage}
\end{table}
\begin{table}[t]\centering
\def\sym#1{\ifmmode^{#1}\else\(^{#1}\)\fi}
\caption{Expert Price Setting Conditional on Investment Decision}
\begin{adjustbox}{max width=\textwidth}
\begin{tabular}{@{}lcccccc|cccccc@{}}
& \multicolumn{6}{c}{Low-Ability Expert} & \multicolumn{6}{c}{High-Ability Expert} \\
\midrule

& \multicolumn{3}{c}{Skill} & \multicolumn{3}{c}{Algorithm} & \multicolumn{3}{c}{Skill} & \multicolumn{3}{c}{Algorithm} \\
\cmidrule(r){2-7} \cmidrule(l){8-13}
& $\vect{P^m}$ & $\vect{P^s}$ & $\vect{P^e}$ & $\vect{P^m}$ & $\vect{P^s}$ & $\vect{P^e}$ & $\vect{P^m}$ & $\vect{P^s}$ & $\vect{P^e}$ & $\vect{P^m}$ & $\vect{P^s}$ & $\vect{P^e}$ \\
\cmidrule(r){2-7} \cmidrule(l){8-13}
Invested & -0.134\sym{*} & 0.114 & 0.019 & -0.173\sym{*} & 0.235\sym{***} & -0.175\sym{***} & -0.119 & 0.207\sym{**} & -0.172\sym{*} & 0.009 & -0.012 & 0.008 \\
& (0.063) & (0.064) & (0.077) & (0.057) & (0.029) & (0.049) & (0.105) & (0.070) & (0.082) & (0.103) & (0.087) & (0.098) \\
\addlinespace
$N$ & 510 & 555 & 540 & 540 & 645 & 630 & 210 & 210 & 255 & 195 & 255 & 210 \\
\bottomrule
\end{tabular}
\label{tab:withinsub_prices}
\end{adjustbox}
\begin{minipage}{0.95\textwidth} 
\vspace*{0.09cm}
{\footnotesize 
This table reports marginal effects of fixed effects panel logistic regressions. The dependent variable is a binary variable that equals 1 if the expert chooses the respective price vector. Observations vary due to the omission of groups where investment shares predict prices perfectly.
 \sym{*} \(p<0.05\), \sym{**} \(p<0.01\), \sym{***} \(p<0.001\)}
\end{minipage}
\end{table}

These results are in line with our general proposition. In \textit{Algorithm}, there appear to be systematic differences between high-ability experts who invest into new diagnostic technology and those who do not. Because high-ability experts can influence consumer perceptions, some of them strategically choose a low-investment path with equal markup prices. Those who invest a lot tend to opt for $\vect{P^s}$, which is again inconsistent with the standard model, but consistent with behavior in Phase 1 and prior empirical evidence about the prevalence of selfish expert preferences \citep{kerschbamer2017social}. In \textit{Skill}, experts cannot differentiate themselves based on their original ability type. Therefore, all experts face the same incentives, and we would not expect two different ``groups'' that strategically choose either a low-investment or a high-investment strategy. Instead, experts should condition their behavior on their \textit{current price menu}. To test this, we opt for a fixed effects logistic panel regression with a binary variable measuring whether an expert invested in a given round. This captures within-subject differences in price setting conditional on the expert's current investment choice. Results from a pooled regression suggest that generally, investing reduces the share of experts choosing $\vect{P^m}$ and $\vect{P^e}$ in favor of $\vect{P^s}$ (see Table \ref{tab:pooled}). Separating regressions with regard to expert type and treatment reveals that these trends are present in all conditions, except high-ability experts in \textit{Algorithm} (Table \ref{tab:withinsub_prices}).

Thus, in our experiment, an investment into novel diagnostic technology that improves accuracy and simultaneously increases prices shifts experts towards the most expensive price menu with the lowest expected value for consumers. This tends to be true for both expert types across the two conditions, \textit{except} for high-ability experts after the introduction of an algorithmic decision aid. In contrast, high-ability experts in \textit{Algorithm} are the only group whose choices are explained by the absolute number of investments, with non- or low-investing experts opting for $\vect{P^e}$ instead of $\vect{P^s}$. Results therefore support the conjunction that high-ability experts diverge into two groups with the introduction of technological decision aids: one group that strategically under-invests to signal their ability type to consumers, and one group that follows the standard model's prediction by purchasing the algorithm. As predicted, this pattern only exists in \textit{Algorithm}, but not in \textit{Skill}.

\begin{table}[t]\centering
\def\sym#1{\ifmmode^{#1}\else\(^{#1}\)\fi}
\caption{Undertreatment Conditional on Expert Price Setting}
\begin{adjustbox}{max width=\textwidth}
\begin{tabular}{@{}lccccccccc@{}}
\toprule
 & \multicolumn{3}{c}{Phase 1} & \multicolumn{3}{c}{Skill} & \multicolumn{3}{c}{Algorithm} \\
\cmidrule(r){1-10} 
Undertreated & (1) & (2) & (3) & (1) & (2) & (3) & (1) & (2) & (3)   \\
\cmidrule(r){1-10}
$\vect{P^m}$ & -0.084 & &  & -0.153\sym{**} &  &  & -0.08 &  &   \\
& (0.054) &  & & (0.054) &  &  & (0.065) &  &  \\

$\vect{P^s}$ &  & 0.221\sym{***} &  &  & 0.252\sym{***} & &  & 0.121 &  \\
&  & (0.041) &  &  & (0.043) &  &  & (0.066) &  \\

$\vect{P^e}$ &  & & -0.152\sym{**} &  &  & -0.131\sym{**} & &  & -0.048  \\
&  &  & (0.049) &  &  & (0.047) &  & & (0.074) \\

\addlinespace
$N$ & 574 & 574 & 574 & 450 & 450 & 450 & 434 & 434 & 434  \\
\bottomrule
\end{tabular}
\label{tab:undertreated_reg}
\end{adjustbox}
\begin{minipage}{0.95\textwidth} 
\vspace*{0.09cm}
{\footnotesize 
This table reports marginal effects of panel logistic regressions using subject-level random effects and a cluster–robust VCE estimator at the matched group level (standard errors in parentheses). The dependent variable is a binary variable that equals 1 if the consumers experiences undertreatment.
 \sym{*} \(p<0.05\), \sym{**} \(p<0.01\), \sym{***} \(p<0.001\)}
\end{minipage}
\end{table}

\noindent
\textbf{Expert Undertreatment and Overtreatment.} The prevalence of $\vect{P^s}$ may be an indication that experts build a reputation of honesty and can therefore offer to split the gains of trade equally. Table \ref{tab:undertreated_reg} shows that this is not generally the case. Consumers who approach a $\vect{P^s}$-expert are more likely to be undertreated, whereas $\vect{P^e}$ is -- as theoretically predicted -- associated with lower rates of undertreatment. This is true for both \textit{Phase 1} and \textit{Skill}. Only in \textit{Algorithm}, the chosen price menu has no significant predictive power. Therefore, consumers cannot plausibly assume experts to treat them honestly irrespective of prices. Investment choices do not predict undertreatment.

\begin{table}[t!]\centering
\def\sym#1{\ifmmode^{#1}\else\(^{#1}\)\fi}
\caption{Expert Intent-To-Undertreat Conditional on Price Setting}
\begin{adjustbox}{max width=\textwidth}
\begin{tabular}{@{}lccccccccc@{}}
\toprule
 & \multicolumn{3}{c}{Phase 1} & \multicolumn{3}{c}{Skill} & \multicolumn{3}{c}{Algorithm} \\
\cmidrule(r){1-10} 
Undertreated & (1) & (2) & (3) & (1) & (2) & (3) & (1) & (2) & (3)   \\
\cmidrule(r){1-10}
$\vect{P^m}$ & -0.112\sym{***} & &  & -0.128\sym{**} &  &  & -0.090\sym{*} &  &   \\
& (0.028) &  & & (0.041) &  &  & (0.039) &  &  \\

$\vect{P^s}$ &  & 0.191\sym{***} &  &  & 0.213\sym{***} & &  & 0.199\sym{***} &  \\
&  & (0.031) &  &  & (0.056) &  &  & (0.066) &  \\

$\vect{P^e}$ &  & & -0.095\sym{**} &  &  & -0.097\sym{*} & &  & -0.149\sym{***}  \\
&  &  & (0.049) &  &  & (0.046) &  & & (0.038) \\

Invested &  & &  & -0.017  & -0.026  & -0.012 & -0.039 & -0.063\sym{*} & -0.047  \\
&  &  &  & (0.027) & (0.026) & (0.027) & (0.028)  & (0.025) & (0.028) \\

\addlinespace
$N$ & 1312 & 1312 & 1312 & 928 & 928 & 928 & 910 & 910 & 910  \\
\bottomrule
\end{tabular}
\label{tab:undertreated_intent_reg}
\end{adjustbox}
\begin{minipage}{0.95\textwidth} 
\vspace*{0.09cm}
{\footnotesize 
This table reports panel regressions using subject-level random effects and a cluster–robust VCE estimator at the matched group level (standard errors in parentheses). The dependent variable is the share of choices where an expert chooses a LQT for a diagnosed big problem.
 \sym{*} \(p<0.05\), \sym{**} \(p<0.01\), \sym{***} \(p<0.001\)}
\end{minipage}
\end{table}

We now look at undertreatment from the expert's perspective. Each round, experts receive three diagnoses, and make three treatment choices. Because of diagnostic uncertainty, an expert may intent to undertreat a consumer, but does not actually do so because of a wrong diagnosis. Therefore, we define the intent-to-undertreat as experts who diagnose a big problem, but prescribe the LQT, irrespective of the consumer's actual underlying condition. The results (Table \ref{tab:undertreated_intent_reg}) confirm that across all treatments, experts cheating behavior is strongly influenced by their chosen price menu. Experts intend to undertreat consumers significantly more often under $\vect{P^s}$, and significantly less often under the two other menus. Investment choices are largely irrelevant. There are no differences between expert types.

For overtreatment, we document all relevant regression analyses in the Appendix. As expected, consumers experience more overtreatment when approaching an expert with $\vect{P^m}$ or $\vect{P^e}$ in Phase 1 (Table \ref{tab:overtreated_reg}). For Phase 2, the results are similar, albeit less strong. In \textit{Skill} but not \textit{Algorithm}, consumers who approach an investing expert are also significantly less likely to be overtreated. Focusing on expert choices confirms the strong negative effect of $\vect{P^s}$ on overtreatment compared to both $\vect{P^m}$ and $\vect{P^e}$. Overall the results are in line with theory such that consumers should expect $\vect{P^s}$-experts to be much more likely to undertreat them. Across all rounds and conditions, intention-to-undertreat rates are 40\% under $\vect{P^s}$, and only 12\% and 11\% for $\vect{P^m}$ and $\vect{P^e}$ respectively. On the other hand, intention-to-overtreat rates are only 22\% under $\vect{P^s}$, and 48\%/43\% for $\vect{P^e}$/$\vect{P^m}$. Investment choices generally do not affect cheating intentions. 

\begin{table}[t!]\centering
\def\sym#1{\ifmmode^{#1}\else\(^{#1}\)\fi}
\caption{Consumer Switching Behavior}
\begin{adjustbox}{max width=\textwidth}
\begin{tabular}{l*{7}{c}}
\toprule
&\multicolumn{1}{c}{\specialcell{Switched Away -- Phase 1}}&\multicolumn{1}{c}{\specialcell{Switched Away -- Phase 1}}&\multicolumn{1}{c}{\specialcell{Switched Away -- Skill}} &\multicolumn{1}{c}{\specialcell{Switched Away -- Skill }} &\multicolumn{1}{c}{\specialcell{Switched Away -- Algorithm }} &\multicolumn{1}{c}{\specialcell{Switched Away -- Algorithm}}\\
\midrule

Undertreated &  0.377\sym{***} &  0.374\sym{***}&  0.256\sym{***}  & 0.259\sym{***} & 0.208\sym{**} & 0.217\sym{**} \\
            & (0.051)         & (0.049)         & (0.067)  & (0.059) & (0.068) & (0.063) \\
Overtreated & 0.046 &  0.049 & 0.042 & 0.048  & 0.042 & 0.049\sym{*}  \\
        &   (0.026)   &  (0.026)    & (0.038)  &  (0.041) & (0.028) & (0.024) \\
$\vect{P^m}$ &  & 0.024 &   & -0.268 & & -0.229 \\
 &       & (0.092)    &   & (0.253)  & & (0.195)\\
$\vect{P^s}$ &  &  -0.004&  & -0.156  & & -0.209 \\
 &        & (0.095)      &   & (0.248)  & & (0.212) \\
$\vect{P^e}$ &  &  0.039&   & -0.364 & & -0.162 \\
 &         & (0.093)      & & (0.238)  & &  (0.204)\\ 
Round &  &  -0.008 &   & -0.011\sym{*}  & & -0.016\sym{***} \\
 &         & (0.006)      & & (0.005)  & & (0.004) \\ 
Risk &  &  0.037\sym{***} &   & 0.027\sym{*} & & 0.039\sym{***}\\
 &         & (0.009)      & & (0.012)  & & (0.011) \\ 
Invested &  &  &   & 0.073\sym{*}  & & -0.006 \\
 &         &       & & (0.034)  & & (0.039)\\
Invested\_LR &  & &   & 0.115\sym{*}  & & 0.021 \\
 &         &      & & (0.046)  & & (0.036)\\
\\

\midrule
$N$          &3600       &3600       &1050   & 1050 & 1050 & 1050\\
\bottomrule
\end{tabular}
\label{tab:consu_sw}
\end{adjustbox}
\begin{minipage}{0.95\textwidth} 
\vspace*{0.09cm}
{\footnotesize 
This table reports marginal effects of panel logistic regressions using subject-level random effects and a cluster–robust VCE estimator at the matched group level (standard errors in parentheses). The dependent variable is a binary variable that equals 1 if the consumer switched to a new expert in the current round. \textit{Undertreated}, \textit{Overtreated} and \textit{Invested\_LR} are lagged variables (one round).
 \sym{*} \(p<0.05\), \sym{**} \(p<0.01\), \sym{***} \(p<0.001\)}
\end{minipage}
\end{table}

\noindent
\textbf{Consumer Choices.} How do consumers react to undertreatment with diagnostic uncertainty? Table \ref{tab:consu_sw} shows that experiencing undertreatment strongly predicts consumer switching in the following round. This holds for both phases and treatments. Overtreatment appears to have no or only little effect on consumer choices. Furthermore, consumers do not appear to condition their switching decision on the expert's chosen price menu. Yet, looking at the number of consumers an expert attracts confirms that the former do exhibit strong preferences against $\vect{P^s}$, and for $\vect{P^m}$ (Table \ref{tab:expert_approached} in the Appendix). In line with consumer switching, undertreating consumers in the previous round also significantly reduces an expert's current number of consumers. Investments do not meaningfully affect an expert's ability to attract consumers. This may be one reason why we find consistent under-investment by experts. Finally, a three-way interaction of expert type, the expert's investment choice, and $\vect{P^e}$ suggests that high-ability experts in \textit{Algorithm} attract significantly more consumers when they choose $\vect{P^e}$ and simultaneously forego the algorithmic decision aid. This result is in line with our theoretical prediction, and does not hold for either \textit{Skill} or low-ability experts, nor for any other price menu. Thus, it provides supportive evidence for the hypothesis that high-ability experts may strategically alter their utilization of costly decision aids under obfuscation to influence consumer beliefs about their skill type.

\section{Experiment 2: The Role of Repeated Interactions and Beliefs \label{sec:e2}}
We conduct a second experiment to replicate our findings for \textit{Algorithm}, demonstrate the importance of reputation building through repeated future interaction, and capture incentivized consumer beliefs. The basic parameters of the game are the same as above, except that we increase investment costs $d$ from $10$ to $12$. This reduces the expected payoff approaching an investing expert to $\pi_{alg,inv} \leq 55 < \pi^e_{ha,ninv} = 57$. The only difference is that we do not strictly rely on uncertainty surrounding the decision aid when making our predictions. Treatment-wise, we differentiate between a \textit{repeated} (25 rounds) and a \textit{one-shot} (11 rounds) paradigm.

In a one-shot setting, where learning has no utility because consumers do not expect their updated beliefs to affect future outcomes, there is no separation between expert types. There are two possible scenarios. If consumers \textit{do not}  (sufficiently) interpret the investment decision as an ability signal ($\tilde{\gamma} < \tilde{\gamma}_{ninv}$), then high-ability experts cannot differentiate themselves and all experts should rent the decision aid as long as uncertainty $w \leq 0.48$ to offer consumers the highest expected payoff.\footnote{Expert ability is fully obfuscated. Experts only forego the decision aid if $\gamma \pi^e_{ninv,ha} + (1-\gamma)\pi^e_{ninv,la} \geq  (1-w)\pi^e_{inv} + w(\pi^e_o - 10)$ where $\gamma = \frac{1}{3}$ because there is no signalling.} If consumers \textit{do} interpret the investment decision as a negative ability signal, then experts condition their investment choice on their beliefs about other experts. To illustrate, we will assume here that the signal is strong enough such that consumers find it profitable to approach the high-ability expert after, ostensibly, identifying them.\footnote{$\, \tilde{\gamma} \pi^e_{ninv,ha} + (1-\tilde{\gamma})\pi^e_{ninv,la} \geq (1-w)\pi^e_{alg,inv} +  w(\tilde{\gamma}\pi^e_{ninv,ha} + (1-\tilde{\gamma})\pi^e_{ninv, la})$} Table \ref{payoff:os_simple} shows a simple payoff matrix relating to the share of consumers an expert can attract under these circumstances. 

\begin{table}[h]
\centering
\caption{Expert 1: Share of attracted consumers conditional on the choices of Expert 2 (E2) and Expert 3 (E3).}
\label{payoff:os_simple}
\renewcommand{\arraystretch}{1.5}
\begin{tabular}{cc}
\begin{tabular}{cc|c|c|}
  & \multicolumn{1}{c}{} & \multicolumn{2}{c}{E3}\\
  & \multicolumn{1}{c}{} & \multicolumn{1}{c}{Invest} & \multicolumn{1}{c}{Not Invest} \\\cline{3-4}
  \multirow{2}{*}{E2} & Invest & $\frac{1}{3}$ & 0 \\\cline{3-4}
  & Not Invest & 0 & 1 \\\cline{3-4}
\end{tabular}
&
\begin{tabular}{cc|c|c|}
  & \multicolumn{1}{c}{} & \multicolumn{2}{c}{E3}\\
  & \multicolumn{1}{c}{} & \multicolumn{1}{c}{Invest} & \multicolumn{1}{c}{Not Invest} \\\cline{3-4}
  \multirow{2}{*}{E2} & Invest & 1 & 0 \\\cline{3-4}
  & Not Invest & 0 & $\frac{1}{3}$ \\\cline{3-4}
\end{tabular}
\\[3em]
Expert 1 Chooses \textit{Invest} & Expert 1 Chooses \textit{Not Invest}
\end{tabular}
\end{table}

The simple intuition behind these payoffs is the same as before: Consumers know that high-ability experts can make a better offer without investments, while low-ability experts cannot. Therefore, if two experts invest, and one expert does not invest, all consumers will identify the not-investing expert as a high-ability expert and approach them (consumer share = 1). Knowing that, low-ability experts are incentivized to \textit{not invest} and create a pooling equilibrium in which consumers must choose randomly, which leads to experts sharing consumers equally in expectation on average (consumer share = $\frac{1}{3}$). Once both other experts forego the investment, the third expert now maximizes their attraction by \textit{investing} -- as long as the uncertainty surrounding the efficacy of the algorithmic decision aid is not exorbitantly large. That is because consumers are uncertain if they approach the not-investing high-ability or the not-investing low-ability expert. In that case, $\pi^e_{alg,ninv} = 0.5*\pi^e_{ha} + 0.5*\pi^e_{la} = 48.5 < \pi^e_{alg,inv}$. Without learning, and under obfuscated abilities, all experts face the same basic decision structure. Hence, there are only mixed equilibria.

Participants first complete 10 rounds if a standard credence goods game with diagnostic uncertainty. Due to ability obfuscation, there is a pooling equilibrium, and each experts might have attracted $x \in \{0, 1, 2, 3\}$ consumers in round 10. From any expert's ($e_1$) perspective, there are three cases. The other two do not invest: $(1-p_{e2})(1-p_{e3}$); The other two do invest: $p_{e2}p_{e3}$; One other invests: $p_{e2}*(1-p_{e3}) + (1-p_{e2})*p_{e3}$. The expected payoff for an expert to (not) invest is:
\begin{equation*}
    \pi_{inv} = [(1-p_{e2})(1-p_{e3})] (3-x) + [p_{e2}p_{e3}]*0 + [p_{e2}*(1-p_{e3}) + (1-p_{e2})*p_{e3}](-x),    
\end{equation*}
\begin{equation*}
    \pi_{ninv} = [(1-p_{e2})(1-p_{e3})]*0 + [p_{e2}p_{e3}] (3-x) + [p_{e2}*(1-p_{e3}) + (1-p_{e2})*p_{e3}](-x).
\end{equation*}
If all three experts choose the same strategy, no consumer has an incentive to switch. The difference is $ \pi_{inv} -  \pi_{ninv} = (1-p_{e2}-p_{e3})(3-x)$, which simplifies to $1-(p_{e2}+p_{e3})$. In the one-shot scenario, the number of current consumers $x$ does not play a role. If $p_{e2}+p_{e3} > 1$, the best response is to not invest. If $p_{e2}+p_{e3} < 1$, the best response is to invest. If $p_{e2}+p_{e3} = 1$, then the expert is indifferent. In this case, all experts have the same best response, thus giving $2p = 1$ and $p = 0.5$.\footnote{Note that this is different with repeated rounds, because consumers strategically approach experts to learn about their type.} The unique mixed equilibrium is $p = 0.5$.

 \textbf{Proposition 2.} \textit{When expert abilities are fully obfuscated at the time of the technological shock, neither consumers nor experts expect repeated future interactions, and experts believe that foregoing the decision sufficiently affects consumer beliefs about expert type, then all experts randomize their investment choice.}

\begin{equation*}
    \left\{\begin{array}{lr}
        \text{Invest} & \,\text{if}\, p_{e2}+p_{e3} < 1,\\
        \text{Not Invest} & \,\text{if}\,p_{e2}+p_{e3} > 1,\\
        \text{Indifferent} & \, \text{if}\,p_{e2}+p_{e3} = 1
        \end{array}\right\}
\end{equation*}\\

Whereas we predict high- and low-ability experts to separate in the \textit{repeated} condition, experts who predict a signaling game should randomize their investment choice in \textit{one-shot}, causing under-investments from low-ability experts and over-investments from high-ability experts.

\subsection{Experimental Design}
We implement a between-subject design with two treatments that are based on the \textit{Algorithm} condition from Experiment 1. In \textit{repeated}, subjects complete the original \textit{Algorithm} procedure, except that investment costs increase by $2$ such that $d = 12$, and that we elicit incentivized consumer and expert beliefs at the end of the experiment. In \textit{one-shot}, subjects only complete 11 rounds, meaning there is only one round after the technological shock. To ensure incentive-compatibility, participants learn that their payoff from the last round will be multiplied by 4 when calculating their bonus. Other than that, nothing changes. For belief elicitation, we follow an (incentivized) introspective approach \citep{charness2021experimental}. We ask consumers (1) the likelihood that an expert who chooses to rent the decision aid does not use it, (2) the likelihood that an expert who chooses to rent the decision aid does not use it correctly and hence does not achieve the best-possible prediction accuracy of 90\%, and (3) the likelihood that an expert who chooses to rent the decision aid is a high-ability expert. For each question, we ask (not incentivized) what the consumer themselves thought in round 11, as well as (incentivized) the average likelihoods across all observations. For the latter, we inform consumers that they earn \$2 for a perfect prediction. For each point that their prediction is off by, that \$2 is reduced by \$0.1. Similarly, experts state two beliefs about consumer beliefs that an expert who chooses to rent the decision aid is of high-ability, one non-incentivized about their own beliefs in round 11, and one incentivized about average consumer beliefs. In addition, experts indicate who they think is more likely to rent the algorithmic decision aid. We rely on simple introspection, as opposed to, e.g., scoring rules, because they have been shown to be effective even without incentives \citep{charness2021experimental,trautmann2015belief} and have the upside of being very simple. Complex rules can distort subject beliefs \citep{danz2022belief}, which may be especially likely after a long and complex main game such as this one. Participants indicate their beliefs after the main game to avoid influencing behavior. While experiences throughout the game might affect subject belief measurement, it is more important to avoid distorting actual behavior. We collected data from 312 participants via Cloud Research's \textit{Connect} platform \citep{hartman2023introducing}. All participants completed at least 10 prior tasks with a minimal approval rating of 90\%, reside in the US, and successfully answered 4 comprehension questions within three trials. The base payment is \$7.

\subsection{Results -- Beliefs}
We first look at consumer and expert beliefs. Experts in \textit{repeated} generally believe that low-ability experts are most likely to rent the decision aid (59\%), followed by ``equally likely'' (35\%), whereas only 6\% opt for the high-ability expert. This is in line with our expectations. Yet, when asked to indicate their incentivized beliefs about how consumer beliefs that an expert is a high-ability expert would change after an expert invests, 80\% indicate a positive probability, suggesting that consumers would be more likely to identify an expert as high-ability after renting the decision aid (Figure \ref{fig:e2_expertbeliefs} in the Appendix). While there may be some measurement error due to the involvement of negative probabilities, results suggest that a majority of experts expects consumers to draw the wrong inference from investments. For consumers, 17\% believe that the investment choice is not an informative signal about expert type (stated likelihood that an investing expert is a high-ability expert = $33\%$), 32\% believe that investing reduces the likelihood of an expert being of high ability type, and 51\% believe that investing increases the likelihood (Figure \ref{fig:e2_expertbeliefs} in the Appendix). Hence, there is substantial heterogeneity in consumer beliefs. Note again, however, that these might be affected by participant experience during the experiment. To approximate the extent to which consumers perceive uncertainty around the decision aid (see Experiment 1), we also ask about the likelihood that experts choose to ignore the decision aid, or do not use the corresponding information from the algorithm correctly. While most consumers judge that to be unlikely, average incentivized consumer beliefs that an expert does not use the system are around 30\% (Figure \ref{fig:e2_cons_beliefs} in the Appendix), corroborating that uncertainty may play a substantial role. The belief results for \textit{one-shot} are similar, suggesting a limited effect of experimental exposure on belief formation. A larger percentage of consumers beliefs that investments have no informative value for ability type (28\%), and the average incentivized indicated likelihood that an expert does not use the decision aid is only 22\%. On the expert side, roughly 36\% (60\%) believe that the effect of investing on consumer high ability beliefs would be negative (positive). Overall, despite some moderate changes, belief patterns between treatment remain comparable.

\subsection{Results -- Repeated vs. One-Shot}

\begin{figure}[t]
    \centering
    \caption{Left: Expert investment shares for \textit{repeated} in the \textit{Algorithm} paradigm over the 15 rounds of Phase 2. Right: High-ability expert investment shares dis-aggregated by direction of incentivized beliefs. }    
    \includegraphics[width=0.45\textwidth]{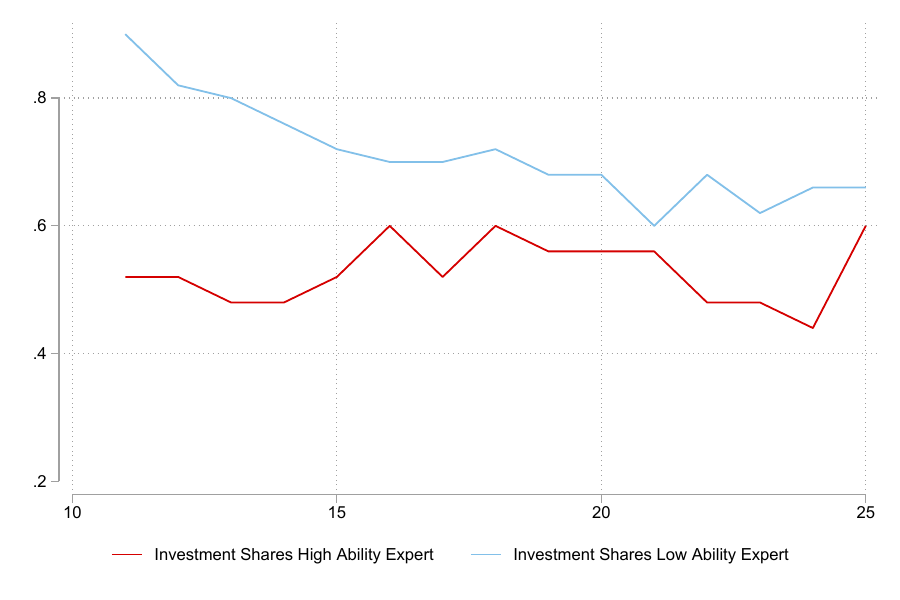}
    \includegraphics[width=0.45\textwidth]{fig/e2_repeated_diff_beliefsincentivized}    
    \label{fig:e2_repeated_investments}
\end{figure}

Figure \ref{fig:e2_repeated_investments} shows investment shares for \textit{repeated}. In line with our prediction and Experiment 1, there is substantial expert type separation when learning through repeated future interactions is possible. With the appearance of the technological shock in round 11, 52\% of high-ability experts and 90\% of low-ability experts rent the decision aid ($\tilde{\chi}^2 = 13.73, p < 0.001$). Over time, that difference becomes significantly smaller, mostly because low-ability experts invest less. This stands in contrast to our results above, where the gap did not meaningfully change over time. Two potential reasons are the increased investment costs $d$ and simple population effects, as we utilize a different subject pool from a different platform, with an experiment that is quite sensitive to a number of individual parameters. Still, a random effects panel logistic regression finds a significantly negative effect of high-ability expert type on the likelihood to invest (Table \ref{tab:reg_invest_e2}).

Theoretically, this result should be driven by the sub-population of high-ability experts who expect consumers to adjust their beliefs about an expert's high-ability type downwards after observing investments. Otherwise, foregoing the decision aid cannot function as a competence signal. The right panel of Figure \ref{fig:e2_repeated_investments} shows high-ability expert investment shares disaggregated by belief category. A negative belief indicates that experts expect consumers to belief that an investing expert is less likely of high ability type, a positive belief indicates the opposite. As predicted, the separation of experts appears almost entirely driven by experts who believe that consumers read not-investing as a competence signal. In contrast, there is almost no difference between high- and low-ability experts if high-ability experts do not think that foregoing the algorithm can convince consumers of their type. Hence, results suggest that expert beliefs about consumer beliefs drive the separation between expert types. In line with our theory, high-ability experts engage in a signaling game by exploiting the appearance of a novel technology if that strategy is consistent with predicted consumer beliefs. Otherwise, investments remain on a similarly high level.

\begin{figure}[t]
    \centering
    \caption{Left: Expert investment shares for \textit{one-shot} in the \textit{Algorithm} paradigm in Round 11. Right: Expert investment shares dis-aggregated by direction of incentivized beliefs. }    
    \includegraphics[width=0.45\textwidth]{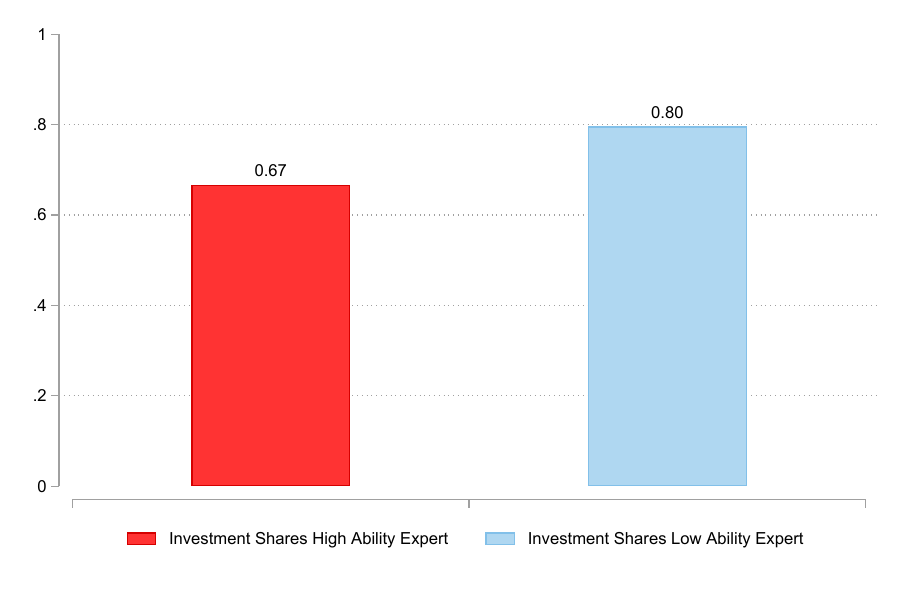}
    \includegraphics[width=0.45\textwidth]{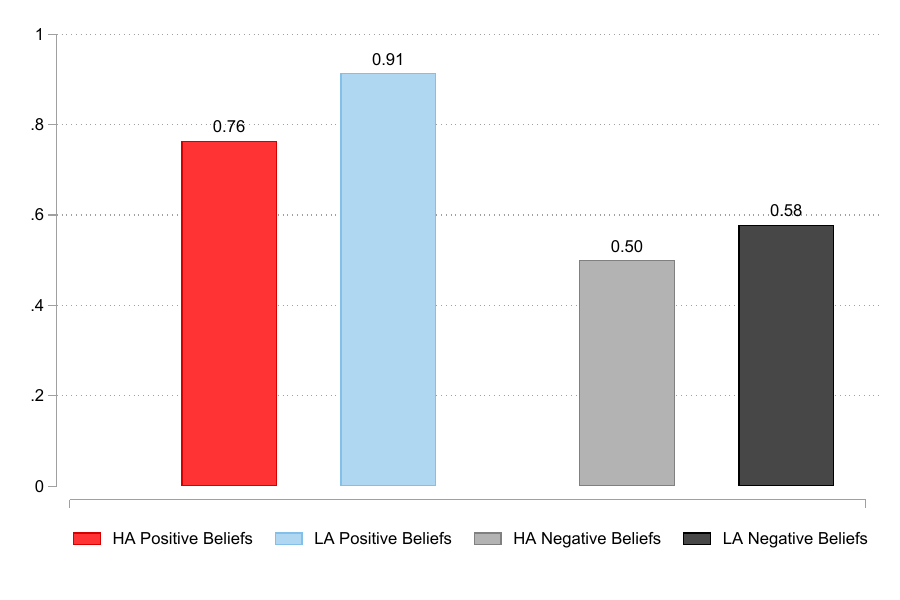}    
    \label{fig:e2_one_investments}
\end{figure}

Second, Figure \ref{fig:e2_one_investments} shows expert investments shares in round 11 for \textit{one-shot}. In contrast to \textit{repeated}, there is no significant difference between high- and low-ability experts (67\% vs. 79\%, $\, \tilde{\chi}^2 = 1.627, p = 0.202$). Moreover, categorizing experts according to their beliefs about consumer beliefs suggests that (1) both high- and low-ability experts tend to rent the algorithm with a probability of around 50\% when they believe that consumers take not-investing as a competence signal, and (2) a large majority of both expert types invests if they believe investing transports no information or suggests that an expert is of high ability type. While sub-sample analyses should be interpret cautiously given the sample size, these results align with expert behavior in \textit{repeated} as well as our theoretical prediction. In the absence of future learning opportunities, experts whose ability differences are obfuscated and who believe that consumers enter a signaling game randomize their investment decision. This generally leads to over-investments from high-ability experts, and under-investments from low-ability experts. On the other hand, experts who do not think that consumers interpret foregoing the algorithm as a competence signal are always incentivized to invest, irrespective of type. In our data, this leads to higher investment shares across the board, resulting in less under-investments from low-ability experts and more over-investments from high-ability experts. On aggregate, high-ability experts invest more than in \textit{repeated}, while low-ability experts invest less -- which, regarding efficiency, is undesirable in both directions. Comparing expert behavior in round 11 between \textit{repeated} and \textit{one-shot} by their beliefs (see Figure \ref{fig:e2_all_beliefs} in the Appendix) suggests that high-ability experts, who belief that consumers take not-investing as a positive ability signal, generally do not rent the decision aid (0\%) with repeated future interactions, while those in \textit{one-shot} randomize (50\%). Similarly, signaling low-ability experts always invest in \textit{repeated} (100\%), but randomize in \textit{one-shot} (58\%). Overall, results from Experiment 2 corroborate all our theoretical predictions and largely replicate Experiment 1.

\section{Concluding Remarks\label{sec:conclusion}}
Technological breakthroughs have the potential to significantly alleviate inefficiencies on credence goods markets. While full automation of expert tasks is usually neither feasible, nor desirable or legal, algorithmic decision aids offer huge potential in complementing or sometimes even substituting human expertise. In this paper, we argue that two prevalent characteristics of real-life credence goods markets can affect the effective adoption of novel diagnostic technologies: diagnostic uncertainty and obfuscated heterogeneous expert abilities. Consumers know that experts cannot perfectly identify their problem, but also that some are better than others. Information asymmetries preclude consumers from accessing that information, hurting both their welfare and the welfare of high-ability experts. Technological shocks, like the invention of a new diagnostic medical test or large language models, offer an opportunity for experts to influence consumer perceptions of their abilities. That is because consumers know that high-ability experts not only derive smaller benefits from decision aids, but are additionally hurt by the lack of skill expression. Once novel technologies allow lower ability experts to costly catch up to higher ability experts, we show that the presence of Bayesian consumers limits the low-ability consumer's capacity for imitation. Then, by strategically ignoring the decision aid, a high-ability expert can signal consumers their ability type, and simultaneously offer a more favorable price.  In reality, the latter may represent (1) lower diagnostic costs for experts or (2) less services for the consumer. In the medical domain, for instance, it is common for patients to pay for additional or advanced tests that are not covered by insurances but increase the chance of a correct diagnosis.

In combination, the obfuscation of heterogeneous abilities with the appearance of a novel technology gives rise to a set of signaling games, many of which, as we show, can lead to a separation of expert strategies, depending on the (1) the mechanism and (2) a number of variables such as the imitation horizon and consumer preferences for a ``safe'' option. These may result in very efficient diagnostic paths where low-ability experts are much more likely to invest than high-ability experts. Yet, they can also imply widespread rejection of novel technologies through low-ability experts' imitation behavior. Under certain conditions, like very noisy and slow consumer learning curves, a small number of repeated future interactions, and a low market share for the high-ability expert, there are even stable equilibria in which only the high-ability expert invests -- maximizing both over- and under-utilization of the decision aid. In addition, we show that without repeated future interactions, all experts who enter a signaling game randomize their investment choice, whereas without signaling, all experts would invest. As randomization simultaneously implies over-utilization from high-ability experts and under-utilization from low-ability experts, theory suggests very inefficient technology adoption paths in domains with limited reputation building, such as highly specialized services which people only very infrequently demand. This may, for example, be one reason why in medicine, both under- overtesting are very common. 

We test these ideas using two online credence goods experiments in which experts experience a technological shock after 10 rounds and thereby receive the option to costly increase their diagnostic precision. The first experiment differentiates between skill investments, and algorithmic decision aids. Skill investments over-ride an expert's prior ability type, representing for example advanced or specialized training. Algorithmic decision aids complement the expert's diagnosis, without affecting their personal ability type. Therefore, experts who do not use or disregard the system may still rely on their own diagnostic skill. Our results show no differences between high-ability and low-ability expert for skill investments. With the algorithmic decision aid, however, high-ability experts are far less likely to pay for increased diagnostic precision. Auxiliary results suggest that investing and non-investing high-ability experts exhibit different pricing strategies, with those largely foregoing the decision aid predominantly relying on price menus that signal honesty. On the other hand, investing experts opt for prices that are more favourable to them. These results offer preliminary evidence that high-ability experts may strategically (under-)utilize technology to differentiate themselves from low-ability experts. They also suggest that the adoption of algorithmic decision aids could be conditional on the distribution of skills within a profession, and that negative consumer attitudes towards professionals with decision aids may be rooted in actual ability differences.

Overall, experts in our experiment substantially under-invest into new diagnostic technologies. This is one reason why efficiency gains are small to negligible. Investments do not meaningfully affect expert fraud, which may be partially driven by very high consumer participation rates. Throughout the experiment, 95\%-98\% of consumers enter the market, which speaks to the potential efficiency of uncertain credence goods markets under verifiability. Consumers react negatively towards undertreatment, but not overtreatment, and generally do not reward expert investments. In contrast to the predictions of the standard model, consumers also do not appear to reward expert prices that signal honesty. Instead, they avoid high prices with expert-incentives to under-treat, but approach lower prices with expert-incentives to over-treat. This strongly reduces investment incentives.

The second experiment provides evidence that there is indeed little to no expert separation without the possibility of reputation building, which leads to strong over-investments from high-ability experts. Furthermore, results suggest that separation is driven by experts who think that foregoing the decision aid is a competence signal, and that this perception leads to highly polarized -- and therefore efficient -- investment patterns. Hence, the inability of consumers to observe expert abilities is very costly, and can, in the absence of reputation building, cause randomized technology adoption. This highlights the constant need of establishing a better information system for consumers, or to allow for de-centralized mechanisms like prices to better internalize dispersed consumer knowledge. This holds specifically of sectors in which (1) consumer demand is rare, e.g., highly specialized services, or (2) sectors in which consumers interact with a mass of potential experts, e.g., hospitals or large insurance firms. On the plus side, technological progress itself can -- under the right circumstances -- endow experts with an information mechanism that allows them to escape inefficient pooling equilibria while simultaneously motivating large-scale algorithmic adoption by low-ability experts -- where it has the largest marginal benefit.

Our findings provide first evidence on the efficacy of algorithmic decision aids on credence goods markets with obfuscated, heterogeneous expert abilities. We hope they provide some useful orientation for future research. In particular, there is much to learn about the relationship between expert investments, technology, and consumer learning/beliefs. For example, identifying institutional or environmental factors that inhibit or promote consumer learning could be highly valuable. It not only reduces information asymmetries and thereby avoids exploitation, but also plays a crucial role in pushing experts towards efficient technology adoption paths. More generally, the role of signalling for creating efficient equilibria in the context of obfuscated information is highly important, with concrete implications for practical interventions that allow high-ability people to differentiate themselves. Finally, carefully isolating and quantifying the mechanism behind (a) what causes experts and consumers to engage in a signalling game, (b) consumer reactions towards experts with decision aids, and (c) the role of ability distributions and market conditions such as excess demand or monopoly actors, is crucial to better understand the transformative impact of technology on credence goods markets. This generally necessitates a better understanding of the beliefs and expectations of experts and consumers alike.

\clearpage
\bibliography{\bib}

\clearpage
\section{Appendix \label{sec:app}}
\begin{figure}[h]
\centering
\caption{Threshold $\bar{r_i}$ condition on uncertainty and consumer beliefs.}
\includegraphics[width=0.8\textwidth]{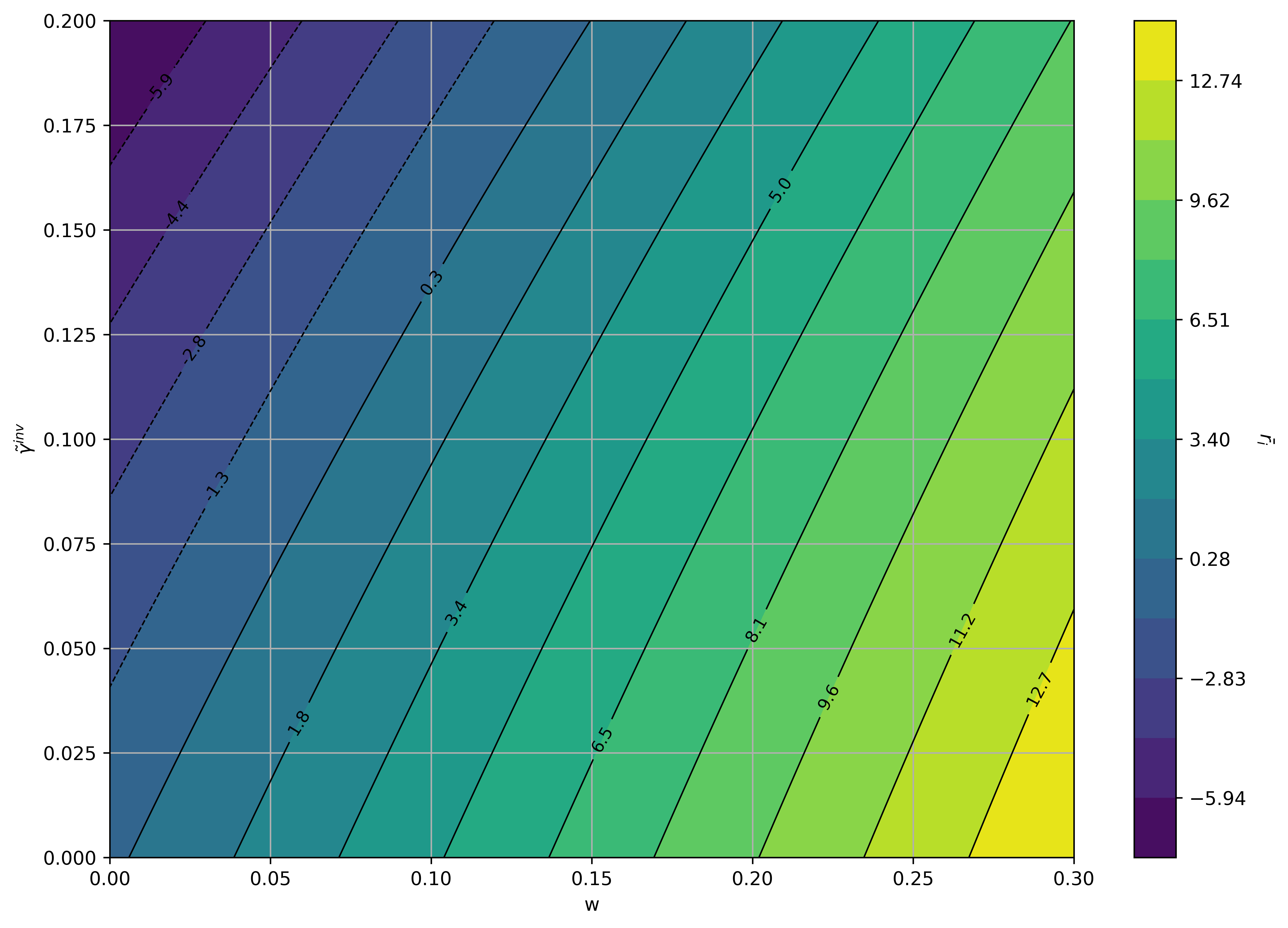}
    \label{fig:contour}
\end{figure}

\clearpage
\subsection{Payoffs}
\begin{table}[h]
\centering
\small
\caption{High-Ability Expert: Expected consumer attraction conditional on the three experts' investment choices in round 11 for \textit{Algorithm} if all consumers coordinated on the high-ability expert, i.e. $x_{ha} = 3$.}
\label{tab:alg_ha_3}
\renewcommand{\arraystretch}{1.2}
\setlength{\tabcolsep}{4pt}
\begin{tabular}{@{}c@{\hspace{0.5em}}c@{}}
\begin{tabular}{cc|c|c|}
  & \multicolumn{1}{c}{} & \multicolumn{2}{c}{LA$_0$}\\
  & \multicolumn{1}{c}{} & \multicolumn{1}{c}{Invest} & \multicolumn{1}{c}{Not Invest} \\\cline{3-4}
  \multirow{2}{*}{LA$_0$} & Invest & 1 & $(1-\alpha)$ \\\cline{3-4}
  & Not Invest & $(1-\alpha)$  & 1 \\\cline{3-4}
\end{tabular}
&
\begin{tabular}{cc|c|c|}
  & \multicolumn{1}{c}{} & \multicolumn{2}{c}{LA$_0$}\\
  & \multicolumn{1}{c}{} & \multicolumn{1}{c}{Invest} & \multicolumn{1}{c}{Not Invest} \\\cline{3-4}
  \multirow{2}{*}{LA$_0$} & Invest & 1 & $0.5\frac{R-r_i}{R}$ \\\cline{3-4}
  & Not Invest & $0.5\frac{R-r_i}{R}$ & $\begin{array}{c} \alpha + \\ (1-\alpha)\frac{R-2r_i}{R} \end{array}$ \\\cline{3-4}
\end{tabular}
\\[4.5em]
HA$_3$ Expert Chooses \textit{Not Invest} & HA$_3$ Expert Chooses \textit{Invest}
\end{tabular}
\end{table}
\begin{table}[h]
\centering
\footnotesize
\caption{Low-Ability Expert: Expected consumer attraction conditional on on the three experts' investment choices in round 11 for \textit{Algorithm} if all consumers coordinated on the high-ability expert, i.e. $x_{ha} = 3$.}
\label{tab:alg_ha_3_la}
\renewcommand{\arraystretch}{1.5}
\begin{tabular}{cc}
\begin{tabular}{cc|c|c|}
  & \multicolumn{1}{c}{} & \multicolumn{2}{c}{HA$_3$}\\
  & \multicolumn{1}{c}{} & \multicolumn{1}{c}{Invest} & \multicolumn{1}{c}{Not Invest} \\\cline{3-4}
  \multirow{2}{*}{LA$_0$} & Invest & $\frac{r_i}{R}$ &$0$ \\\cline{3-4}
  & Not Invest & $\begin{array}{c} (1-\alpha)0.5\frac{r_i}{R} + \\ (1-\alpha)0.5\frac{R-r_i}{R} \end{array}$ & 0 \\\cline{3-4}
\end{tabular}
&
\begin{tabular}{cc|c|c|}
  & \multicolumn{1}{c}{} & \multicolumn{2}{c}{HA$_3$}\\
  & \multicolumn{1}{c}{} & \multicolumn{1}{c}{Invest} & \multicolumn{1}{c}{Not Invest} \\\cline{3-4}
  \multirow{2}{*}{LA$_0$} & Invest & 0 &  0 \\\cline{3-4}
  & Not Invest & $0.5\frac{R-r_i}{R}$ & $\alpha$ \\\cline{3-4}
\end{tabular}
\\[3em]
LA$_0$ Expert Chooses \textit{Not Invest} & LA$_0$ Expert Chooses \textit{Invest}
\end{tabular}
\end{table}

If all consumers coordinate on the high-ability expert (Table \ref{tab:alg_ha_3}), then the high-ability expert always chooses to forego the investment. If nobody invests, consumers do not switch away. Otherwise, not-investing dominates investing. Note that we use $\alpha$ as the share of consumers who prefer the ``safe'' option. The term $\frac{r_i}{R}$ captures the payoff for an imitating low-ability expert if they successfully attract a consumer. Here, $r_i$ describes the number of rounds after a consumer switches away, i.e., after learning about the low-ability expert's true type. LA$_x$ denotes the low-ability expert with $x$ consumers at the beginning of round 11, HA$_x$ the equivalent high-ability expert. We use $r_i$ instead of $\sum_{i=1}^{3} r_i$ for ease of presentation. The payoff term $(1-\alpha)*\frac{R-2r_i}{R}$ always depends on the assumption $r_i\leq 0.5R$. If a low-ability expert attracts consumers by not investing after $\frac{r_i}{R}$ rounds, then the payoff term $\frac{R-r_i}{R}$ becomes $\frac{r_i}{R}$ if $r_i\leq 0.5$.

\begin{table}[h]
\centering
\footnotesize
\caption{High-Ability Expert: Expected consumer attraction conditional on the three experts' investment choices in round 11 for \textit{Algorithm} if all consumers coordinated on one low-ability expert, i.e. $x_{la} = 3$, and $\alpha$ equals share of consumers for whom $r_i \geq \bar{r}_{i}$. All cells show the expected share of approached consumer rounds.}
\label{tab:alg_la_3_ha_alternative}
\renewcommand{\arraystretch}{1.2}
\setlength{\tabcolsep}{4pt}
\begin{tabular}{@{}c@{\hspace{0.5em}}c@{}}
\begin{tabular}{cc|c|c|}
  & \multicolumn{1}{c}{} & \multicolumn{2}{c}{LA$_0$}\\
  & \multicolumn{1}{c}{} & \multicolumn{1}{c}{Invest} & \multicolumn{1}{c}{Not Invest} \\\cline{3-4}
  \multirow{2}{*}{LA$_3$} & Invest & 1 & $\begin{array}{c} 0.5(1-\alpha) + \\ 0.5(1-\alpha)\frac{R-r_i}{R} \end{array}$ \\\cline{3-4}
  & Not Invest & $(1-\alpha)\frac{R-r_i}{R}$ & $\begin{array}{c} 0.5\frac{R-r_i}{R} + \\ 0.5\frac{R-2r_i}{R} \end{array}$ \\\cline{3-4}
\end{tabular}
&
\begin{tabular}{cc|c|c|}
  & \multicolumn{1}{c}{} & \multicolumn{2}{c}{LA$_0$}\\
  & \multicolumn{1}{c}{} & \multicolumn{1}{c}{Invest} & \multicolumn{1}{c}{Not Invest} \\\cline{3-4}
  \multirow{2}{*}{LA$_3$} & Invest & 0 & $0.5\frac{R-r_i}{R}$ \\\cline{3-4}
  & Not Invest & $0.5\frac{R-r_i}{R}$ & $\alpha + (1-\alpha)\frac{R-2r_i}{R}$ \\\cline{3-4}
\end{tabular}
\\[3em]
HA$_0$ Expert Chooses \textit{Not Invest} & HA$_0$ Expert Chooses \textit{Invest}
\end{tabular}
\end{table}

\begin{table}[h]
\centering
\footnotesize
\caption{Low-Ability Expert: Expected consumer attraction conditional on the three experts' investment choices in round 11 for \textit{Algorithm} if all consumers coordinated on one low-ability expert, i.e. $x_{la} = 3$, and $\alpha$ equals share of consumers for whom $r_i \geq \bar{r}_{i}$. All cells show the expected share of approached consumer rounds.}
\label{tab:alg_la_3_la_alternative}
\renewcommand{\arraystretch}{1.5}
\begin{tabular}{cc}
\begin{tabular}{cc|c|c|}
  & \multicolumn{1}{c}{} & \multicolumn{2}{c}{HA$_0$}\\
  & \multicolumn{1}{c}{} & \multicolumn{1}{c}{Invest} & \multicolumn{1}{c}{Not Invest} \\\cline{3-4}
  \multirow{2}{*}{LA$_0$} & Invest & $\frac{r_i}{R}$ & $ (1-\alpha)\frac{r_i}{R}$ \\\cline{3-4}
  & Not Invest & $ (1-\alpha)\frac{r_i}{R}$ & $\frac{r_i}{R}$ \\\cline{3-4}
\end{tabular}
&
\begin{tabular}{cc|c|c|}
  & \multicolumn{1}{c}{} & \multicolumn{2}{c}{HA$_0$}\\
  & \multicolumn{1}{c}{} & \multicolumn{1}{c}{Invest} & \multicolumn{1}{c}{Not Invest} \\\cline{3-4}
  \multirow{2}{*}{LA$_0$} & Invest & 1 & $0$ \\\cline{3-4}
  & Not Invest &  $\begin{array}{c} \alpha + (1-\alpha)\\0.5\frac{R-r_i}{R} \end{array}$  & $\alpha$ \\\cline{3-4}
\end{tabular}
\\[3em]
LA$_3$ Expert Chooses \textit{Not Invest} & LA$_3$ Expert Chooses \textit{Invest}
\end{tabular}
\end{table}

\begin{table}[h]
\centering
\footnotesize
\caption{Low-Ability Expert L$_0$: Expected consumer attraction conditional on the three experts' investment choices in round 11 for \textit{Algorithm} if all consumers coordinated on one low-ability expert, i.e. $x_{la} = 3$, and $\alpha$ equals share of consumers for whom $r_i \geq \bar{r}_{i}$. All cells show the expected share of approached consumer rounds. LA$_3$ does not invest, HA$_0$ invests.}
\label{tab:alg_la_0_la_alternative}
\renewcommand{\arraystretch}{1.5}
\begin{tabular}{cc}
\begin{tabular}{cc|c|c|}
  & \multicolumn{1}{c}{} & \multicolumn{2}{c}{HA$_0$}\\
  & \multicolumn{1}{c}{} & \multicolumn{1}{c}{Invest} & \multicolumn{1}{c}{Not Invest} \\\cline{3-4}
  \multirow{2}{*}{LA$_3$} & Invest & $\frac{r_i}{R}$ &  $(1-\alpha)0.5\frac{r_i}{R}$ \\\cline{3-4}
  & Not Invest & $(1-\alpha)\frac{R-r_i}{R}$  & $0.5\frac{R-r_i}{R}$ \\\cline{3-4}
\end{tabular}
&
\begin{tabular}{cc|c|c|}
  & \multicolumn{1}{c}{} & \multicolumn{2}{c}{HA$_0$}\\
  & \multicolumn{1}{c}{} & \multicolumn{1}{c}{Invest} & \multicolumn{1}{c}{Not Invest} \\\cline{3-4}
  \multirow{2}{*}{LA$_3$} & Invest & $0$ & $0$ \\\cline{3-4}
  & Not Invest &  $0.5\frac{R-r_i}{R}$  & $\alpha$ \\\cline{3-4}
\end{tabular}
\\[3em]
LA$_0$ Expert Chooses \textit{Not Invest} & LA$_0$ Expert Chooses \textit{Invest}
\end{tabular}
\end{table}

\clearpage
\subsection{Mixed Distribution}

\begin{table}[h]
\centering
\footnotesize
\caption{High-Ability Expert: Expected consumer attraction conditional on the three experts' investment choices in round 11 for \textit{Algorithm} if $x_{ha} = 2$. The additional income for attracting new consumers ``x'' has no implications for the dominant choice.}
\label{tab:alg_ha_2}
\renewcommand{\arraystretch}{1.2}
\setlength{\tabcolsep}{4pt}
\begin{tabular}{@{}c@{\hspace{0.5em}}c@{}}
\begin{tabular}{cc|c|c|}
    & \multicolumn{1}{c}{} & \multicolumn{2}{c}{LA$_0$}\\
    & \multicolumn{1}{c}{} & \multicolumn{1}{c}{Invest} & \multicolumn{1}{c}{Not Invest} \\\cline{3-4}
\multirow{2}{*}{LA$_1$} & Invest & 1 & $\begin{array}{c} (1-\alpha)\frac{2}{3} + \\ (1-\alpha)\frac{1}{3}\frac{R-r_i}{R} \end{array}$ \\\cline{3-4}
    & Not Invest & $\begin{array}{c} (1-\alpha)\frac{2}{3} + \\ (1-\alpha)\frac{1}{3}\frac{R-r_i}{R} \end{array}$ & $\begin{array}{c} \frac{2}{3} + \\ \frac{1}{3}\frac{R-r_i}{R} \end{array}$ \\\cline{3-4}
\end{tabular}
&
\begin{tabular}{cc|c|c|}
    & \multicolumn{1}{c}{} & \multicolumn{2}{c}{LA$_0$}\\
    & \multicolumn{1}{c}{} & \multicolumn{1}{c}{Invest} & \multicolumn{1}{c}{Not Invest} \\\cline{3-4}
    \multirow{2}{*}{LA$_1$} & Invest & $\frac{2}{3}$ & $0.5\frac{R-r_i}{R}$ \\\cline{3-4}
    & Not Invest & $0.5\frac{R-r_i}{R}$ & $\begin{array}{c} \alpha + \\ (1-\alpha) \frac{R-2r_i}{R} \end{array}$ \\\cline{3-4}
\end{tabular}
\\[2em]
HA$_2$ Chooses \textit{Not Invest} & HA$_2$ Chooses \textit{Invest}
\end{tabular}
\end{table}

\begin{table}[h]
\centering
\footnotesize
\caption{Low-Ability Expert: Expected consumer attraction conditional on the three experts' investment choices in round 11 for \textit{Algorithm} if $x_{la} = 1$ for this low-ability expert and $x_{ha} = 2$.}
\label{tab:alg_ha_2_la1}
\vspace{-1cm}
\renewcommand{\arraystretch}{1.5}
\begin{tabular}{cc}
\begin{tabular}{cc|c|c|}
  & \multicolumn{1}{c}{} & \multicolumn{2}{c}{HA$_2$}\\
  & \multicolumn{1}{c}{} & \multicolumn{1}{c}{Invest} & \multicolumn{1}{c}{Not Invest} \\\cline{3-4}
  \multirow{2}{*}{LA$_0$} & Invest & $\frac{r_i}{R}$ & $ (1-\alpha) \frac{r_i}{3R}$ \\\cline{3-4}
  & Not Invest & $\begin{array}{c} (1-\alpha) \frac{2}{3} \frac{r_i}{R} + \\ (1-\alpha) \frac{1}{3} \frac{R - r_i}{R} \end{array}$ & $\frac{r_i}{3R}$ \\\cline{3-4}
\end{tabular}
&
\begin{tabular}{cc|c|c|}
  & \multicolumn{1}{c}{} & \multicolumn{2}{c}{HA$_2$}\\
  & \multicolumn{1}{c}{} & \multicolumn{1}{c}{Invest} & \multicolumn{1}{c}{Not Invest} \\\cline{3-4}
  \multirow{2}{*}{LA$_0$} & Invest & $\frac{1}{3}$ & 0 \\\cline{3-4}
  & Not Invest & $0.5\frac{R-r_i}{R}$ & $\alpha$\\\cline{3-4}
\end{tabular}
\\[3em]
LA$_1$ Expert Chooses \textit{Not Invest} & LA$_1$ Expert Chooses \textit{Invest}
\end{tabular}
\end{table}

\begin{table}[h]
\centering
\footnotesize
\caption{Low-Ability Expert: Expected consumer attraction conditional on the three experts' investment choices in round 11 for \textit{Algorithm} if $x_{la} = 0$ for this low-ability expert and $x_{ha} = 2$. }
\label{tab:alg_ha_2_la0}
\vspace{-1cm}
\renewcommand{\arraystretch}{1.5}
\begin{tabular}{cc}
\begin{tabular}{cc|c|c|}
  & \multicolumn{1}{c}{} & \multicolumn{2}{c}{HA$_2$}\\
  & \multicolumn{1}{c}{} & \multicolumn{1}{c}{Invest} & \multicolumn{1}{c}{Not Invest} \\\cline{3-4}
  \multirow{2}{*}{LA$_1$} & Invest & $\frac{r_i}{R}$& $ (1-\alpha) \frac{r_i}{6R}$ \\\cline{3-4}
& Not Invest & $\begin{array}{c} (1-\alpha)\frac{1}{3}\frac{r_i}{R} + \\ (1-\alpha)\frac{2}{3} \frac{R-r_i}{R} \end{array}$ & $\frac{R-r_i}{6R}$ \\\cline{3-4}
\end{tabular}
&
\begin{tabular}{cc|c|c|}
  & \multicolumn{1}{c}{} & \multicolumn{2}{c}{HA$_2$}\\
  & \multicolumn{1}{c}{} & \multicolumn{1}{c}{Invest} & \multicolumn{1}{c}{Not Invest} \\\cline{3-4}
  \multirow{2}{*}{LA$_1$} & Invest & 0 & 0 \\\cline{3-4}
  & Not Invest & $0.5\frac{R-r_i}{R}$ & $\alpha$\\\cline{3-4}
\end{tabular}
\\[3em]
LA$_0$ Expert Chooses \textit{Not Invest} & LA$_0$ Expert Chooses \textit{Invest}
\end{tabular}
\end{table}

\begin{table}[h]
\centering
\footnotesize
\caption{High-Ability Expert: Expected consumer attraction conditional on the three experts' investment choices in round 11 for \textit{Algorithm} if $x_{ha} = 1$, $x_{i, la} = 2$, $x_{j, la} = 0$. }
\label{tab:alg_la_2_ha1}
\renewcommand{\arraystretch}{1.2}
\setlength{\tabcolsep}{4pt}
\begin{tabular}{@{}c@{\hspace{0.5em}}c@{}}
\begin{tabular}{cc|c|c|}
    & \multicolumn{1}{c}{} & \multicolumn{2}{c}{LA$_0$}\\
    & \multicolumn{1}{c}{} & \multicolumn{1}{c}{Invest} & \multicolumn{1}{c}{Not Invest} \\\cline{3-4}
    \multirow{2}{*}{LA$_2$} & Invest & 1 & $\begin{array}{c} (1-\alpha)\frac{2}{3} + \\ (1-\alpha)\frac{1}{3}(\frac{R-r_i}{R}) \end{array}$ \\\cline{3-4}
    & Not Invest & $\begin{array}{c} (1-\alpha)\frac{1}{3} + \\ (1-\alpha)\frac{2}{3}(\frac{R-r_i}{R}) \end{array}$ & $\begin{array}{c} \frac{1}{3} + \frac{1}{3}(\frac{R-r_i}{R}) + \\ \frac{1}{3}(\frac{R-2r_i}{R}) \end{array}$ \\\cline{3-4}
\end{tabular}
&
\begin{tabular}{cc|c|c|}
    & \multicolumn{1}{c}{} & \multicolumn{2}{c}{LA$_0$}\\
    & \multicolumn{1}{c}{} & \multicolumn{1}{c}{Invest} & \multicolumn{1}{c}{Not Invest} \\\cline{3-4}
    \multirow{2}{*}{LA$_2$} & Invest & $\frac{1}{3}$ & $0.5\frac{R-r_i}{R}$ \\\cline{3-4}
    & Not Invest & $0.5\frac{R-r_i}{R}$ & $\alpha$ \\\cline{3-4}
\end{tabular}
\\[4.5em]
HA$_1$ Expert Chooses \textit{Not Invest} & HA$_1$ Expert Chooses \textit{Invest}
\end{tabular}
\end{table}

\begin{table}[h]
\centering
\footnotesize
\caption{Low-Ability Expert: Expected consumer attraction conditional on the three experts' investment choices in round 11 for \textit{Algorithm} if $x_{la} = 2$ and $x_{ha} = 1$. }
\label{tab:alg_la_2_la1}
\renewcommand{\arraystretch}{1.5}
\begin{tabular}{cc}
\begin{tabular}{cc|c|c|}
  & \multicolumn{1}{c}{} & \multicolumn{2}{c}{HA$_1$}\\
  & \multicolumn{1}{c}{} & \multicolumn{1}{c}{Invest} & \multicolumn{1}{c}{Not Invest} \\\cline{3-4}
  \multirow{2}{*}{LA$_0$} & Invest & $\frac{r_i}{R}$ & $ (1-\alpha) \frac{2}{3} \frac{r_i}{R}$ \\\cline{3-4}
& Not Invest & $\begin{array}{c} (1-\alpha) \frac{2}{3} \frac{r_i}{R} + \\ (1-\alpha) \frac{r_i}{6R} + \\ (1-\alpha)\frac{R-r_i}{6R} \end{array}$ & $\frac{2}{3} \frac{r_i}{R}$ \\\cline{3-4}
\end{tabular}
&
\begin{tabular}{cc|c|c|}
  & \multicolumn{1}{c}{} & \multicolumn{2}{c}{HA$_1$}\\
  & \multicolumn{1}{c}{} & \multicolumn{1}{c}{Invest} & \multicolumn{1}{c}{Not Invest} \\\cline{3-4}
\multirow{2}{*}{LA$_0$} & Invest & $\frac{2}{3}$ & 0 \\\cline{3-4}
  & Not Invest & $0.5(\frac{R-r_i}{R})$ & $\alpha$\\\cline{3-4}
\end{tabular}
\\[3em]
LA$_2$ Expert Chooses \textit{Not Invest} & LA$_2$ Expert Chooses \textit{Invest}
\end{tabular}
\end{table}

\begin{table}[h]
\centering
\footnotesize
\caption{Low-Ability Expert: Expected consumer attraction conditional on the three experts' investment choices in round 11 for \textit{Algorithm} if $x_{la} = 0$ for this expert and $x_{ha} = 1$. }
\label{tab:alg_la_2_la0}
\renewcommand{\arraystretch}{1.5}
\begin{tabular}{cc}
\begin{tabular}{cc|c|c|}
  & \multicolumn{1}{c}{} & \multicolumn{2}{c}{HA$_1$}\\
  & \multicolumn{1}{c}{} & \multicolumn{1}{c}{Invest} & \multicolumn{1}{c}{Not Invest} \\\cline{3-4}
  \multirow{2}{*}{LA$_2$} & Invest & $\frac{r_i}{R}$ & $ (1-\alpha) \frac{1}{3} \frac{r_i}{R}$ \\\cline{3-4}
  & Not Invest & $\begin{array}{c} (1-\alpha) \frac{1}{6} \frac{r_i}{R} + \\ (1-\alpha) \frac{5}{6} \frac{R-r_i}{R} \end{array}$ & $\frac{1}{3} \frac{r_i}{R}$ \\\cline{3-4}
\end{tabular}
&
\begin{tabular}{cc|c|c|}
  & \multicolumn{1}{c}{} & \multicolumn{2}{c}{HA$_1$}\\
  & \multicolumn{1}{c}{} & \multicolumn{1}{c}{Invest} & \multicolumn{1}{c}{Not Invest} \\\cline{3-4}
\multirow{2}{*}{LA$_2$} & Invest & $0$ & $0$ \\\cline{3-4}
  & Not Invest & $0.5(\frac{R-r_i}{R})$ & $\alpha$\\\cline{3-4}
\end{tabular}
\\[3em]
LA$_0$ Expert Chooses \textit{Not Invest} & LA$_0$ Expert Chooses \textit{Invest}
\end{tabular}
\end{table}

\clearpage

\begin{table}[h]
\centering
\footnotesize
\caption{High-Ability Expert: Expected consumer attraction conditional on the three experts' investment choices in round 11 for \textit{Algorithm} if $x_{ha} = 0$, $x_{i, la} = 2$, $x_{j, la} = 1$. }
\label{tab:alg_la2_la1_ha0_ha}
\renewcommand{\arraystretch}{1.2}
\setlength{\tabcolsep}{4pt}
\begin{tabular}{@{}c@{\hspace{0.5em}}c@{}}
\begin{tabular}{cc|c|c|}
    & \multicolumn{1}{c}{} & \multicolumn{2}{c}{LA$_2$}\\
    & \multicolumn{1}{c}{} & \multicolumn{1}{c}{Invest} & \multicolumn{1}{c}{Not Invest} \\\cline{3-4}
    \multirow{2}{*}{LA$_1$} & Invest & 1 & $\begin{array}{c} (1-\alpha)\frac{1}{6} + \\ (1-\alpha)\frac{5}{6}\frac{R-r_i}{R} \end{array}$ \\\cline{3-4}
    & Not Invest & $\begin{array}{c} (1-\alpha)\frac{1}{3} + \\ (1-\alpha)\frac{2}{3}\frac{R-r_i}{R} \end{array}$ & $\begin{array}{c} \frac{1}{2} \frac{R-r_i}{R} + \\ \frac{1}{2}\frac{R-2r_i}{R} \end{array}$ \\\cline{3-4}
\end{tabular}
&
\begin{tabular}{cc|c|c|}
    & \multicolumn{1}{c}{} & \multicolumn{2}{c}{LA$_2$}\\
    & \multicolumn{1}{c}{} & \multicolumn{1}{c}{Invest} & \multicolumn{1}{c}{Not Invest} \\\cline{3-4}
    \multirow{2}{*}{LA$_1$} & Invest & 0 & $0.5\frac{R-r_i}{R}$ \\\cline{3-4}
    & Not Invest & $0.5\frac{R-r_i}{R}$ & $\alpha + (1-\alpha)\frac{R-2r_i}{R}$ \\\cline{3-4}
\end{tabular}
\\[4.5em]
HA$_0$ Expert Chooses \textit{Not Invest} & HA$_0$ Expert Chooses \textit{Invest}
\end{tabular}
\end{table}

\begin{table}[h]
\centering
\footnotesize
\caption{Low-Ability Expert: Expected consumer attraction of $LA_2$ conditional on the three experts' investment choices in round 11 for \textit{Algorithm} if $x_{ha} = 0$, $x_{i, la} = 2$, $x_{j, la} = 1$. }
\label{tab:alg_la2_la1_ha0_la2}
\renewcommand{\arraystretch}{1.2}
\setlength{\tabcolsep}{4pt}
\begin{tabular}{@{}c@{\hspace{0.5em}}c@{}}
\begin{tabular}{cc|c|c|}
    & \multicolumn{1}{c}{} & \multicolumn{2}{c}{HA$_0$}\\
    & \multicolumn{1}{c}{} & \multicolumn{1}{c}{Invest} & \multicolumn{1}{c}{Not Invest} \\\cline{3-4}
    \multirow{2}{*}{LA$_1$} & Invest & $\frac{r_i}{R}$ & $(1-\alpha)\frac{5}{6}\frac{r_i}{R}$ \\\cline{3-4}
    & Not Invest & $\begin{array}{c} (1-\alpha)\frac{2}{3}\frac{r_i}{R} + \\ (1-\alpha)\frac{1}{3}\frac{R-r_i}{R} \end{array}$ & $\begin{array}{c} \frac{2}{3} \frac{r_i}{R} + \\ \frac{1}{6}\frac{R-r_i}{R} \end{array}$ \\\cline{3-4}
\end{tabular}
&
\begin{tabular}{cc|c|c|}
    & \multicolumn{1}{c}{} & \multicolumn{2}{c}{HA$_0$}\\
    & \multicolumn{1}{c}{} & \multicolumn{1}{c}{Invest} & \multicolumn{1}{c}{Not Invest} \\\cline{3-4}
    \multirow{2}{*}{LA$_1$} & Invest & $\frac{2}{3}$ & $0$ \\\cline{3-4}
    & Not Invest & $0.5\frac{R-r_i}{R}$ & $\alpha$ \\\cline{3-4}
\end{tabular}
\\[4.5em]
LA$_2$ Expert Chooses \textit{Not Invest} & LA$_2$ Expert Chooses \textit{Invest}
\end{tabular}
\end{table}

\begin{table}[h]
\centering
\footnotesize
\caption{Low-Ability Expert: Expected consumer attraction of $LA_0$ conditional on the three experts' investment choices in round 11 for \textit{Algorithm} if $x_{ha} = 0$, $x_{i, la} = 2$, $x_{j, la} = 1$. }
\label{tab:alg_la2_la1_ha0_la0}
\renewcommand{\arraystretch}{1.2}
\setlength{\tabcolsep}{4pt}
\begin{tabular}{@{}c@{\hspace{0.5em}}c@{}}
\begin{tabular}{cc|c|c|}
    & \multicolumn{1}{c}{} & \multicolumn{2}{c}{HA$_0$}\\
    & \multicolumn{1}{c}{} & \multicolumn{1}{c}{Invest} & \multicolumn{1}{c}{Not Invest} \\\cline{3-4}
    \multirow{2}{*}{LA$_2$} & Invest & $\frac{r_i}{R}$ & $\begin{array}{c} (1-\alpha)\frac{1}{3}\frac{r_i}{R} + \\ (1-\alpha) \frac{1}{3}\frac{R-r_i}{R} \end{array}$ \\\cline{3-4}
    & Not Invest & $\begin{array}{c} (1-\alpha)\frac{1}{3}\frac{r_i}{R} + \\ (1-\alpha)\frac{2}{3}\frac{R-r_i}{R} \end{array}$ & $\begin{array}{c} \frac{1}{3} \frac{r_i}{R} + \\ \frac{1}{3}\frac{R-r_i}{R} \end{array}$ \\\cline{3-4}
\end{tabular}
&
\begin{tabular}{cc|c|c|}
    & \multicolumn{1}{c}{} & \multicolumn{2}{c}{HA$_0$}\\
    & \multicolumn{1}{c}{} & \multicolumn{1}{c}{Invest} & \multicolumn{1}{c}{Not Invest} \\\cline{3-4}
    \multirow{2}{*}{LA$_2$} & Invest & $\frac{1}{3}$ & $0$ \\\cline{3-4}
    & Not Invest & $0.5\frac{R-r_i}{R}$ & $\alpha$ \\\cline{3-4}
\end{tabular}
\\[4.5em]
LA$_1$ Expert Chooses \textit{Not Invest} & LA$_1$ Expert Chooses \textit{Invest}
\end{tabular}
\end{table}

\clearpage
\subsection{Equal Distribution}

\begin{table}[h]
\centering
\footnotesize
\caption{High-Ability Expert: Expected consumer attraction conditional on the three experts' investment choices in round 11 for \textit{Algorithm} if $x_{ha} = 1$, $x_{i, la} = 1$, $x_{j, la} = 1$. }
\label{tab:alg_1ha1}
\renewcommand{\arraystretch}{1.2}
\setlength{\tabcolsep}{4pt}
\begin{tabular}{@{}c@{\hspace{0.5em}}c@{}}
\begin{tabular}{cc|c|c|}
    & \multicolumn{1}{c}{} & \multicolumn{2}{c}{LA$_1$}\\
    & \multicolumn{1}{c}{} & \multicolumn{1}{c}{Invest} & \multicolumn{1}{c}{Not Invest} \\\cline{3-4}
    \multirow{2}{*}{LA$_1$} & Invest & 1 & $\begin{array}{c} (1-\alpha)\frac{1}{3} + \\ (1-\alpha)\frac{2}{3}(\frac{R-r_i}{R}) \end{array}$ \\\cline{3-4}
    & Not Invest & $\begin{array}{c} (1-\alpha)\frac{1}{3} + \\ (1-\alpha)\frac{2}{3}(\frac{R-r_i}{R}) \end{array}$ & $\begin{array}{c} \frac{1}{3} + \frac{1}{3}(\frac{R-r_i}{R}) + \\ \frac{1}{3}(\frac{R-2r_i}{R}) \end{array}$ \\\cline{3-4}
\end{tabular}
&
\begin{tabular}{cc|c|c|}
    & \multicolumn{1}{c}{} & \multicolumn{2}{c}{LA$_1$}\\
    & \multicolumn{1}{c}{} & \multicolumn{1}{c}{Invest} & \multicolumn{1}{c}{Not Invest} \\\cline{3-4}
    \multirow{2}{*}{LA$_1$} & Invest & $\frac{1}{3}$ & $0.5\frac{R-r_i}{R}$ \\\cline{3-4}
    & Not Invest & $0.5\frac{R-r_i}{R}$ & $\alpha$ \\\cline{3-4}
\end{tabular}
\\[4.5em]
HA$_1$ Expert Chooses \textit{Not Invest} & HA$_1$ Expert Chooses \textit{Invest}
\end{tabular}
\end{table}

\begin{table}[h]
\centering
\footnotesize
\caption{Low-Ability Expert: Expected consumer attraction conditional on the three experts' investment choices in round 11 for \textit{Algorithm} if if $x_{ha} = 1$, $x_{i, la} = 1$, $x_{j, la} = 1$. }
\label{tab:alg_1la1}
\renewcommand{\arraystretch}{1.5}
\begin{tabular}{cc}
\begin{tabular}{cc|c|c|}
  & \multicolumn{1}{c}{} & \multicolumn{2}{c}{HA$_1$}\\
  & \multicolumn{1}{c}{} & \multicolumn{1}{c}{Invest} & \multicolumn{1}{c}{Not Invest} \\\cline{3-4}
  \multirow{2}{*}{LA$_1$} & Invest & $\frac{r_i}{R}$ & $ (1-\alpha) \frac{1}{2} \frac{r_i}{R} $ \\\cline{3-4}
  & Not Invest & $\begin{array}{c} (1-\alpha) \frac{1}{2} \frac{r_i}{R} + \\ (1-\alpha) \frac{1}{2}\frac{R-r_i}{R} \end{array}$ & $\begin{array}{c} \frac{1}{3} \frac{r_i}{R} + \\ \frac{1}{6} \frac{R-r_i}{R} \end{array} $ \\\cline{3-4}
\end{tabular}
&
\begin{tabular}{cc|c|c|}
  & \multicolumn{1}{c}{} & \multicolumn{2}{c}{HA$_1$}\\
  & \multicolumn{1}{c}{} & \multicolumn{1}{c}{Invest} & \multicolumn{1}{c}{Not Invest} \\\cline{3-4}
\multirow{2}{*}{LA$_1$} & Invest & $\frac{1}{3}$ & $0$ \\\cline{3-4}
  & Not Invest & $0.5(\frac{R-r_i}{R})$ & $\alpha$\\\cline{3-4}
\end{tabular}
\\[3em]
LA$_1$ Expert Chooses \textit{Not Invest} & LA$_1$ Expert Chooses \textit{Invest}
\end{tabular}
\end{table}

\clearpage
\subsection{Calculations}

\subsubsection{Separating Equilibria -- Only HA Invests}
\begin{figure}[htbp]
    \centering 

    \begin{minipage}[b]{0.45\textwidth} 
        \centering
        \includegraphics[width=\textwidth]{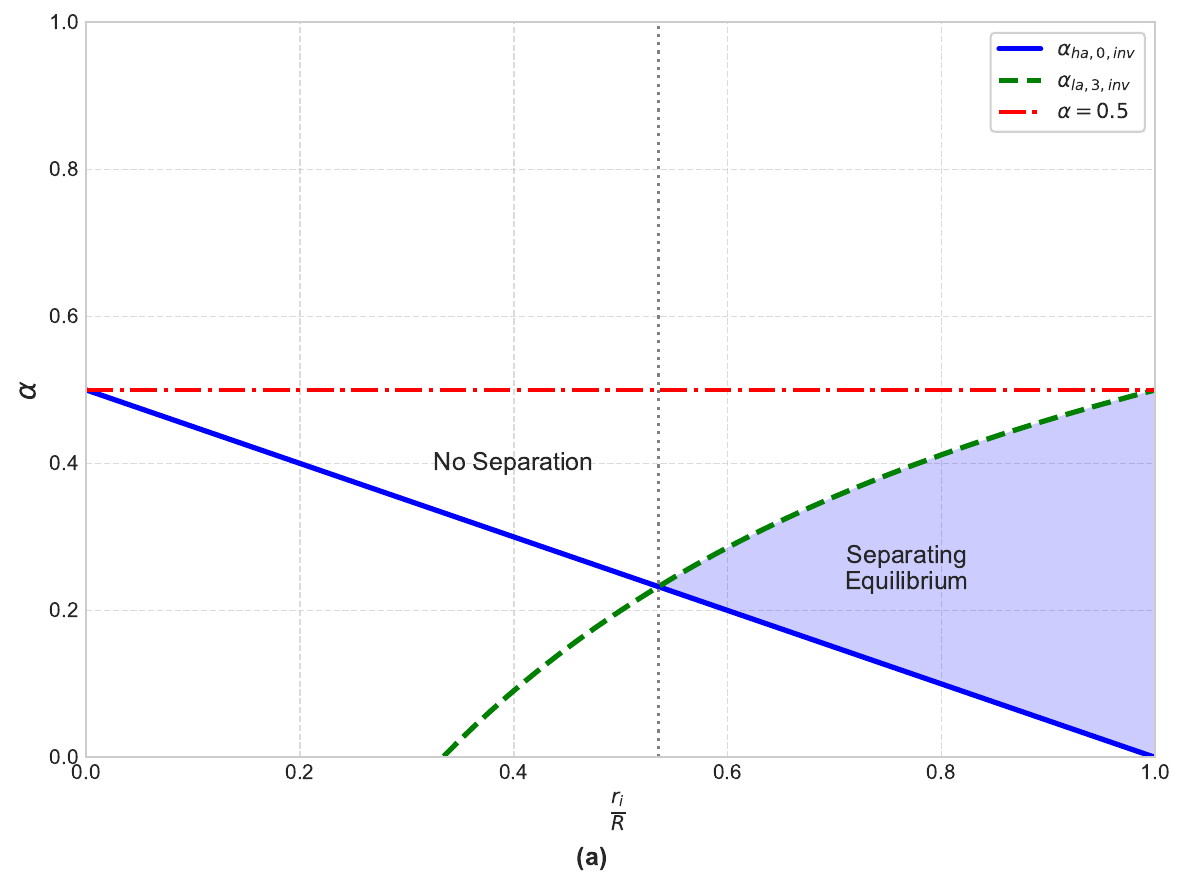}
    \end{minipage}
    \hfill 
    \begin{minipage}[b]{0.45\textwidth}
        \centering
        \includegraphics[width=\textwidth]{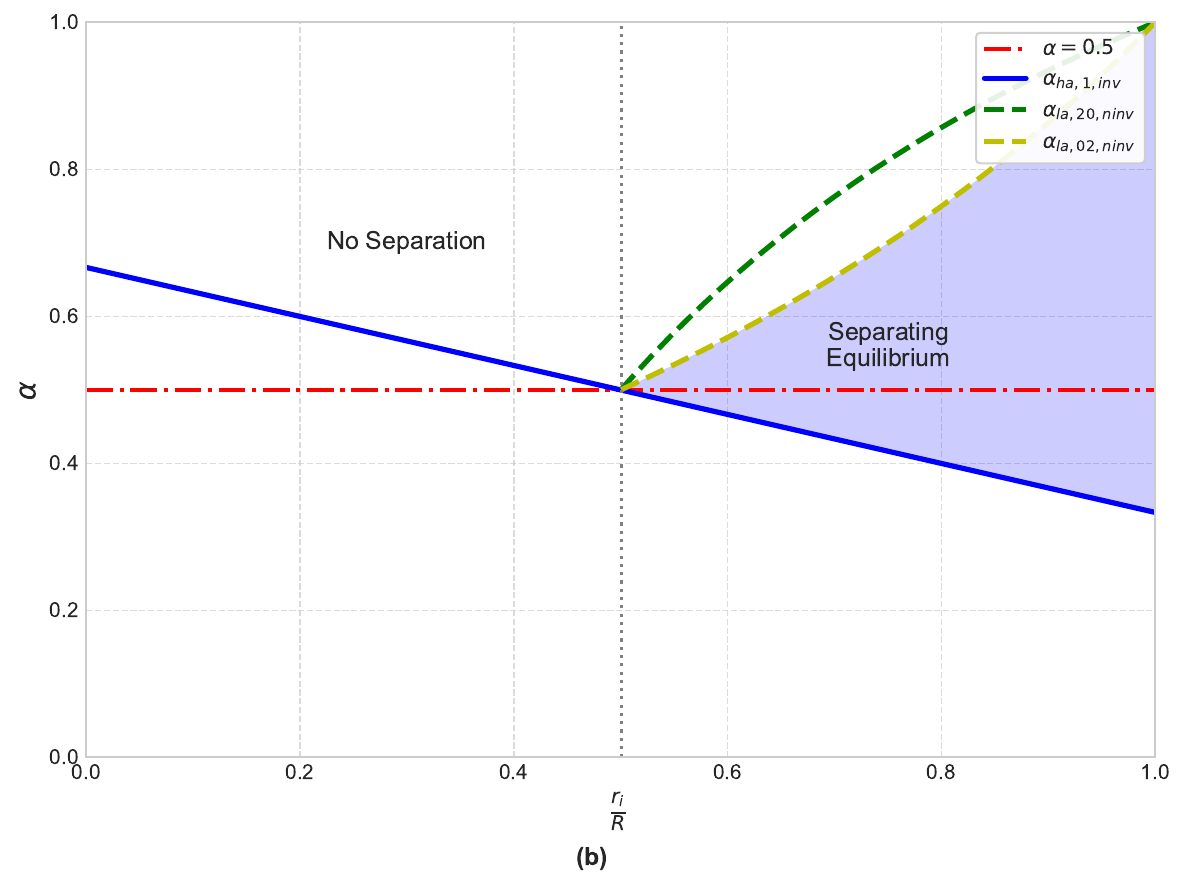}
    \end{minipage}

    \vspace{0.5cm} 

    \begin{minipage}[b]{0.45\textwidth}
        \centering
        \includegraphics[width=\textwidth]{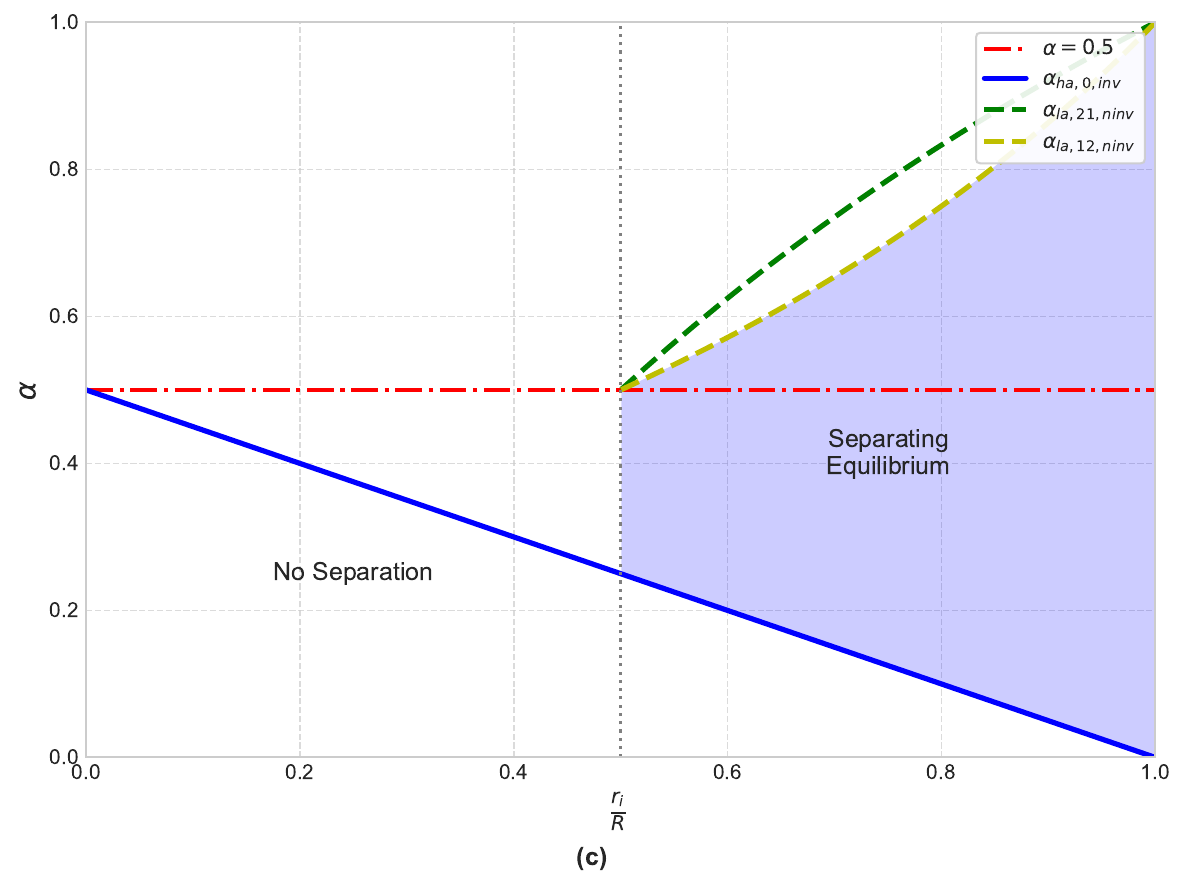}
    \end{minipage}
    \hfill 
    \begin{minipage}[b]{0.45\textwidth}
        \centering
        \includegraphics[width=\textwidth]{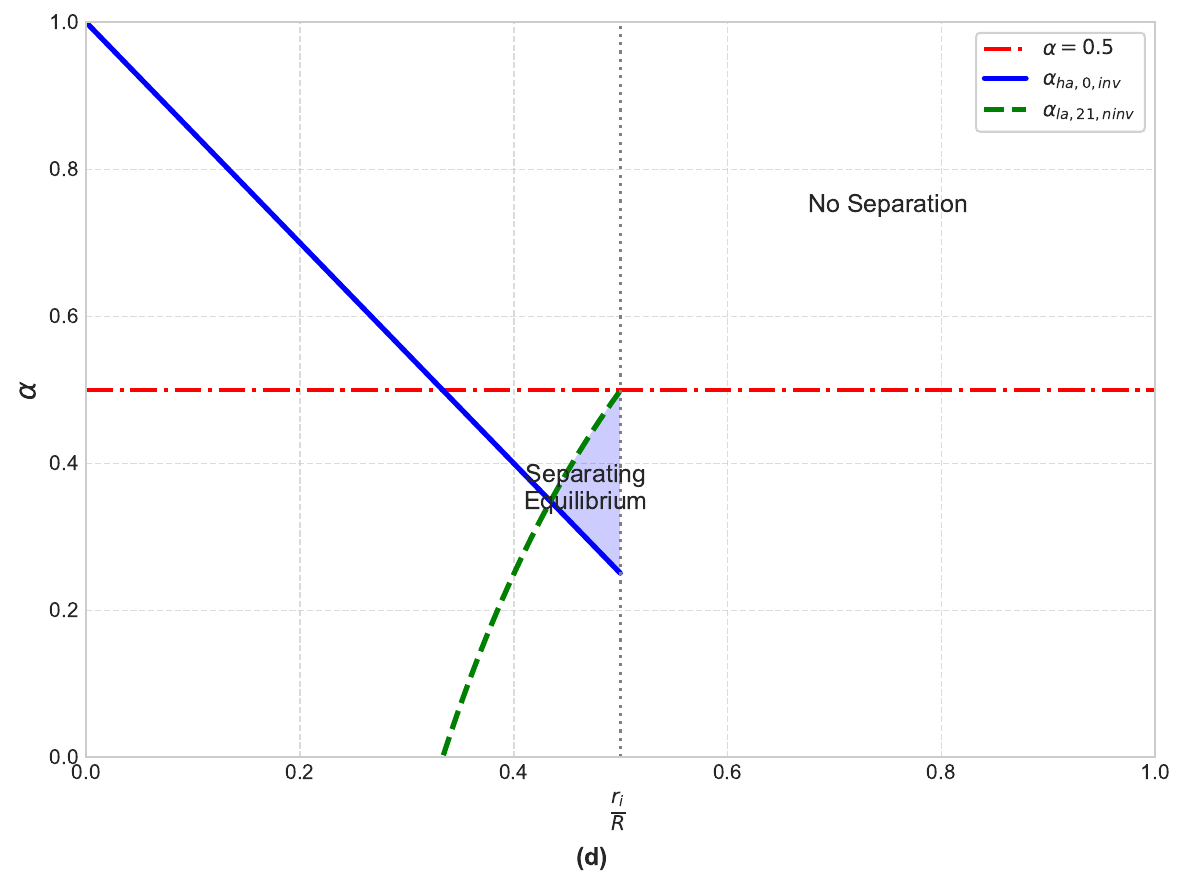}
    \end{minipage}

    \vspace{0.5cm} 

    \begin{minipage}[b]{0.45\textwidth}
        \centering
        \includegraphics[width=\textwidth]{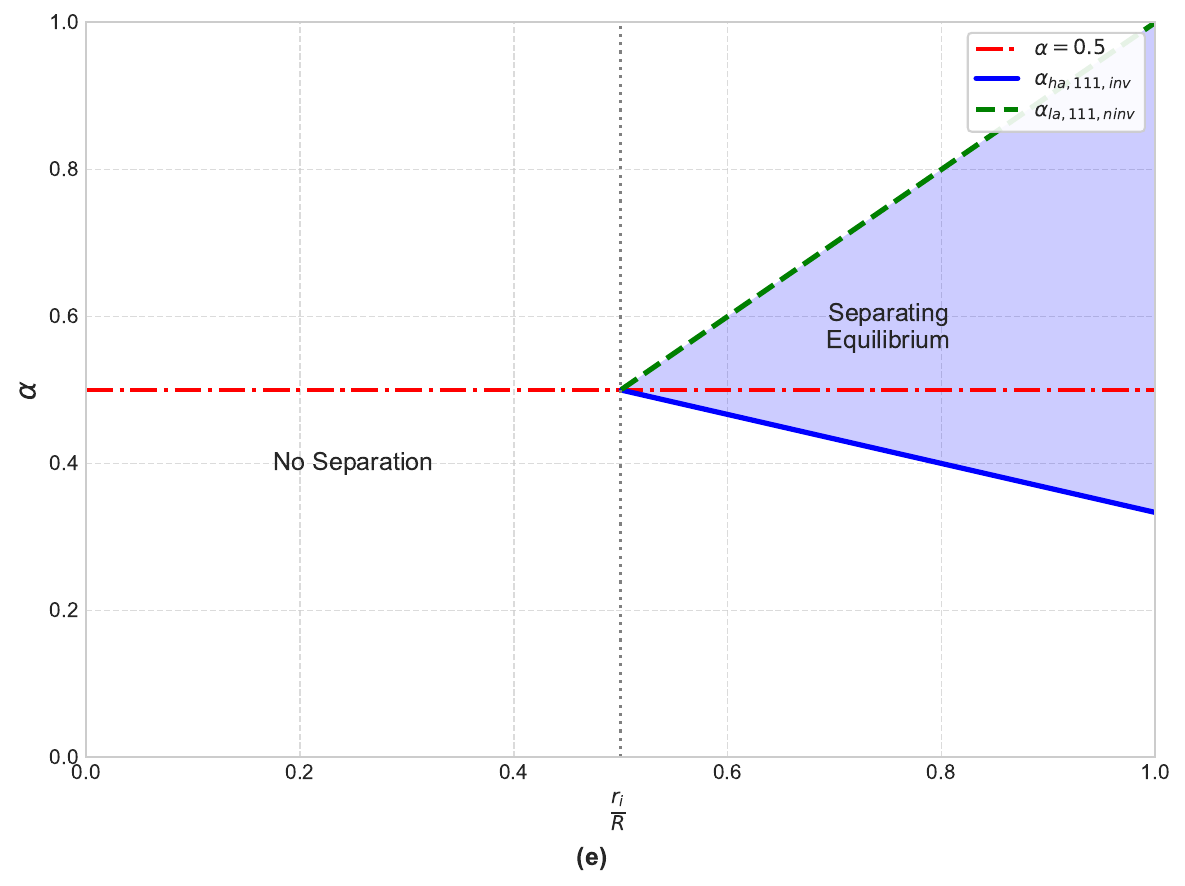}
    \end{minipage}

    \caption{Expert type separating equilibria (PBE). Only the high-ability expert invests. Panel \textbf{(a)} shows the separating space when $x_{la} = 3$. Panel \textbf{(b)} shows the same for $x_{ha} = 1$ and $x_{la} = 2$. Panels \textbf{(c)} and \textbf{(d)} show separation when $x_{ha} = 0$, $x_{la,i} = 2$ and $x_{la,j} = 1$ for $r_i < 0.5R$ and $r_i \geq 0.5R$ respectively. In panel \textbf{(e)}, consumers are distributed equally. }
    \label{fig:separating_la3} 
\end{figure}

If $x_{ha} = 3$, then high-ability experts always forego the decision aid, and specifically never opt for investments if the other two experts choose to not invest because $1 \geq \alpha +(1-\alpha)\frac{R-2r_i}{R}$. Assuming that all consumers coordinated on one low-ability expert, if both low-ability experts do not invest, the high-ability expert only invests if $\alpha \geq 0.25$ and $r_i < 0.5R$ or $\alpha \geq 0.5 - \frac{0.5r_i}{R} \coloneqq \alpha_{ha,0,inv}$ and $r_i \geq 0.5R$. This can be straightforwardly calculated from the payoff matrix (Table \ref{tab:alg_la_3_ha_alternative}). The low-ability expert without consumers chooses to not invest, given investments from the high-ability expert and no investments from the second low-ability expert $LA_3$, if $\alpha \leq 1.5 - \frac{0.5R}{r_i} \coloneqq \alpha_{la,0,ninv}$ and $r_i < 0.5R$ or $\alpha \leq 0.5$ and $r_i \geq 0.5R$. Finally, the coordinated low-ability expert foregoes investments if $r_i > \frac{1}{3}R$ and $\alpha \leq \frac{3r_i -R}{R+3r_i} \coloneqq \alpha_{la,3,inv}$. If $r_i < 0.5R$, then experts can only separate when $0.25 \leq \alpha_{la,3,ninv} \leq \alpha_{la,0,ninv}$. This implies $r_i \geq \frac{5}{9}R$, which is ruled out by $r_i < 0.5R$. There is no separation for $r_i < 0.5R$. When $r_i \geq 0.5R$, then expert types can only separate if $\alpha_{ha,0,inv} \leq \alpha_{la,3,ninv} \leq 0.5$ which holds for $r_i \geq 0.535R$. Figure \ref{fig:separating_la3} illustrates the separating space, given the conditions that $r_i \geq 0.535R$ and $\alpha \in [\alpha_{ha,0,inv}, \alpha_{la,3,inv}]$. For a small set of parameter values, there are equilibria in which \textit{only} the high-ability expert invests, if all consumers coordinate on one low-ability expert.

What if consumers do not coordinate and $x_{ha} = 2$? Again, we look for the two candidate equilibria in which either only the high-ability expert or only the low-ability experts invest (Table \ref{tab:alg_ha_2}). First, when $r_i \geq 0.5R$, the high-ability expert never invests. If $r_i < 0.5R$, the high-ability expert invests given non-investments from the two low-ability experts when $\alpha > \frac{3R-r_i}{5R-6r_i} \coloneqq \alpha_{ha,2,inv}$. Then, for both low-ability experts to invest, $\alpha < 1.5 - \frac{0.5R}{r_i} \coloneqq \alpha_{la,1,inv}$ must hold. However $\alpha_{ha,2,inv} \leq \alpha \leq \alpha_{la,1,inv}$ requires $r_i \geq \frac{5}{6}R$, which violates $r_i < 0.5R$. Hence, there is no separating equilibrium in which only the high-ability expert invests for $x_{ha} = 2$. Similarly, while the high-ability expert never invests if they believe the other two to forego investments, both low-ability experts are then incentivized to also forego investments and imitate the high-ability expert, precluding any strategic separation.

For $x_{ha} = 1$ and $x_{la} = 2$, the high-ability experts separates through investments if $\alpha > 1 - \frac{r_i}{R}$ and $r_i < 0.5R$ or $\alpha > \frac{2R - r_i}{3R} \coloneqq \alpha_{ha,1,inv}$ and $r_i \geq 0.5R$. The equivalent conditions for $LA_2$ are $\alpha < 1.5 - \frac{0.5R}{r_i}$ and $r_i < 0.5R$ or $\alpha < \frac{7r_i - 2R}{4r_i + R} \coloneqq \alpha_{la,20,ninv}$ and $r_i \geq 0.5R$. For $LA_0$, $\alpha < 1.5 - \frac{0.5R}{r_i}$ or $\alpha < \frac{-0.5(R+r_i)}{r_i - 2R} \coloneqq \alpha_{la,02,ninv}$ must hold. Comparing these conditions reveals no possible separation through high-ability expert investments if $r_i < 0.5R$. As before, when learning is slow and $r_i \geq 0.5R$, then separation may arise. Panel (b) in Figure \ref{fig:separating_la3} shows the separation space. When Bayesian consumers take a long time to figure out the imitating low-ability expert's true type, e.g., due to small performance differences or noisy feedback, and a large share of consumers prefers a safe option because of the possibility to accidentally approach an imitating low-ability expert, then high-ability experts may be the only ones to invest. 

After consumers coordinate only on the two low-ability experts with $x_{ha} = 0$, there are separating equilibria in which only the high-ability expert invests for both $r_i < 0.5R$ and $r_i \geq 0.5R$ (Panels \textbf{(c)} and \textbf{(d)} in Figure \ref{fig:separating_la3}). The high-ability expert invests if $\alpha > 1 - \frac{1.5r_i}{R}$ and $r_i < 0.5R$ or $\alpha > \frac{0.5(R-r_i)}{R} \coloneqq \alpha_{ha,0,inv}$ and $r_i \geq 0.5R$. For $LA_2$, $\alpha < 1.5 - \frac{0.5R}{r_i}$ or $\alpha < \frac{2.5r_i - 0.5R}{r_i + R} \coloneqq \alpha_{la,21,ninv}$ must hold under $r_i < 0.5R$ and $r_i \geq 0.5R$ respectively. For $LA_1$, the condition under $r_i \geq 0.5R$ changes to $\alpha < \frac{0.5(R+r_i)}{2R - r_i} \coloneqq \alpha_{la,12,ninv}$. Then, expert types separate in equilibrium if $\alpha_{ha,0,inv} < \alpha < \alpha_{la,12,ninv}$ because $\alpha_{la,12,ninv} \leq \alpha_{la,21,ninv}$.

Under an equal consumer distribution at the time of the technological shock, separation requires again $r_i \geq 0.5R$. Here, the respective conditions are $\alpha > \frac{2R - r_i}{3R}$ for the high-ability expert, and $\alpha < \frac{r_i}{R}$ for each low-ability expert (Panel \textbf{(e)} in Figure \ref{fig:separating_la3}).

Finally, we look at separation behavior when experts are completely uninformed about the distribution of the other consumers who did not approach them. Hence, if an expert has attracted all consumers, they are fully informed. Otherwise, they must incorporate all different possibilities. Here, we assume that experts assign equal probabilities to all possible unobservable states. For the high-ability expert, considerations only change if $r_i \geq 0.5R$. Then, the high-ability expert only separates through investments if $\alpha > \frac{R-r_i}{R + 2r_i} \coloneqq \alpha_{ha0,inv,noinf}$. The low-ability expert's condition when $x_{la,i} = 2$ changes to $\alpha < \frac{14r_i - 3R}{8r_i + 3R} \coloneqq \alpha_{la2,ninv,noinf}$ if $r_i \geq 0.5R$. The low-ability expert's condition when $x_{la,i} = 1$ changes to $\alpha < \frac{r_i}{R} \alpha_{la1,ninv,noinf} $ if $r_i \geq 0.5R$. Finally, if a low-ability expert attracts 0 consumers at the time of the technological shock, separation with an investing high-ability expert requires $\alpha < \frac{8R - 5r_i}{17R-14r_i} \alpha_{la0,ninv,noinf} $ $r_i \geq 0.5R$.

\subsubsection{Mixed Equilibria -- One LA Invests}
If $x_{ha} = 0$, then the respective hyperbolas for the high-ability expert are $\alpha < \frac{1.5R-0.5r_i}{3R-2r_i} \coloneqq \alpha_{ha0,la21,ninv}$ if LA$_2$ invests, $\alpha < \frac{3R-2r_i}{6R-5r_i} \coloneqq \alpha_{ha0,la12,ninv}$ if LA$_1$ invests, $\alpha > \frac{5}{6} \frac{r_i}{R} \coloneqq \alpha_{la2,ha0,inv}$ and $r_i < 0.5R$ or $\alpha > \frac{1.5r_i + 0.5R}{3R} \coloneqq \alpha_{la2,ha0,inv}$ and $r_i \geq 0.5R$ for LA$_2$ to invest, and $\alpha > \frac{2}{3} \frac{r_i}{R} \coloneqq \alpha_{la1,ha0,inv}$ if $r_i < 0.5R$ or $\alpha > \frac{1}{3} \coloneqq \alpha_{la1,ha0,inv}$ if $r_i \geq 0.5R$ when LA$_1$ is the only one to invest. If all consumers coordinate on one low-ability expert, then there are stable mixed equilibria in which only LA${_3}$ invests for $\alpha  > \frac{r_i}{R} \coloneqq \alpha_{la3,ha0la0,inv}$ and $\alpha < \frac{R}{2R-r_i} \coloneqq \alpha_{ha0,la3la0,ninv}$. The second low-ability expert always foregoes the decision aid. Similarly, if only LA$_0$ chooses to rent the decision aid, behavior is stable for $\alpha > 0.5\frac{R-r_i}{R} \coloneqq \alpha_{la0,ha0la3,inv}$ and $\alpha < 0.5 \coloneqq \alpha_{ha0,la0la3,ninv}$, with LA$_3$ always ignoring the algorithm. If $x_{ha} = 1$ and consumers are distributed equally, then the high-ability expert foregoes the decision aid along $\alpha < \frac{1.5R-0.5r_i}{3R-2r_i} \coloneqq \alpha_{ha1,la1la1,ninv}$, and either low-ability expert invests, if the second one does not invest, when $\alpha > \frac{R+r_i}{6R}  \coloneqq \alpha_{la1,ha1la1,inv}$. If consumers are not distributed equally and $x_{ha} = 1$, the corresponding hyperbolas if LA$_2$ invests are $\alpha < \frac{1.5R+0.5r_i}{3R-r_i} \coloneqq \alpha_{ha,la2la0,ninv}$, and $\alpha > \frac{2}{3} \frac{r_i}{R} \coloneqq \alpha_{la2,ha1la0,inv}$. Once again, in the mixed equilibrium, the second low-ability expert never switches because it would reveal the high-ability expert's type. If LA$_0$ invests, the hyperbolas are $\alpha < \frac{1.5R-0.5r_i}{3R-2r_i} \coloneqq \alpha_{ha,la0la2,ninv}$ and $\alpha > \frac{1}{3} \frac{r_i}{R} \coloneqq \alpha_{la0,ha1la2,inv}$. If $\alpha_{la0,ha1la2,inv} \leq \alpha < \alpha_{la2,ha1la0,inv}$, then only LA$_0$ invests in equilibrium, if $\alpha \geq \alpha_{la2,ha1la0,inv}$, either does. For $x_{ha} = 2$, the relevant hyperbola for the high-ability expert is $\alpha < \frac{1.5R+0.5r_i}{3R-r_i} \coloneqq \alpha_{ha2,la1la0,ninv}$ irrespective of which low-ability expert invests. LA$_1$ invests in equilibrium if $\alpha > \frac{1}{3} \frac{r_i}{R} \coloneqq \alpha_{la1,ha2la0,inv}$, LA$_0$ if $\alpha > \frac{1}{6} \frac{R-r_i}{R}$ and $r_i \geq 0.5R \coloneqq \alpha_{la0,ha2la1,inv}$ or $\alpha > \frac{1}{6} \frac{r_i}{R} \coloneqq \alpha_{la0,ha2la1,inv}$ and $r_i < 0.5R$. Finally, if all consumers coordinate on the high-ability expert, there is no satisfying equilibrium because both low-ability experts can only poach consumers if they are the only investing expert, and never attract any consumer by foregoing the decision aid. Hence, the investment decision is largely inconsequential, and if anything, both should individually invest, revealing the high-ability expert. Knowing all individual conditions, we can calculate how they change if experts do not observe the choices of ``other'' consumers. All results are shown in Table \ref{tab:mixed_equilibria_noinf}.

\subsubsection{Level-1-Thinking Experts Pure Separation}

When consumers coordinate on the high-ability expert, they forego the decision aid along $\alpha \leq \frac{R+3r_i}{2(R+r_i)} \, \text{for} \, r_i < 0.5R$ and $\alpha \leq \frac{2R + r_i}{3R} \coloneqq \alpha_{ha3,ninv}  \, \text{for} \, r_i \geq 0.5R$. Each LA$_0$ invests along $\alpha \geq \frac{5r_i -R}{2(R+r_i)} \, \text{for} \, r < 0.5R$ and $\alpha \geq \frac{r_i}{R} \alpha_{la0,inv}  \, \text{for} \, r \geq 0.5R$.

If $x_{ha} = 2$, high-ability experts forego the decision aid along $\alpha \leq \frac{2R + 3r_i}{3R + 2r_i} \, \text{for} \, r < 0.5R$ and $\alpha \leq \frac{7R}{9R - 2r_i} \coloneqq \alpha_{ha2,ninv}  \, \text{for} \, r \geq 0.5R$. The hyperbola for the low-ability expert with $x_{la} = 1$ to switch from investing to not-investing is $ \alpha \geq \frac{19r_i - 5R}{2(3R + 4r_i)} \, \text{for} \, r_i < 0.5R$ and $\alpha \geq \frac{3(5r_i - R)}{4(2R + r_i)} \coloneqq \alpha_{la1,inv}  \, \text{for} \, r_i \geq 0.5R$. For LA$_0$, it is $\alpha \geq \frac{17r_i - 3R}{6R + 7r_i}  \, \text{for} \, r_i < 0.5R$ and $\alpha \geq \frac{7r_i + 2R}{10R - r_i} \coloneqq \alpha_{la0,inv}  \, \text{for} \, r_i \geq 0.5R$. Hence, there are separating strategies for $ \alpha_{la1,inv} < \alpha_{la0,inv}  < \alpha \leq \alpha_{ha2,ninv}$.

For $x_{ha} = 1$ and $x_{la} = 2$, the high-ability expert prefers no investment along $\alpha \leq \frac{8R - 3r_i}{3(3R - r_i)}  \, \text{for} \, r_i < 0.5R$ and $ \alpha \leq \frac{7R - r_i}{3(3R - r_i)} \coloneqq \alpha_{ha1,la2la0,ninv}  \, \text{for} \, r_i \geq 0.5R$. LA$_2$ invests if $\alpha \geq \frac{23r_i - 7R}{2(3R + 5r_i)}  \, \text{for} \, r_i < 0.5R$ and $\alpha \geq \frac{21r_i - 6R}{7R + 8r_i} \coloneqq \alpha_{la2,ha1la0,inv}  \, \text{for} \, r_i \geq 0.5R$. Finally, LA$_0$ invests if $\alpha \geq \frac{19r_i - 3R}{2(3R + 4r_i)}  \, \text{for} \, r_i < 0.5R$ and $\alpha \geq \frac{9r_i+2R}{11R-2r_i} \coloneqq \alpha_{la0,ha1la2,inv}  \, \text{for} \, r_i \geq 0.5R$. Again, strategic separation occurs for $ \alpha_{la2,ha1la0,inv} < \alpha_{la0,ha1la2,inv} < \alpha \leq \alpha_{ha1,ninv} $.

Under equal consumer distribution, we get $\alpha \leq \frac{4(2R-r_i)}{9R - 4r_i}  \, \text{for} \, r_i < 0.5R$ and $ \alpha \leq \frac{7R - 2r_i}{9R - 4r_i} \coloneqq \alpha_{ha1,la1,ninv} \, \text{for} \, r_i \geq 0.5R$ for HA$_1$ as the high-ability expert's hyperbolas, and $\alpha \geq \frac{13r_i - R}{9R + 3r_i} \coloneqq \alpha_{la1,ha1,inv}$ for the low-ability expert LA$_1$.

When all consumers coordinate on the low-ability expert LA$_3$, the high-ability expert foregoes the decision aid along $\alpha \leq \frac{4R}{4R + r_i}  \, \text{for} \, r_i<0.5R$ and $\alpha \leq \frac{5R - 2r_i}{3(2R - r_i)} \coloneqq \alpha_{ha0,la3,ninv}  \, \text{for} \, r_i \geq 0.5R$. LA$_0$ invests if $\alpha \geq \frac{2R + r_i}{4R - r_i} \coloneqq \alpha_{la0,la3,inv}$, and LA$_3$ along $\alpha \geq \frac{3(3r_i - R)}{3R + 5r_i} \coloneqq \alpha_{la3,inv}$.

Finally, under $x_{ha} = 0$, $x_{la,i} = 1$ and $x_{la,j} = 2$, the high-ability expert does not use the decision aid for $\alpha \leq \frac{4(2R - r_i)}{9R - 4r_i}  \, \text{for} \, r < 0.5R$ and $ \alpha  \frac{7R - 2r_i}{9R - 4r_i}  \coloneqq \alpha_{ha0,la2la1,ninv}  \, \text{for} \, r_i \geq 0.5R$. LA$_1$ invests along $\alpha \geq \frac{3R + 7r_i}{2(6R - r_i)} \coloneqq \alpha_{la1,ha0la2,inv}$, and LA$_2$ if $\alpha \geq \frac{19r_i - 4R}{8R + 7r_i} \coloneqq \alpha_{la2,ha0la1,inv}$. Table \ref{tab:conditions_separations_level1} summarizes all conditions for separation depending on the distribution of consumers at the time of the shock.

If experts cannot observe the distribution of other consumers, then the conditions change for low-ability experts with $x \leq 2$ and high-ability experts with $x \leq 1$. This has implications for most distributions. If $x_{ha} = 3$, then the two low-ability experts invest along $\alpha \geq \frac{-144R^{4}+1062R^{3}r_i+2786R^{2}r_i^{2}+1257Rr_i^{3}-353r_i^{4}}
               {8\,(R+r_i)\,(3R+4r_i)\,(4R-r_i)\,(6R+7r_i)}
           \, \text{for} \, r_i < 0.5R \, \text{and} \, LA_0$ and $\alpha \geq  \frac{194R^4 + 545R^3r_i - 257R^2r_i^2 + 32Rr_i^3 - r_i^4}{2R(10R - r_i)(11R - 2r_i)(4R - r_i)} \coloneqq \alpha_{la0,inv,noinf}  \, \text{for} \, r_i \geq 0.5R \, \text{and} \, LA_0$. If $x_{ha} = 2$, then $LA_1$ invests along $\alpha \geq \frac{-225R^3 + 1731R^2r_i + 1091Rr_i^2 - 77r_i^3}{18(3R + 4r_i)(3R + r_i)(6R - r_i)} \, \text{for} \, r < 0.5R$ and $\alpha \geq \frac{-102R^3 + 1733R^2r_i + 584Rr_i^2 - 55r_i^3}{36(2R + r_i)(3R + r_i)(6R - r_i)} \coloneqq  \alpha_{la1,inv,noinf} \, \text{for} \, r \geq 0.5R$. For LA$_0$, the condition is the same as above. If $x_{ha} = 1$, high-ability experts forego along $\alpha \leq \frac{144R^2 - 119Rr_i + 24r_i^2}{6(3R - r_i)(9R - 4r_i)} \, \text{for} \, r_i < 0.5R$ and $\alpha \geq \frac{63R^2 - 38Rr_i + 5r_i^2}{3(3R - r_i)(9R - 4r_i)} \coloneqq \alpha_{ha1,ninv,ninf} \, \text{for} \, r_i \geq 0.5R$. LA$_2$ considers the possibilities of $x_{ha} = 1$ or $x_{la} = 1$, resulting in $\alpha \geq \frac{-80R^2 + 209Rr_i + 351r_i^2}{4(3R + 5r_i)(8R + 7r_i)} \, \text{for} \, r_i < 0.5$ and $\alpha \geq \frac{-76R^2 + 227Rr_i + 299r_i^2}{2(7R + 8r_i)(8R + 7r_i)} \coloneqq \alpha_{la2,inv,ninf} \, \text{for} \, r_i \geq 0.5R$. For LA$_3$, nothing changes because they know that they monopolize the market. Finally, if the high-ability expert has not attracted any consumers, they face $\alpha \leq \frac{2(17R^2 - 6Rr_i - r_i^2)}{(9R - 4r_i)(4R + r_i)} \, \text{for} \, r_i < 0.5R$ and $ \alpha \leq \frac{87R^2 - 71Rr_i + 14r_i^2}{6(2R - r_i)(9R - 4r_i)} \coloneqq \alpha_{ha0,ninv,ninf} \, \text{for} \, r_i \geq 0.5R$. Table X summarizes these conditions. 

\clearpage
\subsection{Figures Expert Separation}

\begin{figure}[htbp]
    \centering 

    \begin{minipage}[b]{0.45\textwidth} 
        \centering
        \includegraphics[width=\textwidth]{fig/conditions_equilibrium_ha0la3_noinf.pdf}
    \end{minipage}
    \hfill 
    \begin{minipage}[b]{0.45\textwidth}
        \centering
        \includegraphics[width=\textwidth]{fig/conditions_equilibrium_ha1la2_noinf.pdf}
    \end{minipage}

    \vspace{0.5cm} 

    \begin{minipage}[b]{0.45\textwidth}
        \centering
        \includegraphics[width=\textwidth]{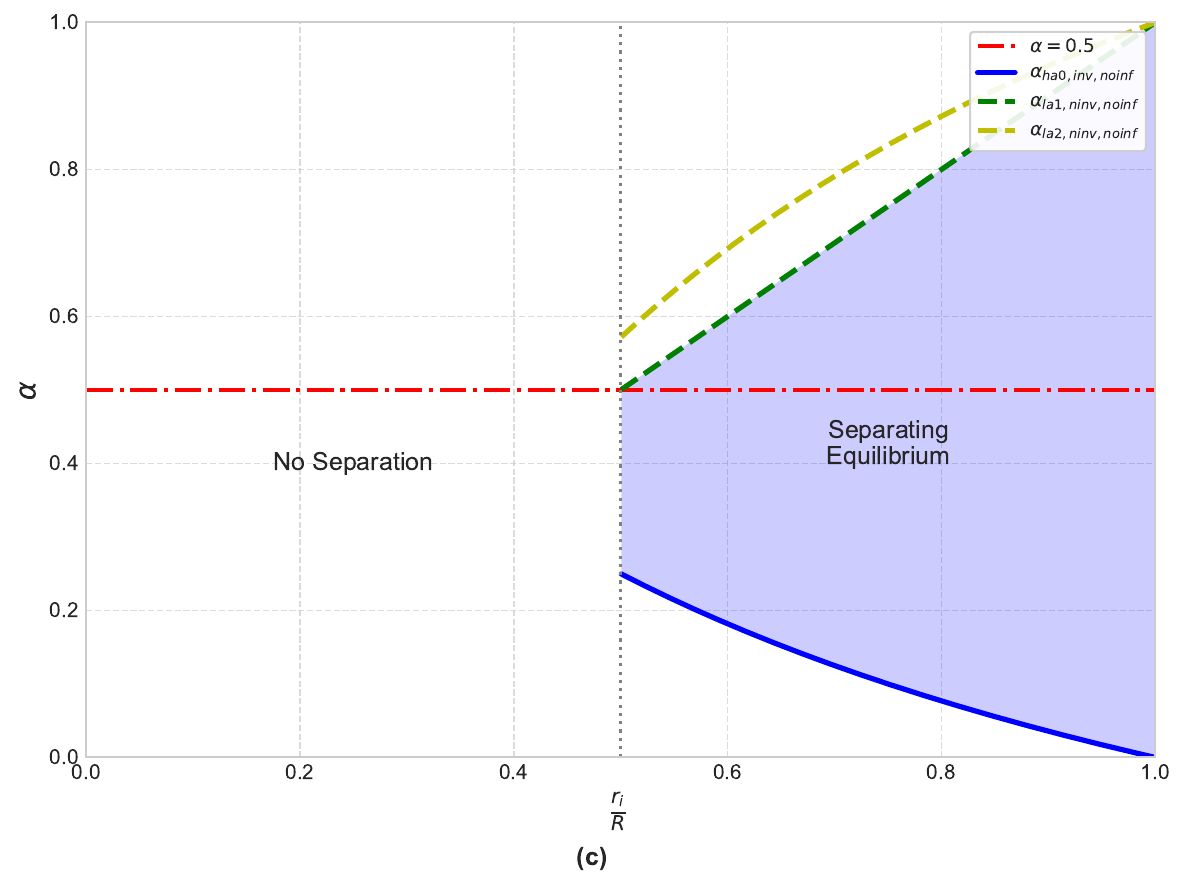}
    \end{minipage}
    \hfill 
    \begin{minipage}[b]{0.45\textwidth}
        \centering
        \includegraphics[width=\textwidth]{fig/new_conditions_equilibrium_la2la1ha0_2.pdf}
    \end{minipage}

    \vspace{0.5cm} 

    \begin{minipage}[b]{0.45\textwidth}
        \centering
        \includegraphics[width=\textwidth]{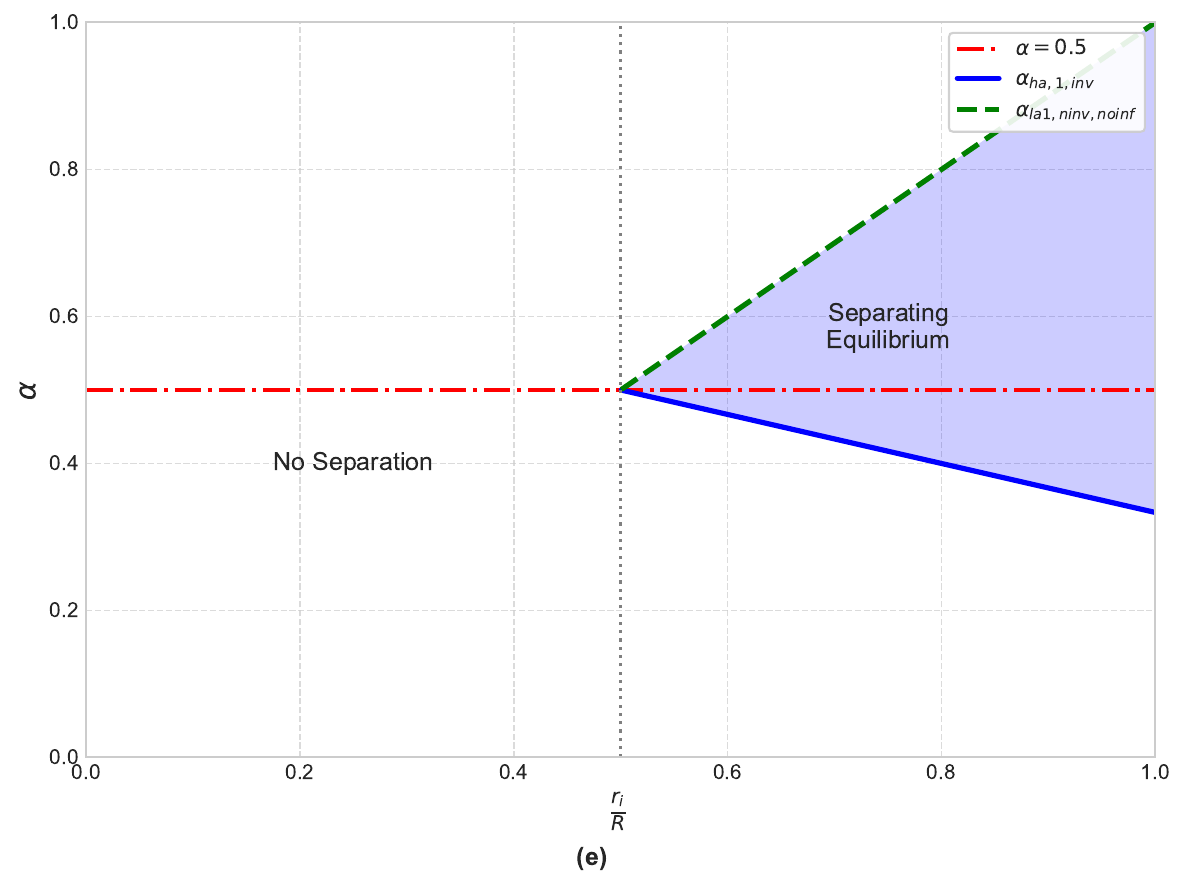}
    \end{minipage}

    \caption{Expert type separating equilibria (PBE) when the distribution of other consumers is unknown. Only the high-ability expert invests. Panel \textbf{(a)} shows the separating space when $x_{la} = 3$. Panel \textbf{(b)} shows the same for $x_{ha} = 1$ and $x_{la} = 2$. Panels \textbf{(c)} and \textbf{(d)} show separation when $x_{ha} = 0$, $x_{la,i} = 2$ and $x_{la,j} = 1$ for $r_i < 0.5R$ and $r_i \geq 0.5R$ respectively. In panel \textbf{(e)}, consumers are distributed equally. }
    \label{fig:separating_noinf} 
\end{figure}

\begin{figure}[htbp]
    \centering 

    \begin{minipage}[b]{0.45\textwidth} 
        \centering
        \includegraphics[width=\textwidth]{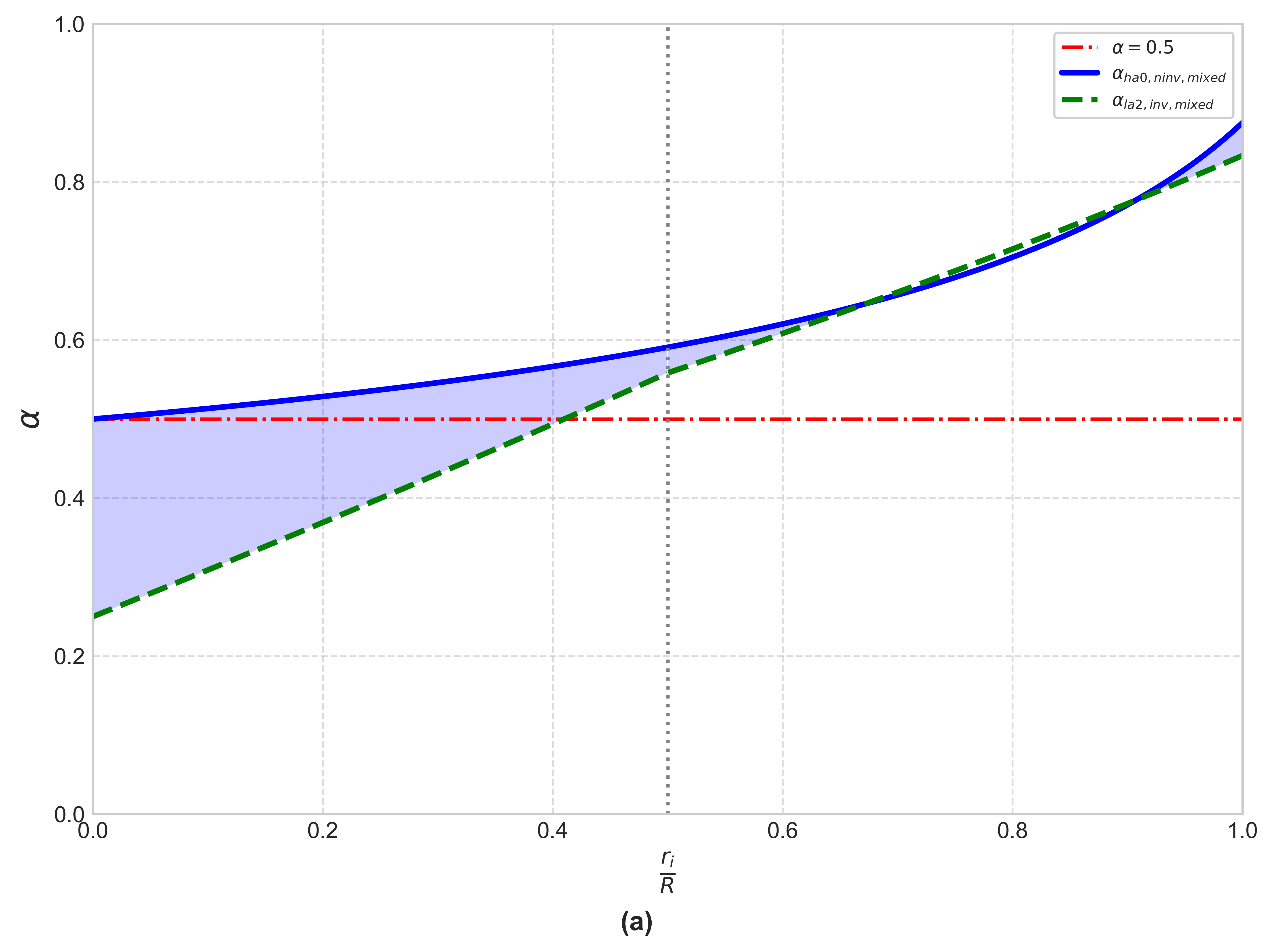}
    \end{minipage}
    \hfill 
    \begin{minipage}[b]{0.45\textwidth}
        \centering
        \includegraphics[width=\textwidth]{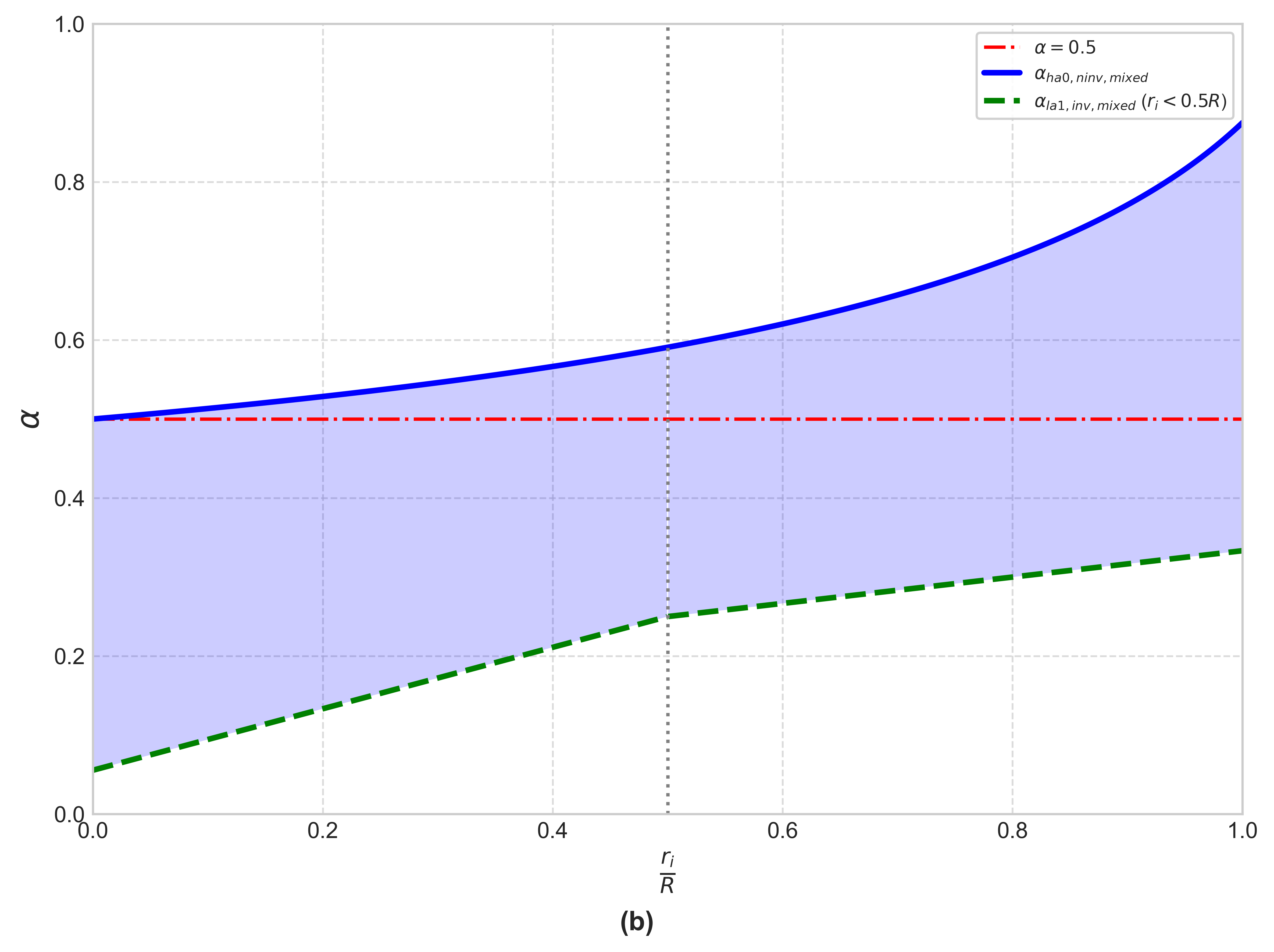}
    \end{minipage}

    \vspace{0.5cm} 

    \begin{minipage}[b]{0.45\textwidth}
        \centering
        \includegraphics[width=\textwidth]{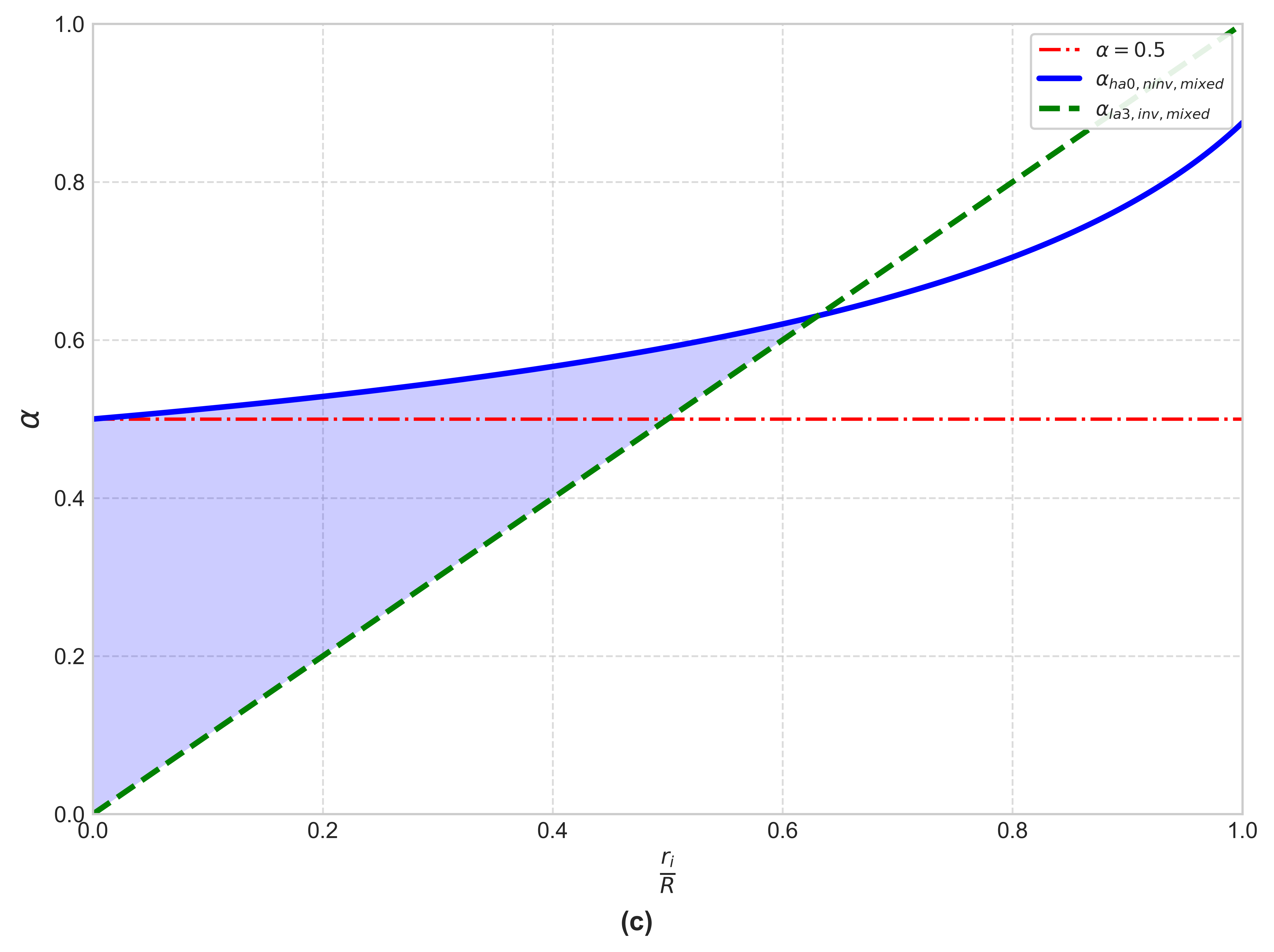}
    \end{minipage}
    \hfill 
    \begin{minipage}[b]{0.45\textwidth}
        \centering
        \includegraphics[width=\textwidth]{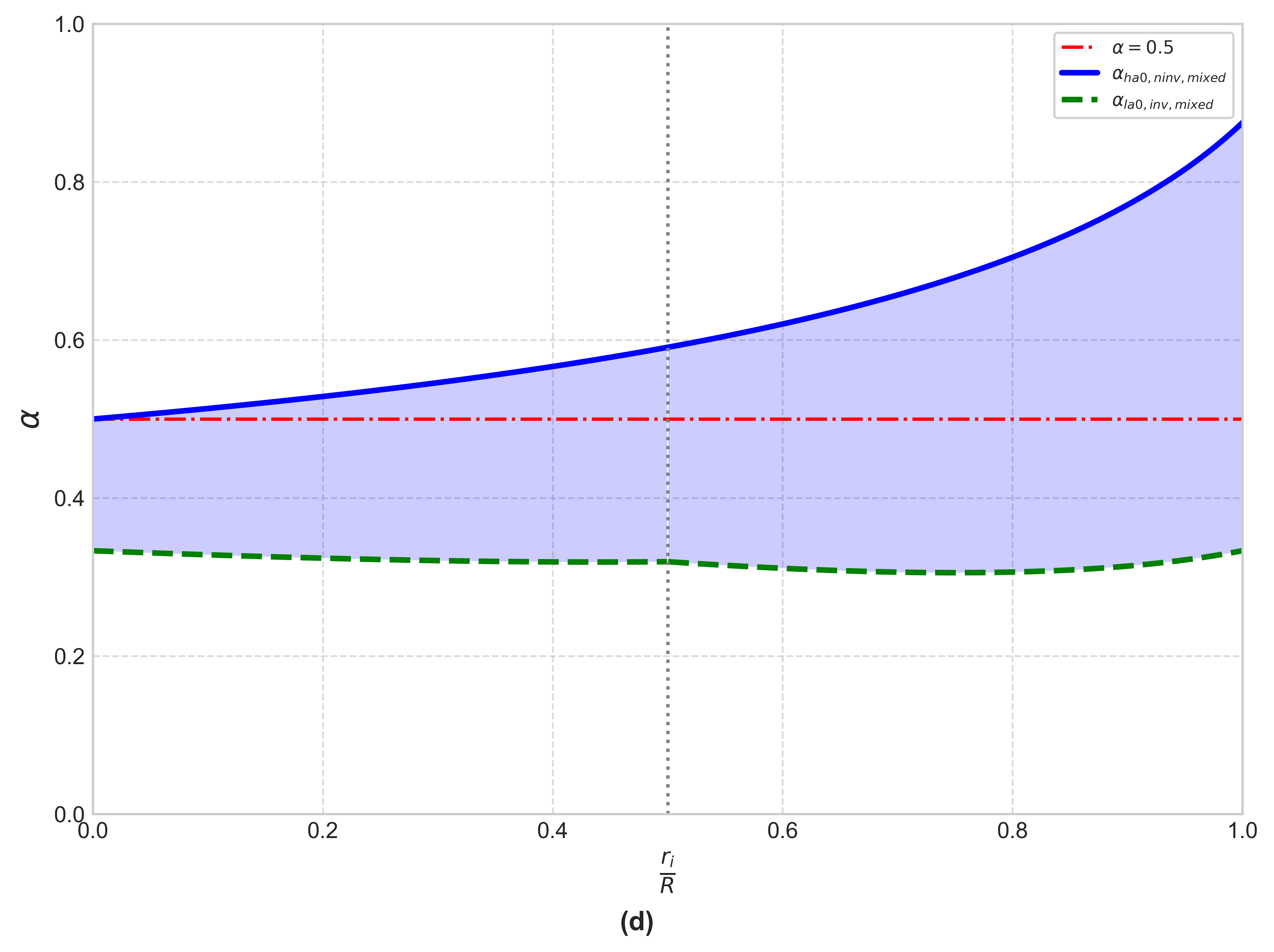}
    \end{minipage}

    \caption{Mixed equilibria in which \textbf{only one low-ability expert invests} when the distribution of other consumers is unknown. Panel \textbf{(a)} shows the mixed space when $x_{ha} = 0$, $x_{la,i} = 2$, $x_{la,j} = 1$, Panel \textbf{(b)} when $x_{ha} = 0$, $x_{la,i} = 2$, $x_{la,j} = 1$, Panel \textbf{(c)} when $x_{ha} = 0$, $x_{la,i} = 3$, $x_{la,j} = 0$, \textbf{(d)} when $x_{ha} = 0$, $x_{la,i} = 0$, $x_{la,j} = 3$.}
    \label{fig:mixed_inf}
\end{figure}

\begin{figure}[htbp]
    \centering 

    \begin{minipage}[b]{0.45\textwidth} 
        \centering
        \includegraphics[width=\textwidth]{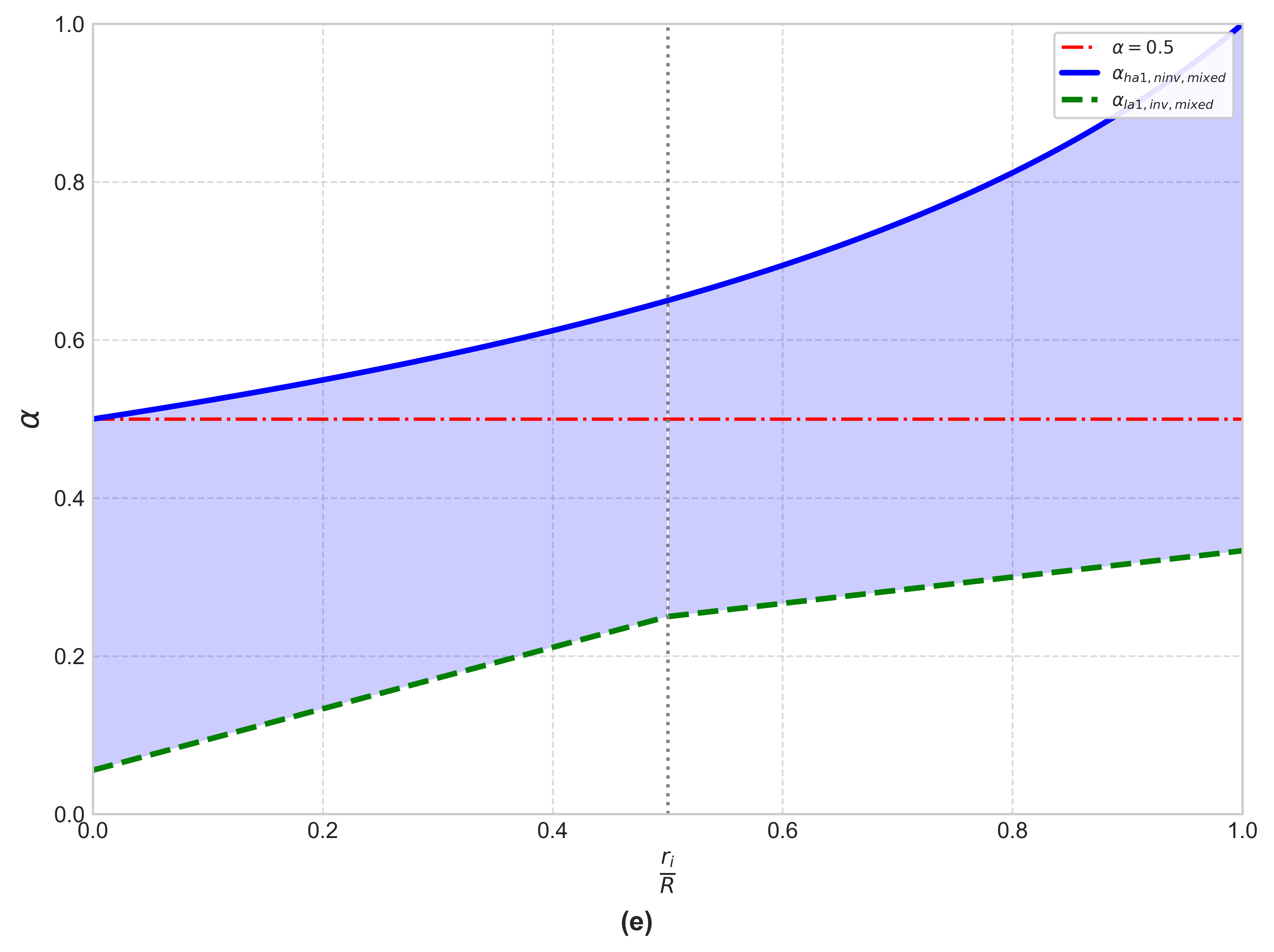}
    \end{minipage}
    \hfill 
    \begin{minipage}[b]{0.45\textwidth}
        \centering
        \includegraphics[width=\textwidth]{fig/equilibrium_ha1la2la0_haninvsep_noinf.png}
    \end{minipage}

    \vspace{0.5cm} 

    \begin{minipage}[b]{0.45\textwidth}
        \centering
        \includegraphics[width=\textwidth]{fig/equilibrium_ha1la0la2_haninvsep_noinf.png}
    \end{minipage}
    \hfill 
    \begin{minipage}[b]{0.45\textwidth}
        \centering
        \includegraphics[width=\textwidth]{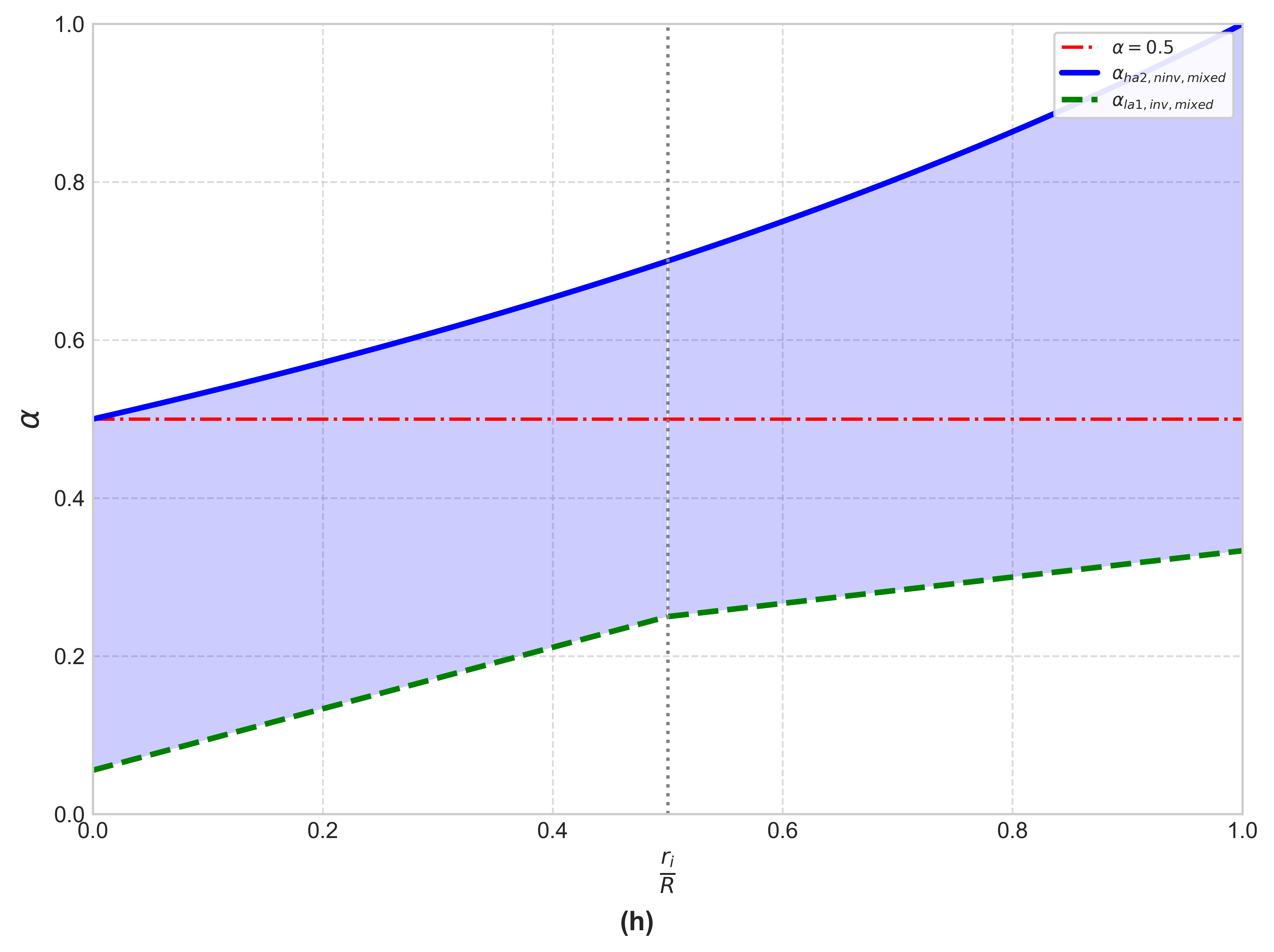}
    \end{minipage}

    \vspace{0.5cm} 

    \begin{minipage}[b]{0.45\textwidth}
        \centering
        \includegraphics[width=\textwidth]{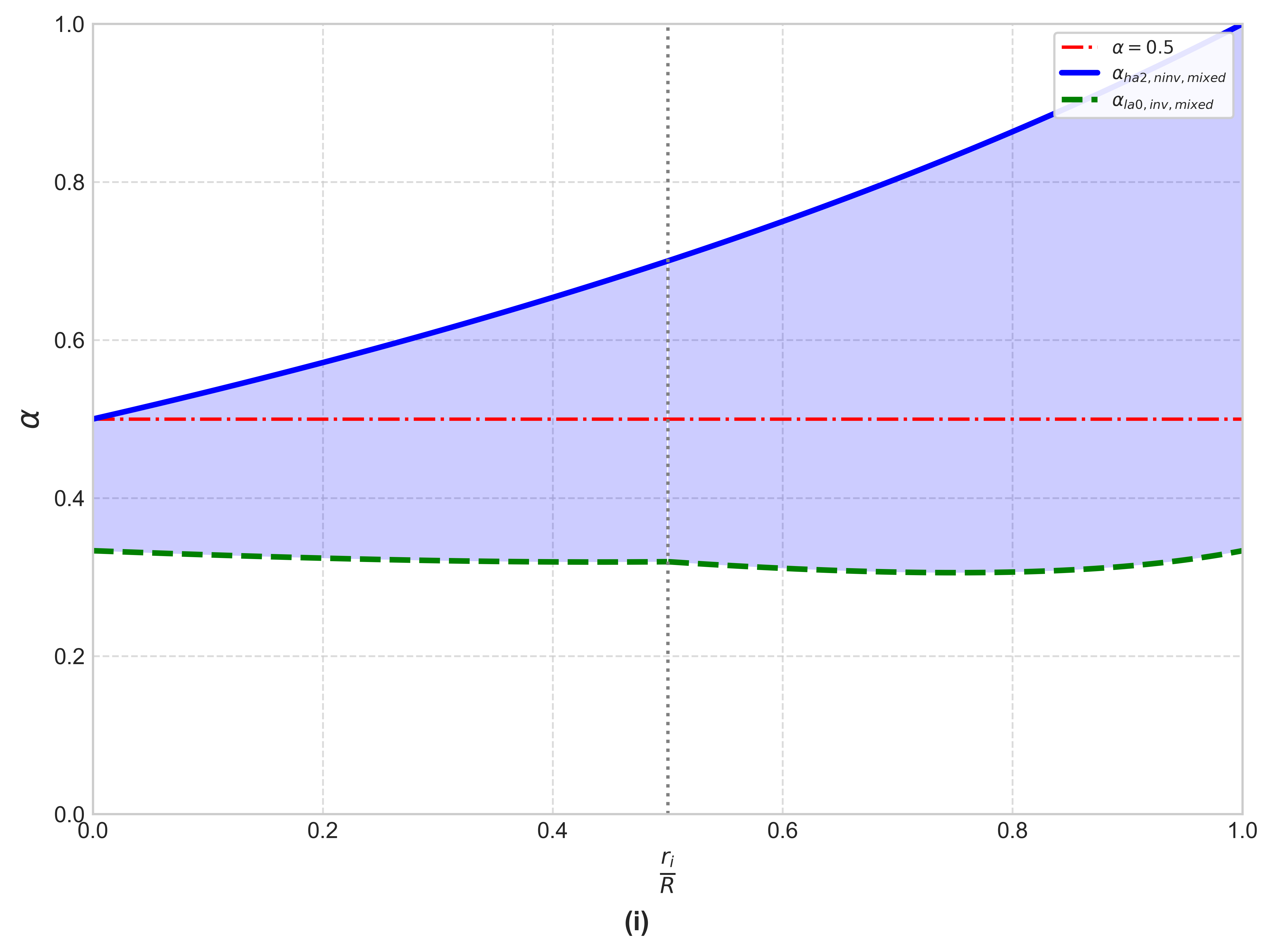}
    \end{minipage}


    \caption{Mixed equilibria in which \textbf{only one low-ability expert invests} when the distribution of other consumers is unknown. Panel \textbf{(a)} shows the mixed space when $x_{ha} = 1$, $x_{la,i} = 1$, $x_{la,j} = 1$, Panel \textbf{(b)} when $x_{ha} = 1$, $x_{la,i} = 2$, $x_{la,j} = 0$, Panel \textbf{(c)} when $x_{ha} = 1$, $x_{la,i} = 0$, $x_{la,j} = 2$, \textbf{(d)} when $x_{ha} = 2$, $x_{la,i} = 1$, $x_{la,j} = 0$, \textbf{(e)} when $x_{ha} = 2$, $x_{la,i} = 0$, $x_{la,j} = 1$.}
    \label{fig:mixed_noinf}
\end{figure}

\clearpage

\clearpage
\subsection{Figures and Tables}
\begin{table}[htbp] 
\centering
\caption{Separating Equilibria Conditions Only HA Invests}
\label{tab:conditions_ha_only_invest_app}
\small 
\begin{tabular}{@{}lll@{}} 
\toprule
\textbf{Equilibrium Description} & \textbf{Conditions ($\alpha$)} & \textbf{Threshold Definitions} \\
\midrule

\multicolumn{3}{@{}l}{\textit{Case: $x_{la}=3$ -- separation possible for $r_i \geq 0.5R$ where $\alpha_{ha,0,inv} \leq \alpha \leq \alpha_{la,3,ninv}$}} \\ \addlinespace

Only HA$_0$ invests    & $\alpha \geq \alpha_{ha,0,inv}$     & $\alpha_{ha,0,inv} \coloneqq \begin{cases} 0.25 & \text{if } r_i < 0.5R \\ 0.5 - \frac{0.5r_i}{R} & \text{if } r_i \geq 0.5R \end{cases}$ \\ \addlinespace
LA$_0$ does not invest & $\alpha \leq \alpha_{la,0,ninv}$ & $\alpha_{la,0,ninv} \coloneqq \begin{cases} 1.5 - \frac{0.5R}{r_i} & \text{if } r_i < 0.5R \\ 0.5 & \text{if } r_i \geq 0.5R \end{cases}$ \\ \addlinespace 
LA$_3$ does not invest & $\alpha \leq \alpha_{la,3,ninv}$ & $\alpha_{la,3,ninv} \coloneqq \frac{3r_i -R}{R+3r_i}$  \\ 
\midrule

\multicolumn{3}{@{}l}{\textit{Case: $x_{ha}=1, x_{la}=2$ -- \textit{separation possible only if $r_i \geq 0.5R$.}}} \\ \addlinespace
Only HA$_1$ invests         & $\alpha > \alpha_{ha,1,inv}$      & $\alpha_{ha,1,inv} \coloneqq \begin{cases} 1 - \frac{r_i}{R} & \text{if } r_i < 0.5R \\ \frac{2R - r_i}{3R} & \text{if } r_i \geq 0.5R \end{cases}$ \\ \addlinespace
LA$_2$ does not invest   & $\alpha < \alpha_{la,20,ninv}$    & $\alpha_{la,20,ninv} \coloneqq \begin{cases} 1.5 - \frac{0.5R}{r_i} & \text{if } r_i < 0.5R \\ \frac{7r_i - 2R}{4r_i + R} & \text{if } r_i \geq 0.5R \end{cases}$ \\ \addlinespace
LA$_0$ does not invest    & $\alpha < \alpha_{la,02,ninv}$    & $\alpha_{la,02,ninv} \coloneqq \begin{cases} 1.5 - \frac{0.5R}{r_i} & \text{if } r_i < 0.5R \\ \frac{-0.5(R+r_i)}{r_i - 2R} & \text{if } r_i \geq 0.5R \end{cases}$ \\
\midrule

\multicolumn{3}{@{}l}{\textit{Case: $x_{ha}=0$ -- separation requires $\alpha_{ha,0,inv} < \alpha < \alpha_{la,12,ninv}$}} \\ \addlinespace
Only HA$_0$ invests      & $\alpha > \alpha_{ha,0,inv}$      & $\alpha_{ha,0,inv} \coloneqq \begin{cases} 1 - \frac{1.5r_i}{R} & \text{if } r_i < 0.5R \\ \frac{0.5(R-r_i)}{R} & \text{if } r_i \geq 0.5R \end{cases}$ \\ \addlinespace 
LA$_2$ does not invest    & $\alpha < \alpha_{la,21,ninv}$    & $\alpha_{la,21,ninv} \coloneqq \begin{cases} 1.5 - \frac{0.5R}{r_i} & \text{if } r_i < 0.5R \\ \frac{2.5r_i - 0.5R}{r_i + R} & \text{if } r_i \geq 0.5R \end{cases}$ \\ \addlinespace
LA$_2$ does not invest      & $\alpha < \alpha_{la,12,ninv}$    & $\alpha_{la,12,ninv} \coloneqq \begin{cases} 1.5 - \frac{0.5R}{r_i} & \text{if } r_i < 0.5R \\ \frac{0.5(R+r_i)}{2R - r_i} & \text{if } r_i \geq 0.5R \end{cases}$ \\
\midrule

\multicolumn{3}{@{}l}{\textit{Case: equal consumer distribution -- \textit{separation possible only if $r_i \geq 0.5R$.}}} \\ \addlinespace
Only HA$_1$ invests          & $\alpha > \alpha_{ha,1,inv}$      & see above for $r_i \geq 0.5R$ \\ \addlinespace
LA$_1$ invests & $\alpha < \alpha_{la,eq,ninv}$    & $\alpha_{la,eq,ninv} \coloneqq \frac{r_i}{R}$ \\ 

\bottomrule
\end{tabular}
\end{table}

\begin{table}[htbp] 
\centering
\caption{Mixed Equilibria Conditions}
\label{tab:mixed_equilibria} 
\small 
\begin{tabular}{@{}lll@{}} 
\toprule
\textbf{Equilibrium Description} & \textbf{Conditions ($\alpha$)} & \textbf{Threshold Definitions} \\
\midrule

\multicolumn{3}{@{}l}{\textit{Case: $x_{ha}=0$}} \\ \addlinespace 

HA$_0$ & $\alpha < \alpha_{ha0,la21,ninv}$ & $\alpha_{ha0,la21,ninv} \coloneqq \frac{1.5R-0.5r_i}{3R-2r_i}$ \\
Only LA$_2$ invests & $\alpha > \alpha_{la2,ha0,inv}$ & $\alpha_{la2,ha0,inv} \coloneqq \begin{cases} \frac{5}{6} \frac{r_i}{R} & \text{if } r_i < 0.5R \\ \frac{1.5r_i + 0.5R}{3R} & \text{if } r_i \geq 0.5R \end{cases}$ \\
\addlinespace

HA$_0$ & $\alpha < \alpha_{ha0,la12,ninv}$ & $\alpha_{ha0,la12,ninv} \coloneqq \frac{3R-2r_i}{6R-5r_i}$ \\
Only LA$_1$ invests & $\alpha > \alpha_{la1,ha0,inv}$ & $\alpha_{la1,ha0,inv} \coloneqq \begin{cases} \frac{2}{3} \frac{r_i}{R} & \text{if } r_i < 0.5R \\ \frac{1}{3} & \text{if } r_i \geq 0.5R \end{cases}$ \\
\addlinespace

\multicolumn{3}{@{}l}{\textit{Consumers coordinate on LA$_3$}} \\ \addlinespace

HA$_0$ & $\alpha < \alpha_{ha0,la3la0,ninv}$ & $\alpha_{ha0,la3la0,ninv} \coloneqq \frac{R}{2R-r_i}$ \\
Only LA$_3$ invests & $\alpha > \alpha_{la3,ha0la0,inv}$ & $\alpha_{la3,ha0la0,inv} \coloneqq \frac{r_i}{R}$ \\
 & & \\
\addlinespace

HA$_0$ & $\alpha < \alpha_{ha0,la0la3,ninv}$ & $\alpha_{ha0,la0la3,ninv} \coloneqq 0.5$ \\
Only LA$_0$ invests & $\alpha > \alpha_{la0,ha0la3,inv}$ & $\alpha_{la0,ha0la3,inv} \coloneqq 0.5 \frac{R-r_i}{R}$ \\
 & & \\
\midrule

\multicolumn{3}{@{}l}{\textit{Case: $x_{ha}=1$}} \\ \addlinespace

\textit{Equal consumer distribution:} & & \\
HA$_1$ & $\alpha < \alpha_{ha1,la1la1,ninv}$ & $\alpha_{ha1,la1la1,ninv} \coloneqq \frac{1.5R-0.5r_i}{3R-2r_i}$ \\
Either LA$_1$ invests & $\alpha > \alpha_{la1,ha1la1,inv}$ & $\alpha_{la1,ha1la1,inv} \coloneqq \frac{R+r_i}{6R}$ \\
\addlinespace

\textit{Unequal consumer distribution:} & & \\
HA$_1$ & $\alpha < \alpha_{ha1,la2la0,ninv}$ & $\alpha_{ha,la2la0,ninv} \coloneqq \frac{1.5R+0.5r_i}{3R-r_i}$ \\
Only LA$_2$ invests & $\alpha > \alpha_{la2,ha1la0,inv}$ & $\alpha_{la2,ha1la0,inv} \coloneqq \frac{2}{3} \frac{r_i}{R}$ \\
\addlinespace

HA$_1$ & $\alpha < \alpha_{ha1,la0la2,ninv}$ & $\alpha_{ha,la0la2,ninv} \coloneqq \frac{1.5R-0.5r_i}{3R-2r_i}$ \\ 
Only LA$_0$ invests$^{a}$ & $\alpha > \alpha_{la0,ha1la2,inv}$ & $\alpha_{la0,ha1la2,inv} \coloneqq \frac{1}{3} \frac{r_i}{R}$ \\
\midrule

\multicolumn{3}{@{}l}{\textit{Case: $x_{ha}=2$}} \\ \addlinespace

HA$_2$ & $\alpha < \alpha_{ha2,la1la0,ninv}$ & $\alpha_{ha2,la1la0,ninv} \coloneqq \frac{1.5R+0.5r_i}{3R-r_i}$ \\
Only LA$_1$ & $\alpha > \alpha_{la1,ha2la0,inv}$ & $\alpha_{la1,ha2la0,inv} \coloneqq \frac{1}{3} \frac{r_i}{R}$ \\
\addlinespace

HA$_2$ & $\alpha < \alpha_{ha2,la1la0,ninv}$ & $(\text{same as above})$ \\
Only LA$_0$ invests & $\alpha > \alpha_{la0,ha2la1,inv}$ & $\alpha_{la0,ha2la1,inv} \coloneqq \begin{cases} \frac{1}{6} \frac{r_i}{R} & \text{if } r_i < 0.5R \\ \frac{1}{6} \frac{R-r_i}{R} & \text{if } r_i \geq 0.5R \end{cases}$ \\
\midrule

\multicolumn{3}{@{}l@{}}{\textit{Case: All consumers coordinate on high-ability expert}$^{b}$} \\ \addlinespace 

\bottomrule
\end{tabular}

\par 
\vspace{1ex} 
\begin{minipage}{\linewidth} 
\small 
$^{a}$ If $\alpha \geq \alpha_{la2,ha1la0,inv}$, either LA can invest. If $\alpha_{la0,ha1la2,inv} \leq \alpha < \alpha_{la2,ha1la0,inv}$, only LA$_0$ invests. \\ 
$^{b}$ No stable mixed equilibrium exists where only one LA invests. Both LAs investing would reveal the HA type. 
\end{minipage}

\end{table}

\begin{table}[htbp] 
\centering
\caption{Mixed Equilibria Conditions if Distribution and Investments Are Not Observable}
\label{tab:mixed_equilibria_noinf} 
\small 
\begin{tabular}{@{}lll@{}} 
\toprule
\textbf{Equilibrium Description} & \textbf{Conditions ($\alpha$)} & \textbf{Threshold Definitions} \\
\midrule

\multicolumn{3}{@{}l}{\textit{Case: $x_{ha}=0$}} \\ \addlinespace

HA$_0$ does not invest & $\alpha < \alpha_{ha0,ninv,mixed}$ & $\alpha_{ha0,ninv,mixed} \coloneqq \frac{144R^3 - 252R^2 r_i + 138R r^{2}_i -23r^{3}_i}{8(3R-2r_i)(6R-5r_i)(2R-r_i)}$ \\
\addlinespace
Only LA$_2$ invests & $\alpha > \alpha_{la2,inv,mixed}$ & $\alpha_{la2,inv,mixed} \coloneqq \begin{cases} \frac{9R^2 + 18Rr_i - 5 r^{2}_i}{12R(3R-r_i)} & \text{if } r_i < 0.5R \\ \frac{12R^2 + 11Rr_i -3r^{2}_i}{12R(3R-r_i)} & \text{if } r_i \geq 0.5R \end{cases}$ \\
\addlinespace

Only LA$_1$ invests & $\alpha > \alpha_{la1,inv,mixed}$ & $\alpha_{la1,inv,mixed} \coloneqq \begin{cases} \frac{R + 7r_i}{18R} & \text{if } r_i < 0.5R \\ \frac{R + r_i}{6R} & \text{if } r_i \geq 0.5R \end{cases}$ \\
\addlinespace

\multicolumn{3}{@{}l}{\textit{Consumers coordinate on one LA}} \\ \addlinespace

Only LA$_3$ invests & $\alpha > \alpha_{la3,inv,mixed}$ & $\alpha_{la3,inv,mixed} \coloneqq \frac{r_i}{R}$ \\
 & & \\
\addlinespace

Only LA$_0$ invests & $\alpha > \alpha_{la0,inv,mixed}$ & $\alpha_{la0,inv,mixed} \coloneqq \begin{cases}\frac{18R^2 - 15Rr_i + 4r_i^2}{18R(3R - 2r_i)} & \text{if } r_i < 0.5R \\ \frac{21R^2 - 23Rr_i + 8r_i^2}{18R(3R - 2r_i)} & \text{if } r_i \geq 0.5R \end{cases}$ \\
 & & \\
\midrule

\multicolumn{3}{@{}l}{\textit{Case: $x_{ha}=1$}} \\ \addlinespace

HA$_1$ does not invest & $\alpha < \alpha_{ha1,inv,mixed}$ & $\alpha_{ha1,inv,mixed} \coloneqq \frac{9R^2 - 5Rr_i}{2(3R - 2r_i)(3R - r_i)}$ \\ \addlinespace

\textit{Equal consumer distribution:} & & \\
Either LA$_1$ invests & $\alpha > \alpha_{la1,inv,mixed}$ & $\alpha_{la1,inv,mixed} \coloneqq \begin{cases} \frac{R + 7r_i}{18R} & \text{if } r_i < 0.5R \\ \frac{R + r_i}{6R} & \text{if } r_i \geq 0.5R \end{cases}$ \\
\addlinespace

\textit{Unequal consumer distribution:} & & \\
Only LA$_2$ invests & $\alpha > \alpha_{la2,inv,mixed}$ & $\alpha_{la2,inv,mixed} \coloneqq \begin{cases} \frac{9R^2 + 18Rr_i - 5 r^{2}_i}{12R(3R-r_i)} & \text{if } r_i < 0.5R \\ \frac{12R^2 + 11Rr_i -3r^{2}_i}{12R(3R-r_i)} & \text{if } r_i \geq 0.5R \end{cases}$ \\
\addlinespace

Only LA$_0$ invests & $\alpha > \alpha_{la0,inv,mixed}$ & $\alpha_{la0,inv,mixed} \coloneqq \begin{cases}\frac{18R^2 - 15Rr_i + 4r_i^2}{18R(3R - 2r_i)} & \text{if } r_i < 0.5R \\ \frac{21R^2 - 23Rr_i + 8r_i^2}{18R(3R - 2r_i)} & \text{if } r_i \geq 0.5R \end{cases}$ \\ 
\midrule

\multicolumn{3}{@{}l}{\textit{Case: $x_{ha}=2$}} \\ \addlinespace

HA$_2$ does not invest & $\alpha < \alpha_{ha2,ninv,mixed}$ & $\alpha_{ha2,la1la0,ninv} \coloneqq \frac{1.5R+0.5r_i}{3R-r_i}$ \\ \addlinespace
Only LA$_1$ invests & $\alpha > \alpha_{la1,inv,mixed}$ & see above \\
\addlinespace

Only LA$_0$ invests & $\alpha > \alpha_{la0,inv,mixed}$ & see above \\
\midrule

\multicolumn{3}{@{}l@{}}{\textit{Case: All consumers coordinate on High-ability expert}$^{b}$} \\ \addlinespace 

\bottomrule
\end{tabular}

\par 
\vspace{1ex} 
\begin{minipage}{\linewidth} 
\small 
$^{a}$ If $\alpha \geq \alpha_{la2,ha1la0,inv}$, either LA can invest. If $\alpha_{la0,ha1la2,inv} \leq \alpha < \alpha_{la2,ha1la0,inv}$, only LA$_0$ invests. \\ 
$^{b}$ No stable mixed equilibrium exists where only one LA invests. Both LAs investing would reveal the HA type. 
\end{minipage}

\end{table}

\begin{table}[htbp] 
\centering
\caption{Conditions for Strategic Expert Separation under Expert Level-1 Thinking} 
\label{tab:conditions_separations_level1} 
\small 
\begin{tabular}{@{}lll@{}} 
\toprule
\textbf{Equilibrium Description} & \textbf{Conditions ($\alpha$)} & \textbf{Threshold Definitions} \\
\midrule

\multicolumn{3}{@{}l}{\textit{Case: $x_{ha}=3$}} \\ \addlinespace

HA No Investment      & $\alpha \leq \alpha_{ha3,ninv}$    & $\alpha_{ha3,ninv} \coloneqq \begin{cases} \frac{R+3r_i}{2(R+r_i)} & \text{if } r_i < 0.5R \\ \frac{2R + r_i}{3R} & \text{if } r_i \geq 0.5R \end{cases}$ \\ \addlinespace
LA$_0$ Investment     & $\alpha \geq \alpha_{la0,inv}$     & $\alpha_{la0,inv} \coloneqq \begin{cases} \frac{5r_i -R}{2(R+r_i)} & \text{if } r_i < 0.5R \\ \frac{r_i}{R} & \text{if } r_i \geq 0.5R \end{cases}$ \\
\midrule

\multicolumn{3}{@{}l}{\textit{Case: $x_{ha}=2$}} \\ \addlinespace

HA No Investment      & $\alpha \leq \alpha_{ha2,ninv}$    & $\alpha_{ha2,ninv} \coloneqq \begin{cases} \frac{2R + 3r_i}{3R + 2r_i} & \text{if } r_i < 0.5R \\ \frac{7R}{9R - 2r_i} & \text{if } r_i \geq 0.5R \end{cases}$ \\ \addlinespace
LA$_1$ Investment     & $\alpha \geq \alpha_{la1,inv}$     & $\alpha_{la1,inv} \coloneqq \begin{cases} \frac{19r_i - 5R}{2(3R + 4r_i)} & \text{if } r_i < 0.5R \\ \frac{3(5r_i - R)}{4(2R + r_i)} & \text{if } r_i \geq 0.5R \end{cases}$ \\ \addlinespace
LA$_0$ Investment     & $\alpha \geq \alpha_{la0,inv}$     & $\alpha_{la0,inv} \coloneqq \begin{cases} \frac{17r_i - 3R}{6R + 7r_i} & \text{if } r_i < 0.5R \\ \frac{7r_i + 2R}{10R - r_i} & \text{if } r_i \geq 0.5R \end{cases}$ \\
\midrule

\multicolumn{3}{@{}l}{\textit{Case: $x_{ha}=1, x_{la}=2$}} \\ \addlinespace

HA No Investment      & $\alpha \leq \alpha_{ha1,la2la0,ninv}$ & $\alpha_{ha1,la2la0,ninv} \coloneqq \begin{cases} \frac{8R - 3r_i}{3(3R - r_i)} & \text{if } r_i < 0.5R \\ \frac{7R - r_i}{3(3R - r_i)} & \text{if } r_i \geq 0.5R \end{cases}$ \\ \addlinespace
LA$_2$ Investment     & $\alpha \geq \alpha_{la2,ha1la0,inv}$ & $\alpha_{la2,ha1la0,inv} \coloneqq \begin{cases} \frac{23r_i - 7R}{2(3R + 5r_i)} & \text{if } r_i < 0.5R \\ \frac{21r_i - 6R}{7R + 8r_i} & \text{if } r_i \geq 0.5R \end{cases}$ \\ \addlinespace
LA$_0$ Investment & $\alpha \geq \alpha_{la0,ha1la2,inv}$ & $\alpha_{la0,ha1la2,inv} \coloneqq \begin{cases} \frac{19r_i - 3R}{2(3R + 4r_i)} & \text{if } r_i < 0.5R \\ \frac{9r_i+2R}{11R-2r_i} & \text{if } r_i \geq 0.5R \end{cases}$ \\
\midrule

\multicolumn{3}{@{}l}{\textit{Case: Equal Consumer Distribution}} \\ \addlinespace

HA$_1$ No Investment  & $\alpha \leq \alpha_{ha1,la1,ninv}$ & $\alpha_{ha1,la1,ninv} \coloneqq \begin{cases} \frac{4(2R-r_i)}{9R - 4r_i} & \text{if } r_i < 0.5R \\ \frac{7R - 2r_i}{9R - 4r_i} & \text{if } r_i \geq 0.5R \end{cases}$ \\ \addlinespace
LA$_1$ Investment     & $\alpha \geq \alpha_{la1,ha1,inv}$ & $\alpha_{la1,ha1,inv} \coloneqq \frac{13r_i - R}{9R + 3r_i}$ \\ 
\midrule

\multicolumn{3}{@{}l}{\textit{Case: $x_{la}=3$}} \\ \addlinespace

HA No Investment      & $\alpha \leq \alpha_{ha0,la3,ninv}$ & $\alpha_{ha0,la3,ninv} \coloneqq \begin{cases} \frac{4R}{4R + r_i} & \text{if } r_i < 0.5R \\ \frac{5R - 2r_i}{3(2R - r_i)} & \text{if } r_i \geq 0.5R \end{cases}$ \\ \addlinespace
LA$_0$ Investment     & $\alpha \geq \alpha_{la0,la3,inv}$ & $\alpha_{la0,la3,inv} \coloneqq \frac{2R + r_i}{4R - r_i}$ \\ 
\addlinespace
LA$_3$ Investment     & $\alpha \geq \alpha_{la3,inv}$ & $\alpha_{la3,inv} \coloneqq \frac{3(3r_i - R)}{3R + 5r_i}$ \\ 
\midrule

\multicolumn{3}{@{}l}{\textit{Case: $x_{ha}=0, x_{la,1}=1, x_{la,2}=2$}} \\ \addlinespace

HA No Investment      & $\alpha \leq \alpha_{ha0,la2la1,ninv}$ & $\alpha_{ha0,la2la1,ninv} \coloneqq \begin{cases} \frac{4(2R - r_i)}{9R - 4r_i} & \text{if } r_i < 0.5R \\ \frac{7R - 2r_i}{9R - 4r_i} & \text{if } r_i \geq 0.5R \end{cases}$ \\ \addlinespace
LA$_1$ Investment     & $\alpha \geq \alpha_{la1,ha0la2,inv}$ & $\alpha_{la1,ha0la2,inv} \coloneqq \frac{3R + 7r_i}{2(6R - r_i)}$ \\ 
\addlinespace
LA$_2$ Investment     & $\alpha \geq \alpha_{la2,ha0la1,inv}$ & $\alpha_{la2,ha0la1,inv} \coloneqq \frac{19r_i - 4R}{8R + 7r_i}$ \\ 
\bottomrule
\end{tabular}

\end{table}

\clearpage

\begin{table}[htbp] 
\centering
\caption{Conditions for Strategic Expert Separation under Expert Level-1 Thinking if Consumer Distribution Is Not Observable} 
\label{tab:conditions_separations_level1_noinf} 
\small 
\begin{tabular}{@{}lll@{}} 
\toprule
\textbf{Equilibrium Description} & \textbf{Conditions ($\alpha$)} & \textbf{Threshold Definitions} \\
\midrule

\multicolumn{3}{@{}l}{\textit{Case: $x_{ha}=3$}} \\ \addlinespace

HA$_3$ No Investment      & $\alpha \leq \alpha_{ha3,ninv}$    & $\coloneqq \begin{cases} \frac{R+3r_i}{2(R+r_i)} & \text{if } r_i < 0.5R \\ \frac{2R + r_i}{3R} & \text{if } r_i \geq 0.5R \end{cases}$ \\ \addlinespace
LA$_0$ Investment     &$\alpha \geq \alpha_{la0,inv,noinf}$    & $\coloneqq \begin{cases} \frac{-144R^{4}+1062R^{3}r_i+2786R^{2}r_i^{2}+1257Rr_i^{3}-353r_i^{4}}{8(R+r_i)(3R+4r_i)(4R-r_i)(6R+7r_i)} & \text{if } r_i < 0.5R \\ \frac{194R^4 + 545R^3r_i - 257R^2r_i^2 + 32Rr_i^3 - r_i^4}{2R(10R - r_i)(11R - 2r_i)(4R - r_i)} & \text{if } r_i \geq 0.5R \end{cases}$ \\
\midrule

\multicolumn{3}{@{}l}{\textit{Case: $x_{ha}=2$}} \\ \addlinespace

HA$_2$ No Investment      & $\alpha \leq \alpha_{ha2,ninv}$    & $ \coloneqq \begin{cases} \frac{2R + 3r_i}{3R + 2r_i} & \text{if } r_i < 0.5R \\ \frac{7R}{9R - 2r_i} & \text{if } r_i \geq 0.5R \end{cases}$ \\ \addlinespace
LA$_1$ Investment     & $\alpha \geq \alpha_{la1,inv,noinf}$  & $ \coloneqq \begin{cases} \frac{-225R^3 + 1731R^2r_i + 1091Rr_i^2 - 77r_i^3}{18(3R + 4r_i)(3R + r_i)(6R - r_i)} & \text{if } r_i < 0.5R \\ \frac{-102R^3 + 1733R^2r_i + 584Rr_i^2 - 55r_i^3}{36(2R + r_i)(3R + r_i)(6R - r_i)} & \text{if } r_i \geq 0.5R \end{cases}$ \\ \addlinespace
LA$_0$ Investment     & $\alpha \geq \alpha_{la0,inv,noinf}$   & see above \\
\midrule

\multicolumn{3}{@{}l}{\textit{Case: $x_{ha}=1, x_{la}=2$}} \\ \addlinespace

HA$_1$ No Investment      & $\alpha \leq \alpha_{ha1,ninv,noinf}$ & $  \coloneqq \begin{cases} \frac{144R^2 - 119Rr_i + 24r_i^2}{6(3R - r_i)(9R - 4r_i)} & \text{if } r_i < 0.5R \\ \frac{63R^2 - 38Rr_i + 5r_i^2}{3(3R - r_i)(9R - 4r_i)} & \text{if } r_i \geq 0.5R \end{cases}$ \\ \addlinespace 
LA$_2$ Investment     & $\alpha \geq \alpha_{la2,inv,noinf}$  & $\coloneqq \begin{cases} \frac{-80R^2 + 209Rr_i + 351r_i^2}{4(3R + 5r_i)(8R + 7r_i)} & \text{if } r_i < 0.5R \\ \frac{-76R^2 + 227Rr_i + 299r_i^2}{2(7R + 8r_i)(8R + 7r_i)} & \text{if } r_i \geq 0.5R \end{cases}$ \\ \addlinespace
LA$_0$ Investment     & $\alpha \geq \alpha_{la0,inv,noinf}$   & see above \\
\midrule

\multicolumn{3}{@{}l}{\textit{Case: Equal Consumer Distribution}} \\ \addlinespace

HA$_1$ No Investment  & $\alpha \leq \alpha_{ha1,ninv,noinf}$ & see above \\ \addlinespace
LA$_1$ Investment     &  $\alpha \geq \alpha_{la1,inv,noinf}$ & see above \\ 
\midrule

\multicolumn{3}{@{}l}{\textit{Case: $x_{la}=3$}} \\ \addlinespace

HA$_0$ No Investment      & $\alpha \leq \alpha_{ha0,ninv,noinf}$ & $\coloneqq \begin{cases} \frac{2(17R^2 - 6Rr_i - r_i^2)}{(9R - 4r_i)(4R + r_i)} & \text{if } r_i < 0.5R \\ \frac{87R^2 - 71Rr_i + 14r_i^2}{6(2R - r_i)(9R - 4r_i)} & \text{if } r_i \geq 0.5R \end{cases}$ \\ \addlinespace
LA$_0$ Investment     & $\alpha \geq \alpha_{la0,inv,noinf}$   & see above \\
\addlinespace
LA$_3$ Investment     & $\alpha \geq \alpha_{la3,inv}$ & $ \coloneqq \frac{3(3r_i - R)}{3R + 5r_i}$ \\ 
\midrule

\multicolumn{3}{@{}l}{\textit{Case: $x_{ha}=0, x_{la,1}=1, x_{la,2}=2$}} \\ \addlinespace

HA$_0$ No Investment      & $\alpha \leq \alpha_{ha0,ninv,noinf}$ & see above \\ \addlinespace
LA$_1$ Investment     &  $\alpha \geq \alpha_{la1,inv,noinf}$ & see above \\ 
\addlinespace
LA$_2$ Investment     & $\alpha \geq \alpha_{la2,inv,noinf}$  & see above \\ 
\bottomrule
\end{tabular}

\end{table}

\clearpage

\begin{figure}[htpb]
    \centering
\includegraphics[width=0.45\textwidth]{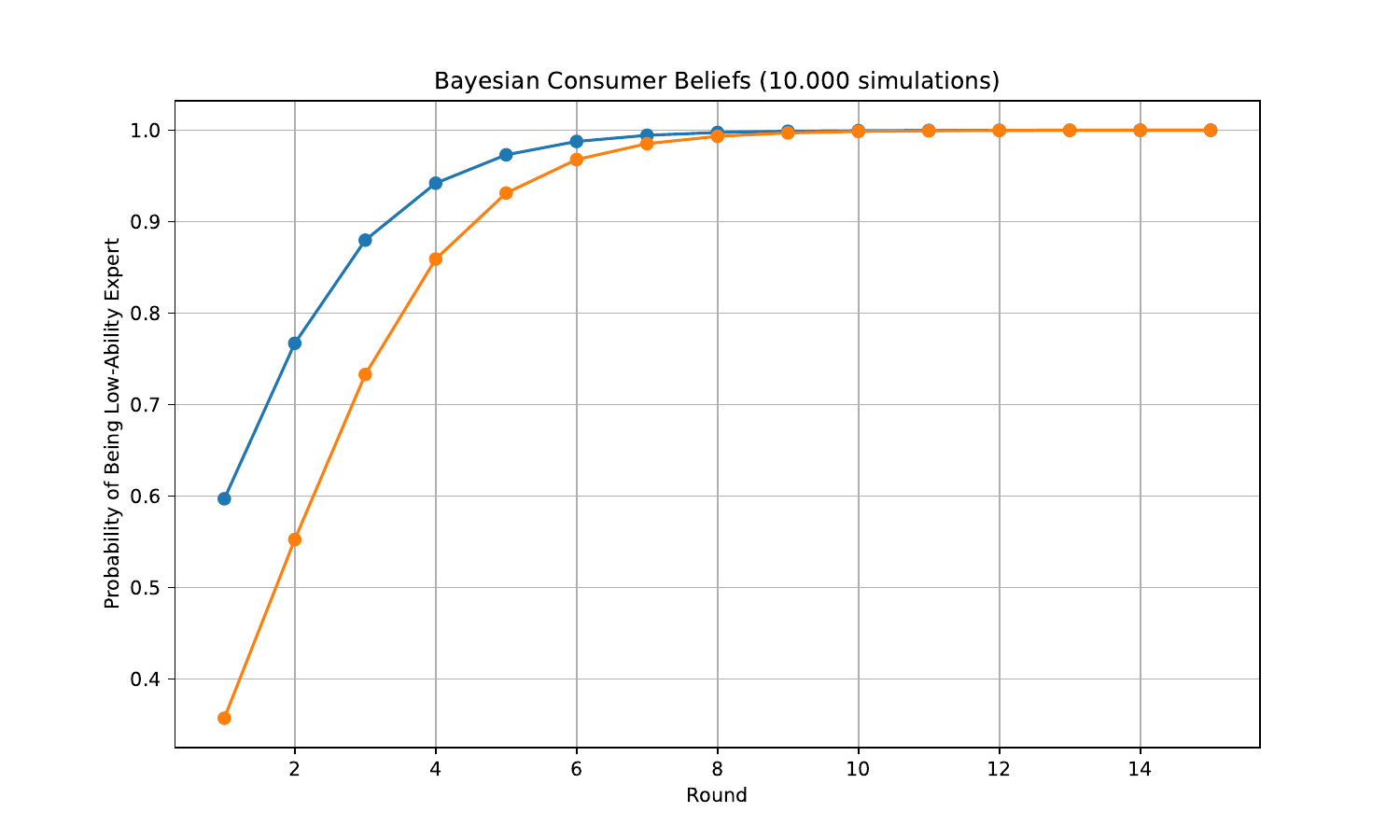}
\includegraphics[width=0.45\textwidth]{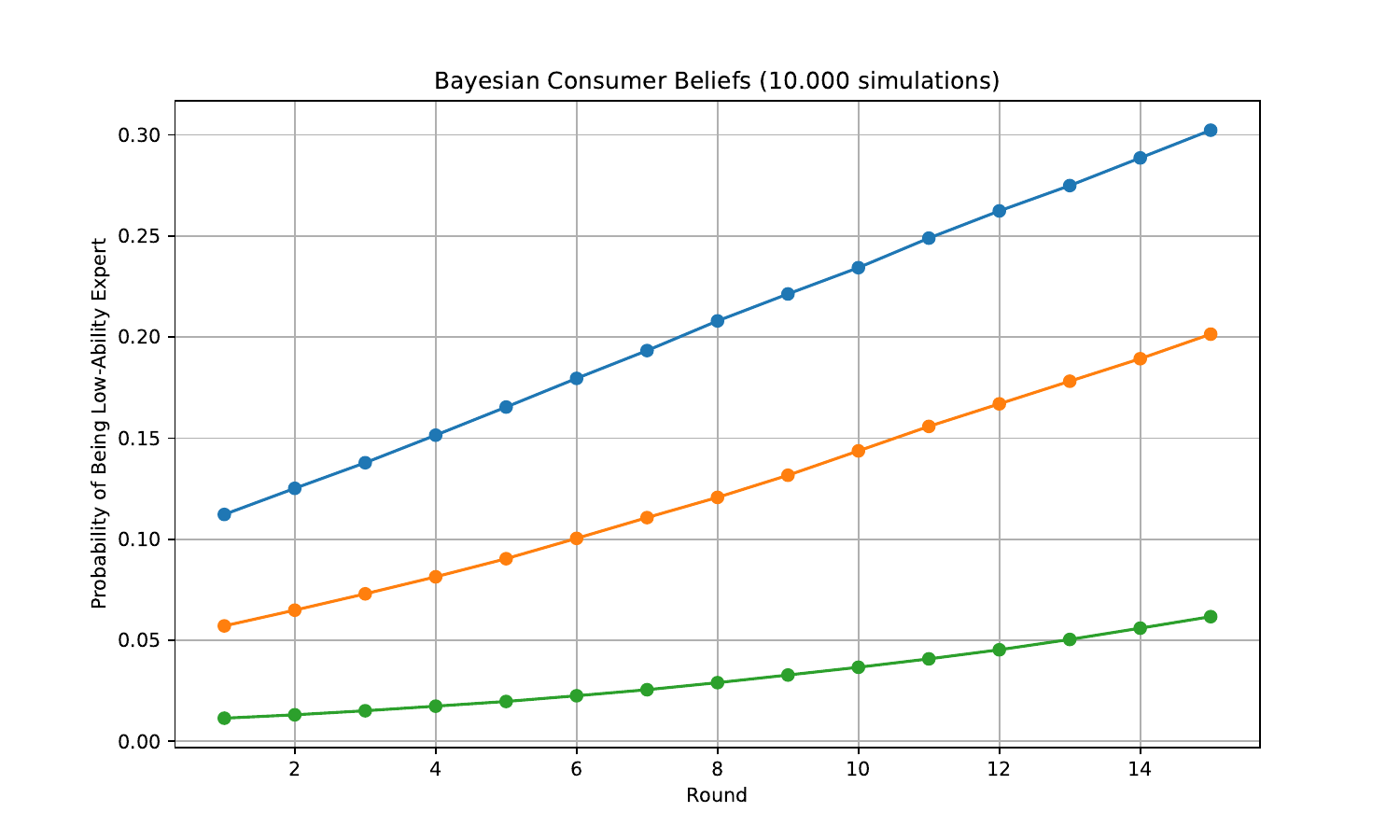}

    \caption{Left: Consumer Belief Updating with $Pr(L) = 0.4$ (blue) and $Pr(L) = 0.2$ (orange) when Low-Ability Experts only choose the HQT. Right: $Pr(L) \in \{0.1, 0.05, 0.01\}$. }
    \label{fig:bay_belief_hqt}
\end{figure}

\begin{figure}[h]
\centering
\includegraphics[width=0.3\textwidth]{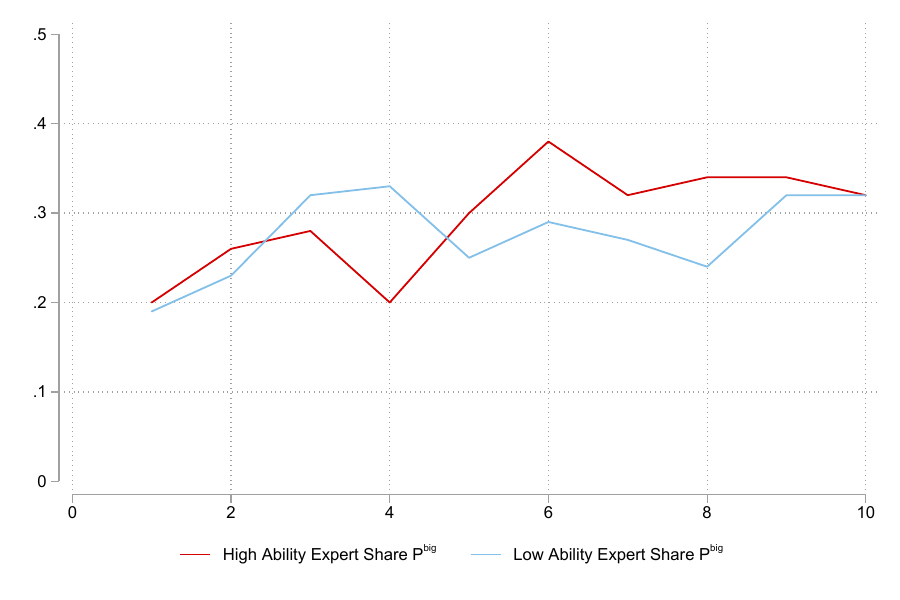}
\includegraphics[width=0.3\textwidth]{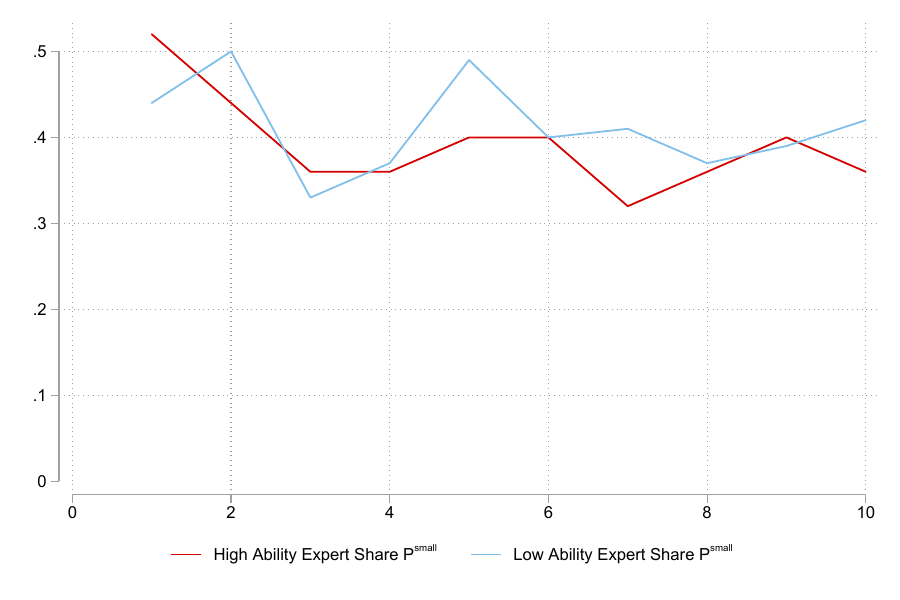}
\includegraphics[width=0.3\textwidth]{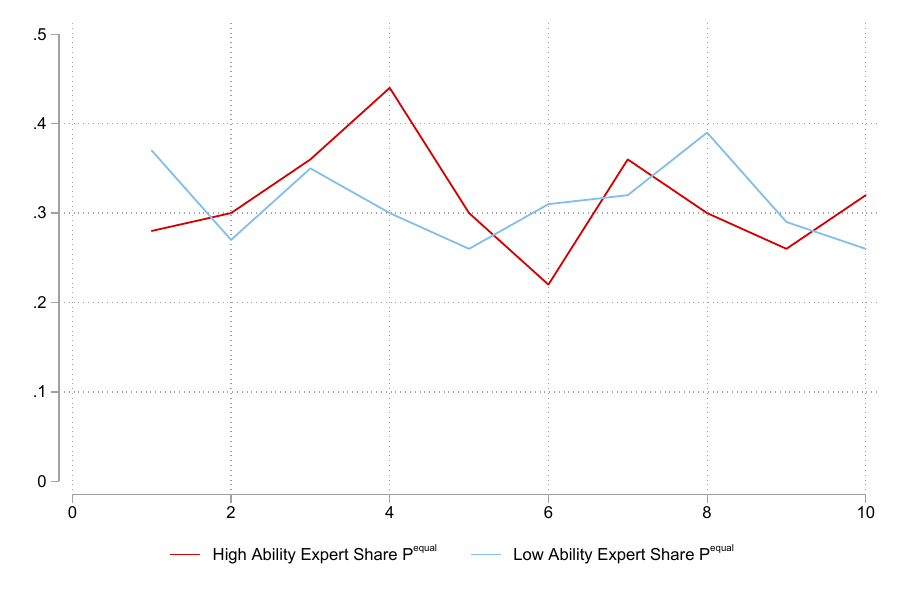}
\caption{Expert price setting $\vect{P^m}$, $\vect{P^s}$ and  $\vect{P^e}$ in Phase 1.}
\label{fig_prices_1}
\end{figure}

\begin{table}[H]\centering
\def\sym#1{\ifmmode^{#1}\else\(^{#1}\)\fi}
\caption{Consumer Approaching Behavior}
\begin{adjustbox}{max width=\textwidth}
\begin{tabular}{l*{3}{c}}
\toprule
&\multicolumn{1}{c}{\specialcell{$\vect{P^m}$}}&\multicolumn{1}{c}{\specialcell{$\vect{P^s}$}}&\multicolumn{1}{c}{\specialcell{$\vect{P^e}$}}  \\
\midrule

Invested &  -0.122\sym{**} &  0.159\sym{***}&  -0.088\sym{*}  \\
            & (0.039)         & (0.031)         & (0.040) \\

\midrule
$N$          &1,455       &1,665       &1,635   \\
\bottomrule
\end{tabular}
\label{tab:pooled}
\end{adjustbox}
\begin{minipage}{0.95\textwidth} 
\vspace*{0.09cm}
{\footnotesize 
This table reports marginal effects of a pooled fixed effects panel logistic regression. The dependent variable is a binary variable that equals 1 if the expert chooses the respective price vector. 
 \sym{*} \(p<0.05\), \sym{**} \(p<0.01\), \sym{***} \(p<0.001\)}
\end{minipage}
\end{table}

\begin{table}[h]
\centering
\caption{Frequencies of expert prices in Phase 2.}

\begin{tabular}{@{}p{2cm}p{3cm}p{3cm}p{3cm}p{3cm}@{}}
\toprule
          & \multicolumn{4}{c}{Expert} \\
\cmidrule(r){2-5}
& HA Invested & HA Not Invested & LA Invested & LA Not Invested \\
\cmidrule(r){2-5}
Algorithm & \begin{tabular}[c]{@{}l@{}} $\vect{P^m}$: 18.6\%\\  $\vect{P^s}$: 61.1\%\\ $\vect{P^e}$: 20.4\%\end{tabular} & \begin{tabular}[c]{@{}l@{}} $\vect{P^m}$: 20.2\%\\  $\vect{P^s}$: 31.2\%\\  $\vect{P^e}$: 48.6\%\end{tabular} & \begin{tabular}[c]{@{}l@{}} $\vect{P^m}$: 26.3\%\\ $\vect{P^s}$: 40.3\%\\ $\vect{P^e}$: 33.3\%\end{tabular} & \begin{tabular}[c]{@{}l@{}} $\vect{P^m}$: 39.5\%\\  $\vect{P^s}$: 29\%\\  $\vect{P^e}$: 31.5\%\end{tabular} \\
\addlinespace
Skill     & \begin{tabular}[c]{@{}l@{}} $\vect{P^m}$: 44.9\%\\  $\vect{P^s}$: 35.7\%\\  $\vect{P^e}$: 19.3\%\end{tabular} & \begin{tabular}[c]{@{}l@{}}$\vect{P^m}$: 50.4\%\\  $\vect{P^s}$: 24.1\%\\  $\vect{P^e}$: 25.5\%\end{tabular} & \begin{tabular}[c]{@{}l@{}} $\vect{P^m}$: 25.9\%\\  $\vect{P^s}$: 46.2\%\\  $\vect{P^e}$: 27.9\%\end{tabular} & \begin{tabular}[c]{@{}l@{}} $\vect{P^m}$: 27.1\%\\  $\vect{P^s}$: 44.9\%\\  $\vect{P^e}$: 27.9\%\end{tabular} \\
\addlinespace
\bottomrule
\end{tabular}
\label{tab_pvs}
\end{table}

\clearpage
\begin{table}[h]\centering
\def\sym#1{\ifmmode^{#1}\else\(^{#1}\)\fi}
\caption{Overtreatment Conditional on Expert Price Setting}
\begin{adjustbox}{max width=\textwidth}
\begin{tabular}{@{}lccccccccc@{}}
\toprule
 & \multicolumn{3}{c}{Phase 1} & \multicolumn{3}{c}{Skill} & \multicolumn{3}{c}{Algorithm} \\
\cmidrule(r){1-10} 
Overtreated & (1) & (2) & (3) & (1) & (2) & (3) & (1) & (2) & (3)   \\
\cmidrule(r){1-10}
$\vect{P^m}$ & 0.094\sym{*} & &  & 0.120 &  &  & 0.111 &  &   \\
& (0.047) &  & & (0.054) &  &  & (0.084) &  &  \\

$\vect{P^s}$ &  & -0.228\sym{***} &  &  & -0.232\sym{**} & &  & -0.199\sym{**} &  \\
&  & (0.041) &  &  & (0.054) &  &  & (0.066) &  \\

$\vect{P^e}$ &  & & 0.119\sym{*} &  &  & 0.081 & &  & 0.055  \\
&  &  & (0.046) &  &  & (0.084) &  & & (0.082) \\

\addlinespace
$N$ & 856 & 856 & 856 & 652 & 652 & 652 &  664 &  664 &  664  \\
\bottomrule
\end{tabular}
\label{tab:overtreated_reg}
\end{adjustbox}
\begin{minipage}{0.95\textwidth} 
\vspace*{0.09cm}
{\footnotesize 
This table reports marginal effects of panel logistic regressions using subject-level random effects and a cluster–robust VCE estimator at the matched group level (standard errors in parentheses). The dependent variable is a binary variable that equals 1 if the consumers experiences overtreatment.
 \sym{*} \(p<0.05\), \sym{**} \(p<0.01\), \sym{***} \(p<0.001\)}
\end{minipage}
\end{table}
\begin{table}[H]\centering
\def\sym#1{\ifmmode^{#1}\else\(^{#1}\)\fi}
\caption{Consumer Approaching Behavior}
\begin{adjustbox}{max width=\textwidth}
\begin{tabular}{l*{5}{c}}
\toprule
&\multicolumn{1}{c}{\specialcell{Phase 1}}&\multicolumn{1}{c}{\specialcell{Skill}}&\multicolumn{1}{c}{\specialcell{Algorithm}} &\multicolumn{1}{c}{\specialcell{Algorithm}} \\
\midrule

Undertreated &  -0.519\sym{**} &  -0.539\sym{***}&  -0.465\sym{*}  & -0.473\sym{*}  \\
            & (0.169)         & (0.212)         & (0.220)  & (0.221)  \\
Overtreated & 0.089 &  0.047 & 0.075 & 0.088 \\
        &   (0.098)   &  (0.097)    & (0.038)  &  (0.095) \\
 &         &     & &  \\         
$\vect{P^e}$ & Baseline &  Baseline & Baseline  & Baseline  \\    
 &         &     & &  \\ 
$\vect{P^m}$ & 0.574\sym{**} & 0.620\sym{**} &  1.174\sym{***} & 1.297\sym{**}  \\
 &    (0.174)   & (0.225)    &  (0.283) & (0.402)  \\
$\vect{P^s}$ & -0.770\sym{***} & -0.803\sym{***}&  -0.748\sym{***} & -0.661\sym{**}  \\
 &    (0.167)    & (0.215)      &  (0.138)  & (0.229)   \\
Not\_Invested &  & 0.059 & -0.262  &  -0.386 \\
 &         &   (0.276)   & (0.166) &  (0.209) \\ 
High-Ability &  &  &   &  0.905  \\
 &         &     & & (0.503) \\ 
Not\_Invested $\times$ $\vect{P^e}$ &  &  &   & 0.393 \\
 &         &       & & (0.375)  \\
High-Ability $\times$ $\vect{P^e}$ &  &  &   & -0.683 \\
 &         &       & & (0.363)  \\ 
Not\_Invested $\times$ High-Ability $\times$ $\vect{P^e}$ &  & &   & 0.849\sym{*} \\
 &         &      & & (0.421) \\
\\

\midrule
$N$          &1,350       &1,125       &1,125   & 1,125 \\
\bottomrule
\end{tabular}
\label{tab:expert_approached}
\end{adjustbox}
\begin{minipage}{0.95\textwidth} 
\vspace*{0.09cm}
{\footnotesize 
This table reports results of a panel ordered logistic regressions using subject-level random effects and a cluster–robust VCE estimator at the matched group level (standard errors in parentheses). The dependent variable is an ordinal variable that captures the number of consumers (0 - 3) who approached the expert in the current round. \textit{Undertreated} and \textit{Overtreated} are lagged variables (one round).
 \sym{*} \(p<0.05\), \sym{**} \(p<0.01\), \sym{***} \(p<0.001\)}
\end{minipage}
\end{table}

\clearpage
\begin{figure}[h]
    \centering
    \caption{Left: High-Ability Expert Investment Behavior conditional on incentivized beliefs that investments decrease consumer beliefs in high ability type (negative) or increase beliefs in high ability type (positive). Right: Non-incentivized beliefs.}
    \includegraphics[width=0.45\textwidth]{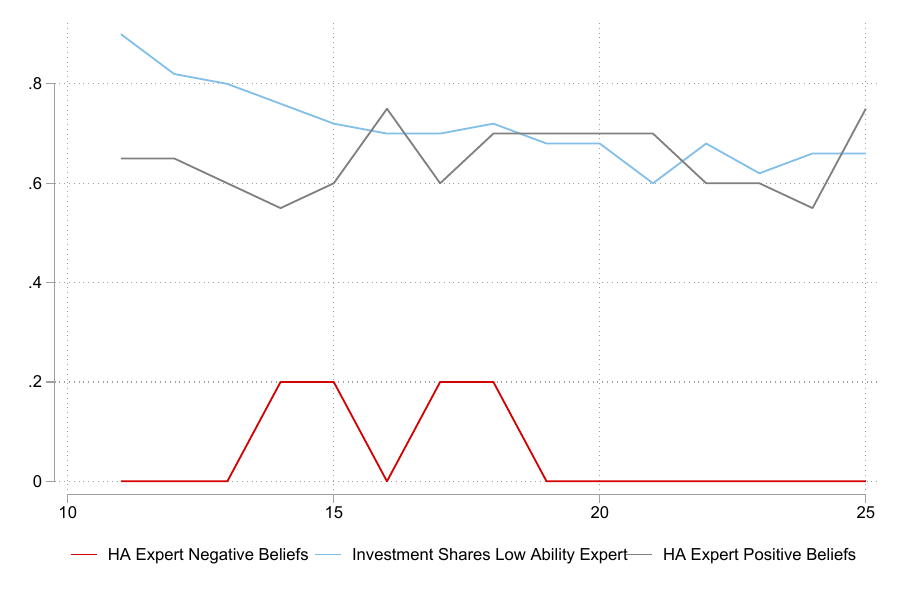}
    \includegraphics[width=0.45\textwidth]{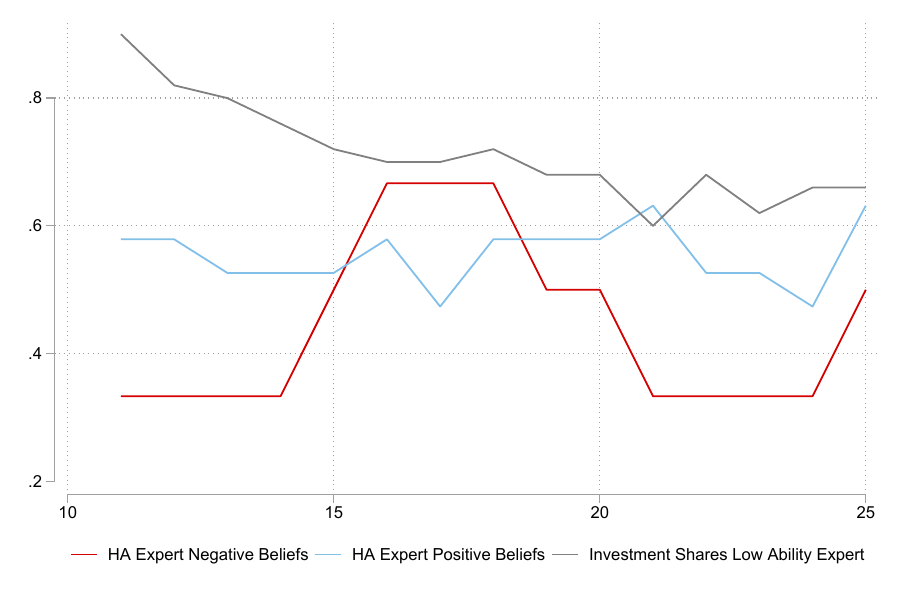}    
    \label{fig:e2_repeated_beliefs}
\end{figure}

\begin{figure}[h]
    \centering
    \caption{Left: Expert beliefs about how consumer beliefs about the expert's ability type change after the expert chooses to invest. Right: Consumer beliefs that an expert is high-ability after investing.}
    \includegraphics[width=0.45\textwidth]{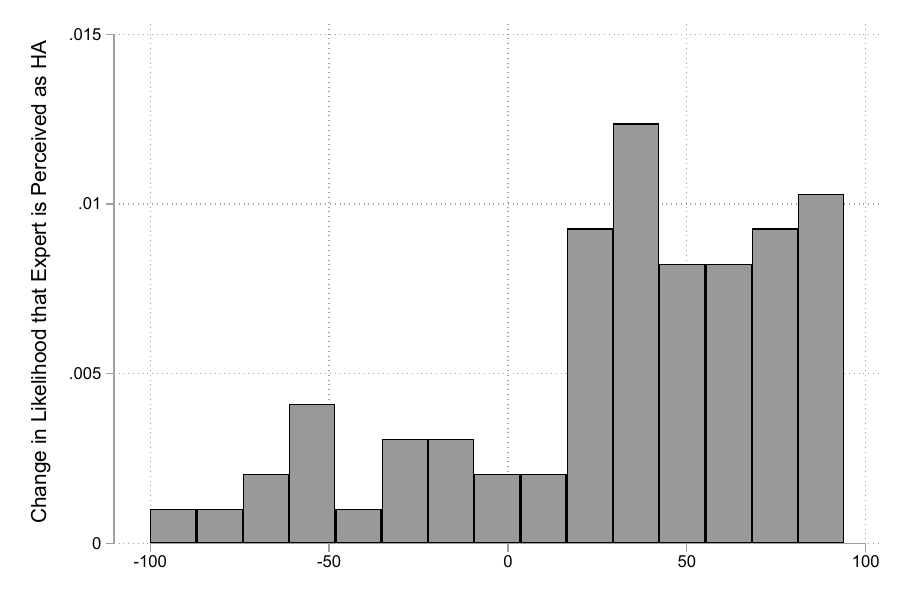}
    \includegraphics[width=0.45\textwidth]{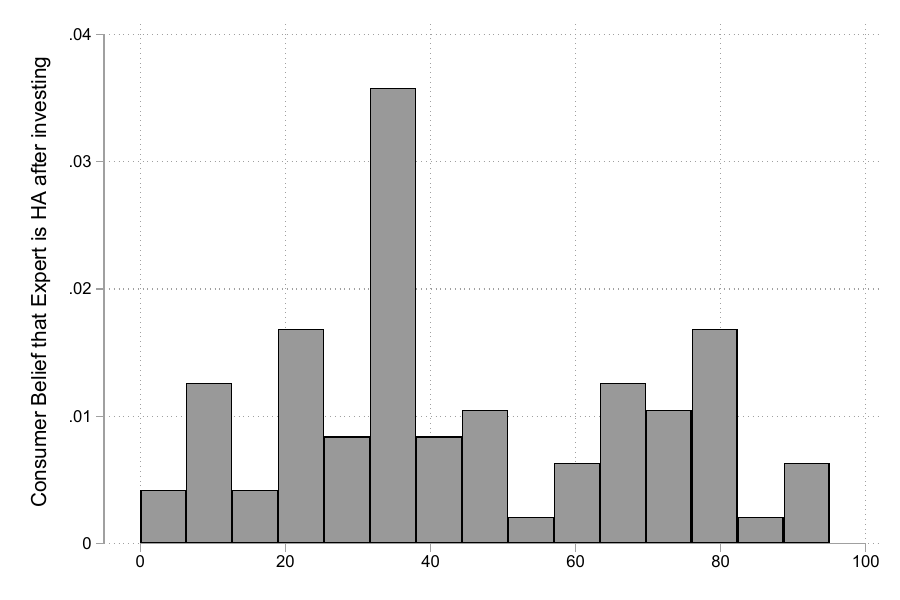}    
    \label{fig:e2_expertbeliefs}
\end{figure}

\begin{figure}[h]
    \centering
    \caption{Left: Consumer beliefs about the likelihood that an expert does not use the decision aid after renting it. Right: Consumer beliefs that investing expert does not use the decision aid correctly.}
    \includegraphics[width=0.45\textwidth]{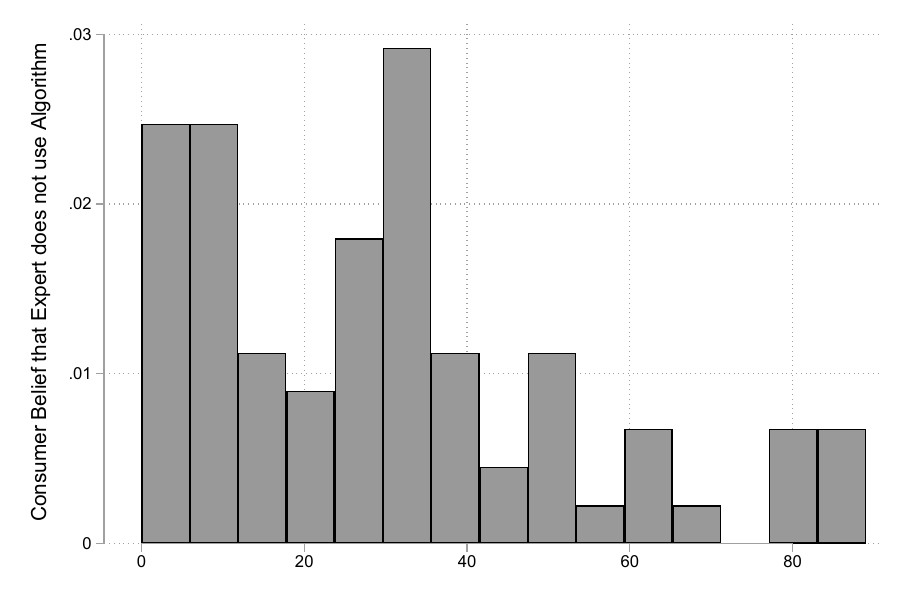}
    \includegraphics[width=0.45\textwidth]{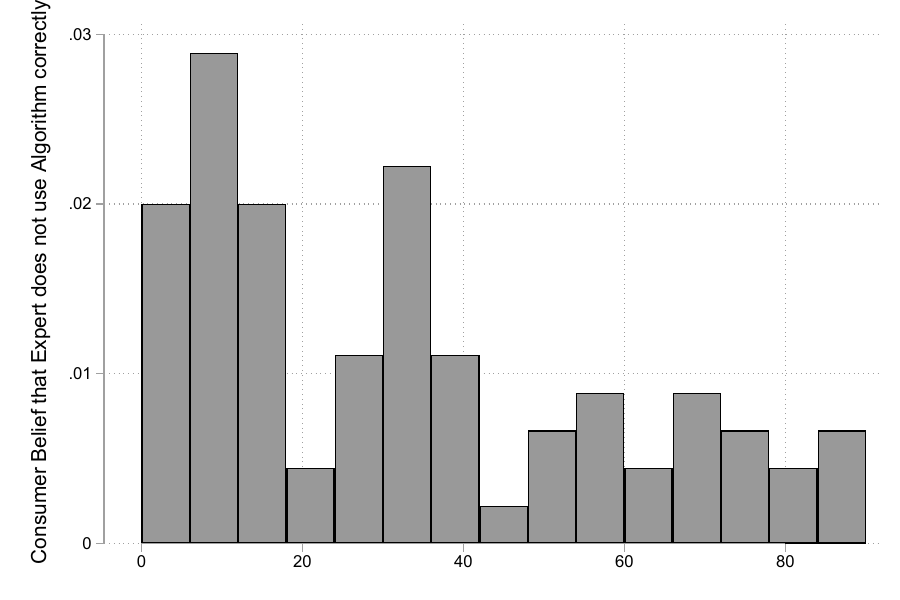}    
    \label{fig:e2_cons_beliefs}
\end{figure}

\begin{table}[h!]\centering
\def\sym#1{\ifmmode^{#1}\else\(^{#1}\)\fi}
\caption{Experiment 2: Expert Investment Behavior in Phase 2}
\begin{adjustbox}{max width=\textwidth}
\begin{tabular}{l*{5}{c}}
\toprule
&\multicolumn{1}{c}{\specialcell{Repeated}}&\multicolumn{1}{c}{\specialcell{Repeated}}&\multicolumn{1}{c}{\specialcell{One-Shot}} &\multicolumn{1}{c}{\specialcell{One-Shot}}\\
\midrule

High-Ability &  -0.181\sym{*}&  -0.179\sym{*}&  -0.1222  & -0.111 \\
 & (0.082)         & (0.079)         & (0.092)  & (0.093)  \\
Round &  &  -0.095\sym{*} & &    \\
 &     &  (0.037)      &  &   \\
Risk &  &  0.015&   & -0.000\\
 &       & (0.015)    &   & (0.017) \\
Female &  &  0.008&  & -0.095 \\
 &        & (0.074)      &   & (0.113)  \\
Age &  &  -0.004&   & 0.001\\
 &         & (0.004)      & & (0.004)  \\ 
\\

\midrule
$N$          &1125       &1125       &150& 150\\
\bottomrule
\end{tabular}
\label{tab:reg_invest_e2}
\end{adjustbox}
\begin{minipage}{0.95\textwidth} 
\vspace*{0.09cm}
{\footnotesize 
This table reports marginal effects of panel logistic regressions using subject-level random effects and a cluster–robust VCE estimator at the matched group level (for \textit{repeated}, standard errors in parentheses) and marginal effects of equivalent logistic regressions (\textit{one-shot}). The dependent variable is a binary variable that equals 1 if the expert invests into the new diagnostic technology.
 \sym{*} \(p<0.05\), \sym{**} \(p<0.01\), \sym{***} \(p<0.001\)}
\end{minipage}
\end{table}

\begin{figure}[h]
    \centering
    \caption{Expert investment shares in \textit{repeated} and \textit{one-shot} by ability type and belief type. A negative belief indicates that experts belief consumers will find an investing expert less likely to be of high ability type.}
    \includegraphics[width=\textwidth]{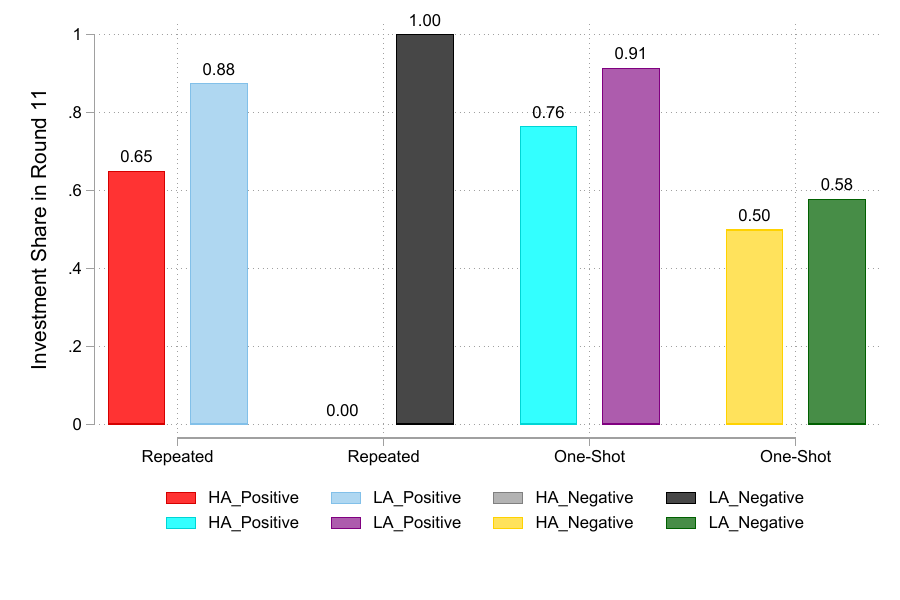}
    \label{fig:e2_all_beliefs}
\end{figure}

\begin{figure}[h]
    \centering
    \caption{\textit{One-Shot}. Left: Expert beliefs about how consumer beliefs about the expert's ability type change after an expert chooses to invest. Right: Consumer beliefs that an expert is high-ability after investing.}
    \includegraphics[width=0.45\textwidth]{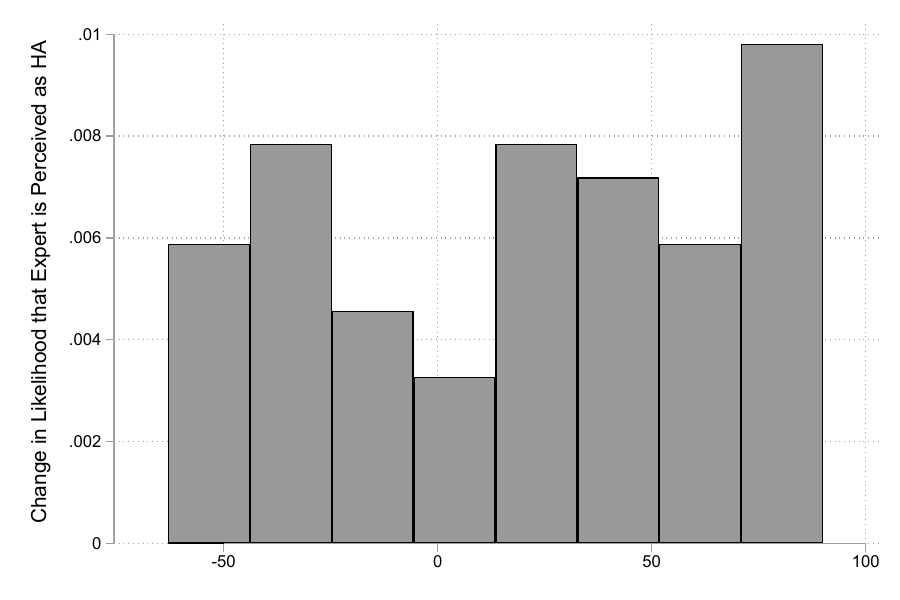}
    \includegraphics[width=0.45\textwidth]{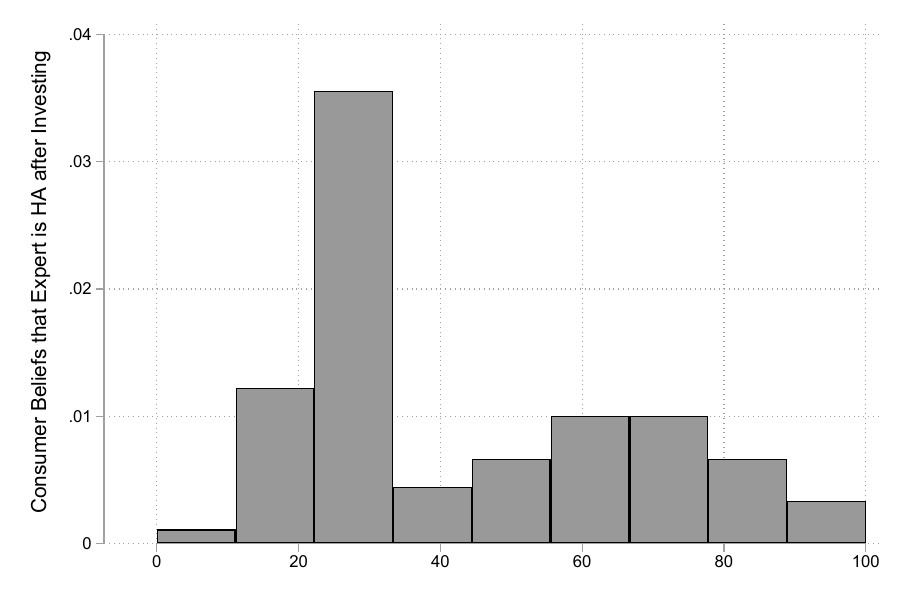}    
    \label{fig:e2_expertbeliefs_one}
\end{figure}

\begin{figure}[h]
    \centering
    \caption{\textit{One-Shot}. Left: Consumer beliefs about the likelihood that an expert does not use the decision aid after renting it. Right: Consumer beliefs that investing expert does not use the decision aid correctly.}
    \includegraphics[width=0.45\textwidth]{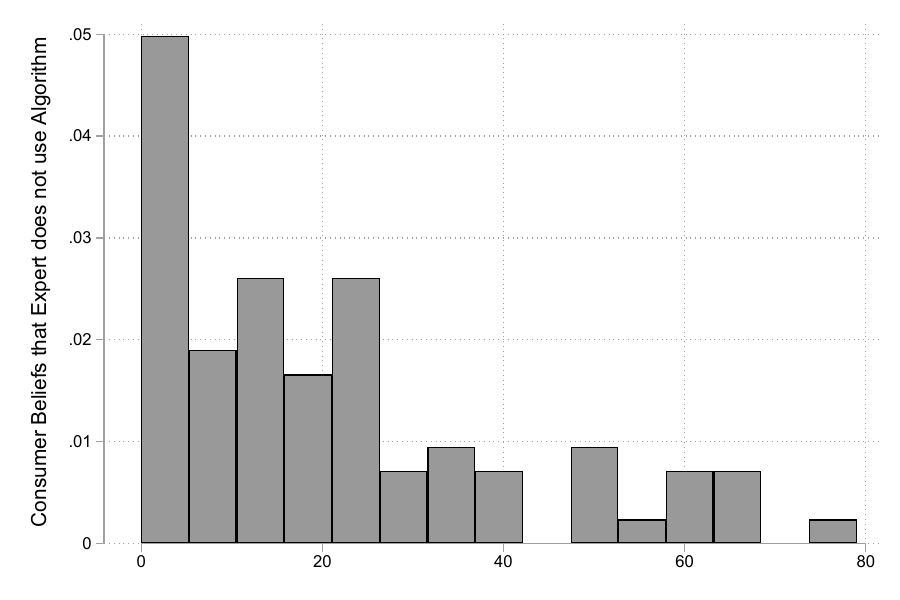}
    \includegraphics[width=0.45\textwidth]{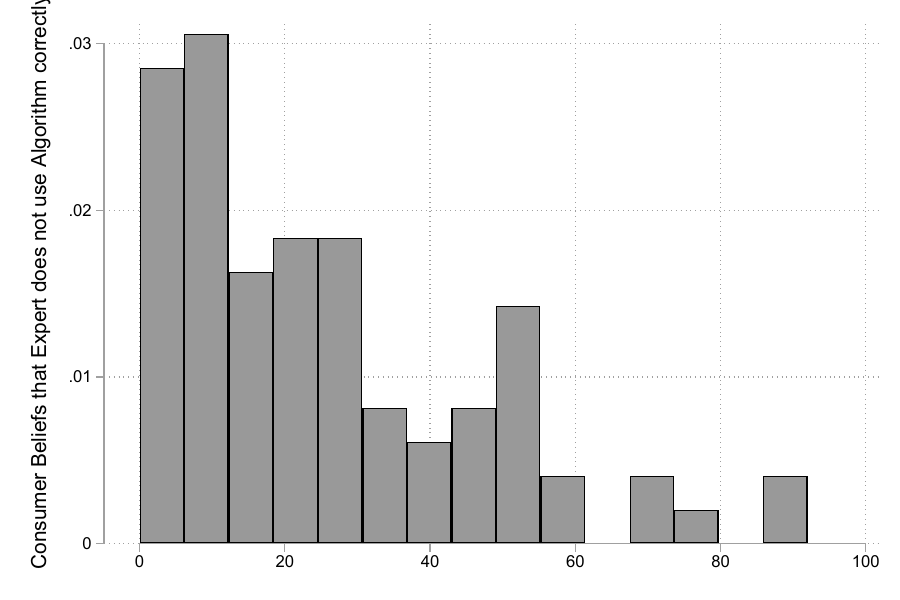}    
    \label{fig:e2_cons_beliefs_one}
\end{figure}

\end{document}